\documentclass[10pt]{article}

\usepackage[latin1]{inputenc}
\usepackage{latexsym}
\usepackage{amssymb}
\usepackage{a4}
\usepackage{epsfig,graphics,graphicx}
\usepackage{tabularx}
\usepackage{amsfonts}
\usepackage{amsmath}
\usepackage{pdflscape}

\usepackage{float}
\usepackage[dvipsnames]{xcolor}
\usepackage{colortbl}
\usepackage{color}
\usepackage{bm}

\usepackage[a4paper,top=2cm,bottom=2cm,left=1.5cm,right=1.5cm]{geometry}
\usepackage{graphicx}
\usepackage{rotating}

\newcommand{\footmsg}[1]{%
  \let\temp\thempfn%
  \def\thempfs{}
  \footnotetext{#1}
  \let\tempfn\temp}

\usepackage{tikz}

\usepackage{enumitem}

\usepackage[at]{easylist}

\usepackage{lipsum,graphicx,subcaption}
\usepackage{dutchcal}

\usepackage{fancyhdr}            
\fancypagestyle{plain}{%
  \fancyhead{}                                     
  \fancyfoot[CE,CO]{\thepage}       
  \fancyhead[CE,CO]{\textcolor[rgb]{0.64,0.15,0.15}{Published in \emph{International Journal of Solids and Structures} (2019) \textbf{176-177}, 1-18
  \\
doi: https://doi.org/10.1016/j.ijsolstr.2019.07.008}}
\setlength{\headheight}{14.5pt}}

\begin{document}

\newcommand{\beq}{\begin{equation}}
\newcommand{\eeq}{\end{equation}}
\newcommand{\lb}{\label}
\newcommand{\ph}{\phantom}
\newcommand{\bqear}{\begin{eqnarray}}
\newcommand{\eeqar}{\end{eqnarray}}
\newcommand{\barr}{\begin{array}}
\newcommand{\earr}{\end{array}}
\newcommand{\jump}{\parallel}
\newcommand{\Ehat}{\hat{E}}
\newcommand{\That}{\hat{\bf T}}
\newcommand{\Ahat}{\hat{A}}
\newcommand{\chat}{\hat{c}}
\newcommand{\shat}{\hat{s}}
\newcommand{\khat}{\hat{k}}
\newcommand{\muhat}{\hat{\mu}}
\newcommand{\mc}{M^{\scriptscriptstyle C}}
\newcommand{\mei}{M^{\scriptscriptstyle M,EI}}
\newcommand{\mec}{M^{\scriptscriptstyle M,EC}}
\newcommand{\hbeta}{{\hat{\beta}}}
\newcommand{\rec}[2]{\left( #1 #2 \ds{\frac{1}{#1}}\right)}
\newcommand{\rep}[2]{\left( {#1}^2 #2 \ds{\frac{1}{{#1}^2}}\right)}
\newcommand{\derp}[2]{\ds{\frac {\partial #1}{\partial #2}}}
\newcommand{\derpn}[3]{\ds{\frac {\partial^{#3}#1}{\partial #2^{#3}}}}
\newcommand{\dert}[2]{\ds{\frac {d #1}{d #2}}}
\newcommand{\dertn}[3]{\ds{\frac {d^{#3} #1}{d #2^{#3}}}}

\def\c{{\circ}}
\def\bob{{\, \underline{\overline{\otimes}} \,}}
\def\ob{{\, \underline{\otimes} \,}}
\def\scalp{\mbox{\boldmath$\, \cdot \, $}}
\def\scalpp{\mbox{\boldmath$\, : \, $}}
\def\gdp{\makebox{\raisebox{-.215ex}{$\Box$}\hspace{-.778em}$\times$}}
\def\daa{\makebox{\raisebox{-.050ex}{$-$}\hspace{-.550em}$: ~$}}
\def\mK{\mbox{${\mathcal{K}}$}}
\def\cK{\mbox{${\mathbb {K}}$}}

\def\Xint#1{\mathchoice
   {\XXint\displaystyle\textstyle{#1}}%
   {\XXint\textstyle\scriptstyle{#1}}%
   {\XXint\scriptstyle\scriptscriptstyle{#1}}%
   {\XXint\scriptscriptstyle\scriptscriptstyle{#1}}%
   \!\int}
\def\XXint#1#2#3{{\setbox0=\hbox{$#1{#2#3}{\int}$}
     \vcenter{\hbox{$#2#3$}}\kern-.5\wd0}}
\def\ddashint{\Xint=}
\def\fpint{\Xint=}
\def\dashint{\Xint-}
\def\cpvint{\Xint-}
\def\intl{\int\limits}
\def\cpvintl{\cpvint\limits}
\def\fpintl{\fpint\limits}
\def\ointl{\oint\limits}
\def\bA{{\bf A}}
\def\bB{{\bf B}}
\def\bC{{\bf C}}
\def\bD{{\bf D}}
\def\bE{{\bf E}}
\def\bF{{\bf F}}
\def\bG{{\bf G}}
\def\bH{{\bf H}}
\def\bI{{\bf I}}
\def\bJ{{\bf J}}
\def\bK{{\bf K}}
\def\bL{{\bf L}}
\def\bM{{\bf M}}
\def\bN{{\bf N}}
\def\bO{{\bf O}}
\def\b0{{\bf 0}}
\def\bP{{\bf P}}
\def\bQ{{\bf Q}}
\def\bR{{\bf R}}
\def\bS{{\bf S}}
\def\bT{{\bf T}}
\def\bU{{\bf U}}
\def\bV{{\bf V}}
\def\bW{{\bf W}}
\def\bX{{\bf X}}
\def\bY{{\bf Y}}
\def\bZ{{\bf Z}}

\def\ba{{\bf a}}
\def\bb{{\bf b}}
\def\bc{{\bf c}}
\def\bd{{\bf d}}
\def\be{{\bf e}}
\def\bbf{{\bf f}}
\def\bg{{\bf g}}
\def\bh{{\bf h}}
\def\bi{{\bf i}}
\def\bj{{\bf j}}
\def\bk{{\bf k}}
\def\bl{{\bf l}}
\def\bm{{\bf m}}
\def\bn{{\bf n}}
\def\bo{{\bf o}}
\def\bp{{\bf p}}
\def\bq{{\bf q}}
\def\br{{\bf r}}
\def\bs{{\bf s}}
\def\bt{{\bf t}}
\def\bu{{\bf u}}
\def\bv{{\bf v}}
\def\bw{{\bf w}}
\def\bx{{\bf x}}
\def\by{{\bf y}}
\def\bz{{\bf z}}

\def\bxi{\mbox{\boldmath${\xi}$}}
\def\balpha{\mbox{\boldmath${\alpha}$}}
\def\bbeta{\mbox{\boldmath${\beta}$}}
\def\bgamma{\mbox{\boldmath${\gamma}$}}
\def\bepsilon{\mbox{\boldmath${\epsilon}$}}
\def\bvarepsilon{\mbox{\boldmath${\varepsilon}$}}
\def\bomega{\mbox{\boldmath${\omega}$}}
\def\bphi{\mbox{\boldmath${\phi}$}}
\def\bsigma{\mbox{\boldmath${\sigma}$}}
\def\bfeta{\mbox{\boldmath${\eta}$}}
\def\bDelta{\mbox{\boldmath${\Delta}$}}
\def\bdelta{\mbox{\boldmath${\delta}$}}
\def\btau{\mbox{\boldmath $\tau$}}
\def\bmu{\mbox{\boldmath $\mu$}}
\def\bchi{\mbox{\boldmath $\chi$}}
\def\bnabla{\mbox{\boldmath $\nabla$}}
\def\tr{{\rm tr}}
\def\dev{{\rm dev}}
\def\div{{\rm div}}
\def\Div{{\rm Div}}
\def\Grad{{\rm Grad}}
\def\grad{{\rm grad}}
\def\Lin{{\rm Lin}}
\def\Sym{{\rm Sym}}
\def\Skw{{\rm Skew}}
\def\abs{{\rm abs}}
\def\Re{{\rm Re}}
\def\Im{{\rm Im}}
\def\capB{\mbox{\boldmath${\mathsf B}$}}
\def\capC{\mbox{\boldmath${\mathsf C}$}}
\def\capD{\mbox{\boldmath${\mathsf D}$}}
\def\capE{\mbox{\boldmath${\mathsf E}$}}
\def\capG{\mbox{\boldmath${\mathsf G}$}}
\def\tcapG{\tilde{\capG}}
\def\capH{\mbox{\boldmath${\mathsf H}$}}
\def\capK{\mbox{\boldmath${\mathsf K}$}}
\def\capL{\mbox{\boldmath${\mathsf L}$}}
\def\capM{\mbox{\boldmath${\mathsf M}$}}
\def\capR{\mbox{\boldmath${\mathsf R}$}}
\def\capW{\mbox{\boldmath${\mathsf W}$}}

\def\i{\mbox{${\mathrm i}$}}
\def\mC{\mbox{\boldmath${\mathcal C}$}}
\def\mB{\mbox{${\mathcal B}$}}
\def\mE{\mbox{${\mathcal{E}}$}}
\def\mL{\mbox{${\mathcal{L}}$}}
\def\mK{\mbox{${\mathcal{K}}$}}
\def\mV{\mbox{${\mathcal{V}}$}}
\def\C{\mbox{\boldmath${\mathcal C}$}}
\def\E{\mbox{\boldmath${\mathcal E}$}}

\def\AAM{{\it Advances in Applied Mechanics }}
\def\ACME{{\it Arch. Comput. Meth. Engng.}}
\def\ARMA{{\it Arch. Rat. Mech. Analysis}}
\def\AMR{{\it Appl. Mech. Rev.}}
\def\ASCEEM{{\it ASCE J. Eng. Mech.}}
\def\ACTA{{\it Acta Mater.}}
\def\CMAME {{\it Comput. Meth. Appl. Mech. Engrg.}}
\def\CRAS{{\it C. R. Acad. Sci. Paris}}
\def\CRM{{\it Comptes Rendus M\'ecanique}}
\def\EFM{{\it Eng. Fracture Mechanics}}
\def\EJMA{{\it Eur.~J.~Mechanics-A/Solids}}
\def\IJES{{\it Int. J. Eng. Sci.}}
\def\IJF{{\it Int. J. Fracture}}
\def\IJMS{{\it Int. J. Mech. Sci.}}
\def\IJNAMG{{\it Int. J. Numer. Anal. Meth. Geomech.}}
\def\IJP{{\it Int. J. Plasticity}}
\def\IJSS{{\it Int. J. Solids Structures}}
\def\IngA{{\it Ing. Archiv}}
\def\JAM{{\it J. Appl. Mech.}}
\def\JAP{{\it J. Appl. Phys.}}
\def\JAE{{\it J. Aerospace Eng.}}
\def\JE{{\it J. Elasticity}}
\def\JM{{\it J. de M\'ecanique}}
\def\JMPS{{\it J. Mech. Phys. Solids}}
\def\JSV{{\it J. Sound and Vibration}}
\def\MACRO{{\it Macromolecules}}
\def\MMT{{\it Mech. Mach. Th.}}
\def\MOM{{\it Mech. Materials}}
\def\MMS{{\it Math. Mech. Solids}}
\def\MMT{{\it Metall. Mater. Trans. A}}
\def\MPCPS{{\it Math. Proc. Camb. Phil. Soc.}}
\def\MSE{{\it Mater. Sci. Eng.}}
\def\NATURE{{\it Nature}}
\def\NATUREM{{\it Nature Mater.}}
\def\PHIL{{\it Phil. Trans. R. Soc.}}
\def\PMPS{{\it Proc. Math. Phys. Soc.}}
\def\PNAS{{\it Proc. Nat. Acad. Sci.}}
\def\PRE{{\it Phys. Rev. E}}
\def\PRL{{\it Phys. Rev. Letters}}
\def\PRSL{{\it Proc. R. Soc.}}
\def\ROCK{{\it Rock Mech. and Rock Eng.}}
\def\QAM{{\it Quart. Appl. Math.}}
\def\QJMAM{{\it Quart. J. Mech. Appl. Math.}}
\def\SCIENCE{{\it Science}}
\def\SCRMAT{{\it Scripta Mater.}}
\def\SM{{\it Scripta Metall.}}
\def\ZAMM{{\it Z. Angew. Math. Mech.}}
\def\ZAMP{{\it Z. Angew. Math. Phys.}}
\def\ZVDI{{\it Z. Verein. Deut. Ing.}}

\title{Identification of second-gradient elastic materials from planar hexagonal lattices. Part I: Analytical derivation of equivalent constitutive tensors} 

\author{G. Rizzi, F. Dal Corso, D. Veber, and D. Bigoni$^1$ \\ 
DICAM, University of Trento\\
via Mesiano 77, I-38123 Trento, Italy
}
\date{}
\maketitle
\footnotetext[1]{Corresponding author: Davide Bigoni 
fax: +39 0461 282599; tel.: +39 0461 282507; web-site:
http://www.ing.unitn.it/$\sim$bigoni/; e-mail:
bigoni@ing.unitn.it}

\begin{abstract}
A second-gradient elastic (SGE) material is identified as the homogeneous solid equivalent to a periodic planar lattice characterized by  a hexagonal unit cell, which is made up of three different linear elastic bars ordered in a way that the hexagonal symmetry is preserved and hinged at each node, so that the lattice bars are subject to pure axial strain while bending is excluded.
Closed form-expressions for the identified non-local constitutive parameters are obtained by imposing the elastic energy equivalence between the lattice and the continuum solid, under remote displacement conditions having a dominant quadratic component.
In order to generate equilibrated stresses, in the absence of body forces, the applied remote displacement has to be constrained, thus leading to the identification in a \lq condensed' form of a higher-order solid, so that imposition of further constraints becomes necessary to fully quantify the equivalent continuum. The identified SGE material reduces to an equivalent Cauchy material only in the limit of vanishing side length of hexagonal unit cell. 
The analysis of positive definiteness and symmetry of the equivalent constitutive tensors, the derivation of the second-gradient elastic properties from those of the higher-order solid in the \lq condensed' definition, and a numerical validation of the identification scheme are  deferred to Part II of this study.
\end{abstract}

\vspace{10 mm}
Keywords: Strain gradient elasticity; non-local material; non-centrosymmetric material; internal length; homogenization

\section{Introduction}

Research on the equivalence between spring networks and continuous bodies was initiated by Cauchy~\cite{cauchy1828} and later continued by Born~\cite{born1954dynamical}, with the purpose of determining the overall elastic properties of crystalline materials subject to small strain. Considering a linear interaction between atoms, a material is modelled as a three-dimensional linear elastic lattice, with elements only subject to axial deformation. This is the so-called \lq Cauchy-Born rule', which yields the \lq rari-constant' theory of elasticity, relating the elastic property of a solid to the interactions between its atoms or molecules. 

\begin{figure}[H]
	\centering
	\includegraphics[width=1\textwidth]{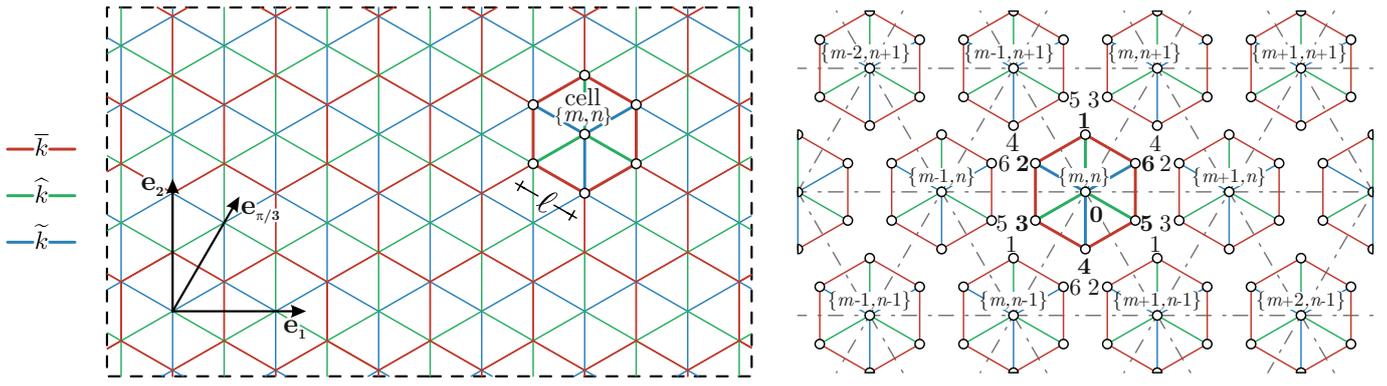}
	\caption{(Left) A planar  lattice obtained as the periodic repetition of a hexagonal unit cell (with side length $\ell$) made up of linear elastic bars, characterized by three stiffnesses $\overline{k}$ (red bars), $\widehat{k}$ (green bars), and $\widetilde{k}$ (blue bars). The bars are connected through hinge joints, so that 
	only axial strain is present and bending is excluded. Reference systems are also reported. (Right) Explosion of the hexagonal lattice displaying the cell and node nomenclature and highlighting how the perimeter nodes are shared among adjacent cells.}	\label{fig:intro_lattice}
\end{figure}

Over the years, the approach has been extended to evaluate mechanical characteristics such as Young modulus, Poisson's ratio and normal modes of vibration for a number of geometrically different networks   \cite{Genoese2018,keating1966effect,kirkwood1939skeletal,latture,neumann1975equations}. With reference to a hexagonal lattice, composed of linearly elastic bars pinned to each other (so that bending effects are excluded) and characterized by three different values of stiffness, as reported in Fig. \ref{fig:intro_lattice}, Day et al.~\cite{day1992elastic,snyder1992elastic} have shown that the overall behaviour of this lattice may be modelled through an  equivalent isotropic Cauchy linear elastic solid defined by the elastic bulk $K$ and shear $\mu$ moduli given by
\begin{equation}
K =\dfrac{\overline{k} + \widehat{k} + \widetilde{k}}{\sqrt{12}}, ~~~~ \mu = \sqrt{\dfrac{27}{16}}\left( \dfrac{1}{\overline{k}} + \dfrac{1}{\widehat{k}} + \dfrac{1}{\widetilde{k}} \right)^{-1},
\label{eq:const_day}
\end{equation}
where $\overline{k}$, $\widehat{k}$ and $\widetilde{k}$ are the three in-plane bars' stiffnesses (so that their dimension is a force per unit out-of-plane thickness divided by a length) defining the hexagonal lattice. 

The goal of the present research is to extend the theory developed by Day et al.~\cite{day1992elastic,snyder1992elastic} towards a higher-order  approximation for the elastic material equivalent to the hexagonal lattice, showing nonlocal effects related  to the four parameters defining the lattice properties at the micro-scale, the hexagon side length $\ell$ and the   stiffnesses $\overline{k}$, $\widehat{k}$ and $\widetilde{k}$. 

Phenomenological constitutive theories, used to model materials of engineering relevance, were traditionally assumed to be local, or, in other words, did not comprise any internal characteristic length. 
Recently, experimental observations at the micro- and nano-scale have evidenced size-effects \cite{BEVERIDGE2013246, buechner2003size, lakes1986experimental, WASEEM2013148}, which cannot be described 
with local constitutive models. Therefore, an enhanced modelling has been introduced, which becomes particularly useful when large strain gradient are involved, as in contact mechanics \cite{gourgiotis2016analysis,zisis2015contact} indentation processes \cite{begley1998mechanics,danas2012size}, fracture \cite{gourgiotis2014steady,piccolroaz2012mode}, and shear band formation \cite{ dal2011stability, sluys1993wave}.

Several authors \cite{Seppecher2018,askar1968structural,
bacigalupo2012computational,bacigalupo2014second2,le2013homogenization,
ostoja2002lattice,Shi1995,SPADONI2012156,warren2002three}  have proposed non-classical continuum models to treat lattice structures involving beam-type interactions.
For these lattices, non-local effects emerge as the response to non simple interactions between material points, generated, for example, when rotational springs are used~\cite{suiker2001comparison}.

The primary goal of the present study is the determination of the non-local response of lattices (having elements only subject to axial forces), which has been scarcely considered so far (an example is the case of pantographic trusses \cite{Seppecher2011}).  In particular, it will be shown that a hexagonal lattice structure with axially-deformable bars can be identified with a \lq form I' Mindlin elastic material, a special type of second-gradient elastic law \cite{mindlin1964micro}. 

%

The present article is organized as follows. 
After the kinematics and the equilibrium of the hexagonal lattice (Fig. \ref{fig:intro_lattice})  is introduced (Sect. \ref{hexa}), the quadratic remote displacement conditions, plus the additional terms needed to enforce equilibrium, are presented in Sect. \ref{qqua}. The   homogeneous Second Gradient Elastic ($\mathsf{SGE}$) solid equivalent to lattice is identified in Sect. \ref{identi}. In particular, by imposing an elastic energy matching, closed-form expressions for the higher-order tensors are derived. 
As a consequence of the fact that the energy matching is imposed under the condition that the applied displacement field 
generates equilibrated stress states, only a \lq condensed' form of the constitutive equations is determined for the $\mathsf{SGE}$ solid. 
As a conclusion, it is shown that the elastic second-gradient solid equivalent to the lattice structure exhibits 
non-locality, anisotropy, and non-centro-symmetry (despite the fact that the equivalent Cauchy material, derived on linear displacement fields, is local, isotropic, and centro-symmetric). 
Important issues related to:  the analysis of (i.) positive definiteness and (ii.) symmetry of the equivalent material, (iii.) the  derivation of the full $\mathsf{SGE}$ solid from the properties of the \lq condensed' one, and (iv.) the validation of the derived second-gradient model are deferred to Part II \cite{rizzipt2} of this study.

\section{The hexagonal lattice}\label{hexa}

\subsection{Preliminaries: the periodic structure and its elastic equilibrium}

An infinite periodic lattice (Fig.~\ref{fig:intro_lattice}, left), defined in the plane containing the orthonormal basis $\be_1$--$\be_2$,  is considered as the repetition of a hexagonal unit cell, which will eventually be 
identified with a representative volume element (RVE) of an equivalent continuum.
The hexagonal cell is regular and has side of length $\ell$, it is characterized by linear elastic bars with three different values of axial stiffnesses,
namely, $\overline{k}$,$ \widehat{k}$, and $\widetilde{k}$, distributed according to the scheme reported in Fig.~\ref{fig:intro_lattice}, which preserves the hexagonal symmetry.
Therefore,  a total of six bars (two groups of three bars having the same stiffness) converge at each hinge node of the lattice.

Among the three tessellations equivalent for the realization of the periodic lattice, the one is chosen for which the unit cell has its center defined by the convergence of the bars of stiffness $\widehat{k}$ and $\widetilde{k}$, while the other bars of stiffness $\overline{k}$ define the hexagon perimeter. 
Each node of the cell is denoted by the index $i=\left\{0,1,2,3,4,5,6\right\}$ and each cell is singled out by the integers $\left\{m,n\right\}\in\mathbb{Z}$, which determine the cell  position with reference respectively  to the non-orthogonal directions $\be_1$ and $\be_{\pi/3}=1/2\be_1+\sqrt{3}/2\be_2$, see Fig.~\ref{fig:intro_lattice} .
It follows that the position $\bx^{(m,n|i)}$ of the  $i$-th node of the $\left\{m,n\right\}$ cell 
can be described with reference to the central node ($i=0$) position $\textbf{x}^{(m,n|0)}$ through the following expression
\begin{equation}
\bx^{(m,n|i)} = \bx^{(m,n|0)} + \ell \bg^{(i)},
\label{eq:Posizio1} 
\end{equation}
where $\bg^{(i)} $ defines the direction spanning from the central node to the $i$-th node,
\begin{equation}
\bg^{(i)} = (1-\delta_{i0})\left\{-\sin\left[\frac{\pi(i-1)}{3} \right]\be_1+\cos\left[\frac{\pi(i-1)}{3}\right] \be_2\right\},
\label{eq:unitevectorG}
\end{equation}
in which the index $i$ is not summed and   the Kronecker delta $\delta_{i0}$ is defined to include the null index value, so that $\delta_{00}=1$ while $\delta_{i0}=0$ 
for every $i\neq 0$. 
From the definition expressed by Eq.~(\ref{eq:unitevectorG}), it follows that the vector $\bg^{(i)}$ has unit modulus for every $i\neq0$, while it vanishes when $i=0$ (central node),
\begin{equation}
\begin{array}{lll}
\bg^{(0)}=\b0, \quad|\bg^{(i)}|=1, \qquad \mbox{for} \qquad i=1,2,...,6.
\end{array}
\end{equation}
 Furthermore, due to the RVE symmetry, the unit vectors  $\bg^{(i)} $  satisfy the following property 
\begin{equation}
\bg^{(i)} = -\bg^{(i+3)},\qquad i=1,2,3,
\end{equation}
and  the following combination of the unit vectors $\bg^{(1)}$, $\bg^{(5)}$, and $\bg^{(6)}$ provides the unit vectors $\be_1$ and $\be_{\pi/3}$
\begin{equation}
\be_1=\frac{\bg^{(5)}+\bg^{(6)}}{\sqrt{3}},\qquad
\be_{\pi/3}=\frac{\bg^{(1)}+\bg^{(6)}}{\sqrt{3}}.
\end{equation}

Considering the definition of the unit vector $\bg^{(i)}$, Eq.~(\ref{eq:unitevectorG}), the position $\textbf{x}^{(m,n|0)}$  of the  central node of the cell $\left\{m,n\right\}$  can be expressed with reference to the position $\bx^{(0,0|0)}$
of the central node of the cell $\left\{m,n\right\}=\left\{0,0\right\}$ as 
\begin{equation}
\bx^{(m,n|0)}=\bx^{(0,0|0)}+ \ell\,\left[m \, \left(\bg^{(5)}+\bg^{(6)}\right) +n\, \left(\bg^{(1)}+\bg^{(6)}\right)\right],
\label{eq:36}
\end{equation}
so that the position $\bx^{(m,n|i)}$ of each node $i$ of every $\left\{m,n\right\}$ cell, expressed by Eq.(\ref{eq:Posizio1}), can be finally reduced to 
\begin{equation}
\bx^{(m,n|i)}=\bx^{(0,0|0)}+ \ell\left[\textbf{g}^{(i)} + m \, \left(\bg^{(5)}+\bg^{(6)}\right) +n\, \left(\bg^{(1)}+\bg^{(6)}\right)  \right]. 
\label{eq:37}
\end{equation}

All the perimeter nodes ($i=\left\{1,2,...,6\right\}$) join three adjacent hexagonal cells, Fig.~\ref{fig:intro_lattice} (right), so that the following identities hold
\begin{equation}
\begin{array}{llllll}
&\bx^{(m,n|1)}  =  \bx^{(m,n+1|3)}   = \bx^{(m-1,n+1|5)},\qquad
&\bx^{(m,n|2)}  =  \bx^{(m-1,n+1|4)} = \bx^{(m-1,n|6)}, \\[3mm]
&\bx^{(m,n|3)}  =  \bx^{(m-1,n|5)}   = \bx^{(m,n-1|1)},\qquad
&\bx^{(m,n|4)}  =  \bx^{(m,n-1|6)}   = \bx^{(m+1,n-1|2)}, \\[3mm]
&\bx^{(m,n|5)}  =  \bx^{(m+1,n-1|1)} = \bx^{(m+1,n|3)},\qquad
&\bx^{(m,n|6)}  =  \bx^{(m+1,n|2)}   = \bx^{(m,n+1|4)}.
\end{array}
\label{eq:38}
\end{equation}
Introducing $\bu^{(m,n|i)}$ as the (small) displacement of the $i$-th node belonging to the cell $\left\{m,n\right\}$, which according to Eq. Eq.~(\ref{eq:38}) satisfies 
\begin{equation}
\begin{array}{llllll}
&\bu^{(m,n|1)}  =  \bu^{(m,n+1|3)}   = \bu^{(m-1,n+1|5)},\qquad
&\bu^{(m,n|2)}  =  \bu^{(m-1,n+1|4)} = \bu^{(m-1,n|6)}, \\[3mm]
&\bu^{(m,n|3)}  =  \bu^{(m-1,n|5)}   = \bu^{(m,n-1|1)},\qquad
&\bu^{(m,n|4)}  =  \bu^{(m,n-1|6)}   = \bu^{(m+1,n-1|2)}, \\[3mm]
&\bu^{(m,n|5)}  =  \bu^{(m+1,n-1|1)} = \bu^{(m+1,n|3)},\qquad
&\bu^{(m,n|6)}  =  \bu^{(m+1,n|2)}   = \bu^{(m,n+1|4)},
\end{array}
\label{eq:39}
\end{equation}
the elongation $E^{(m,n|i,j)}$ of the bar connecting the nodes $i$ and $j$ (with $i\neq j$) is given by
\begin{equation}
E^{(m,n|i,j)}=\left(\bu^{(m,n|i)} - \bu^{(m,n|j)}\right)\scalp
\left(\bg^{(i)} - \bg^{(j)}\right), \qquad i\neq j,
\label{eq:ElonGeneral}
\end{equation}
which is insensitive to a permutation of the node indexes $i$ and $j$,
\begin{equation}
E^{(m,n|i,j)}=E^{(m,n|j,i)}.
\end{equation}
Considering that the bars have a linear elastic response, the force $\bF^{(m,n|i,j)}$ (positive if tensile and negative if compressive) acting on the $i$-th node of the cell $\left\{m,n\right\}$ and generated by the elongation $E^{(m,n|i,j)}$ of the bar with stiffness $k^{(i,j)}$ is given by
\begin{equation}
\bF^{(m,n|i,j)}=-k^{(i,j)}\,E^{(m,n|i,j)}\left(\bg^{(i)} - \bg^{(j)}\right),
\label{eq:ForceGeneral}
\end{equation}
which, according to the second Newton's law, is also the opposite of that acting at the $j$-th node and due to the elongation $E^{(m,n|i,j)}$ of the same bar
\begin{equation}
\bF^{(m,n|j,i)}=-\bF^{(m,n|i,j)}.
\end{equation}

Independently of the cell indexes $\{m,n\}$, the stiffness $k^{(i,j)}$ related to the bar connecting the nodes $i$ and $j$ is defined as (Fig.~\ref{fig:intro_lattice}, left)
\begin{equation}
k^{(i,j)}=
\left\{\begin{array}{lllll}
\overline{k},\qquad \qquad &i\neq 0 \,\,\mbox{and}\,\, j\neq 0,\\
\widetilde{k},\qquad \qquad &i= 0\,\, \mbox{and}\,\, j \,\,\mbox{even}\qquad  
&\mbox{or}
\qquad  
&i\,\,\mbox{even}\, \mbox{and}\, j = 0,\\
\widehat{k},\qquad \qquad &i= 0\,\, \mbox{and}\,\, j \,\,\mbox{odd}\qquad  
&\mbox{or}
\qquad  
&i\,\,\mbox{odd}\,\, \mbox{and}\,\, j = 0.
\end{array}
\right.
\end{equation}

\begin{figure}[H]
	\centering
	\includegraphics[width=0.7\linewidth]{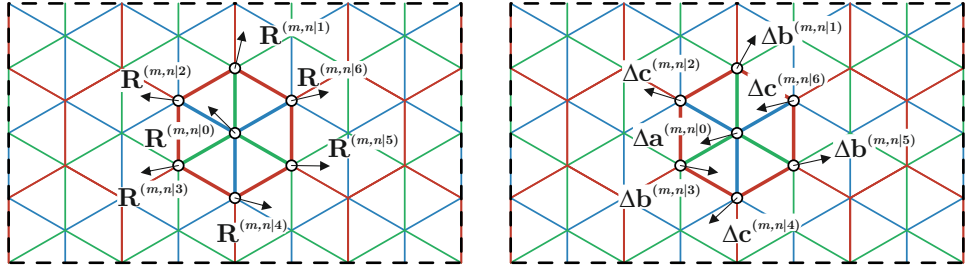}
	\caption{Resultant forces $\bR^{(m,n|i)}$ (left) and additional displacements $\bDelta\bu^{(m,n|i)}$ (right) associated with the node $i$ ($i=0,...,6$) belonging to the  cell 	$\{m,n\}$  within the lattice drawn in its undeformed configuration.}
	\label{fig:CorrEForceNodal}
\end{figure}

The sum of all the forces $\bF^{(m,n|i,j)}$, acting on the node $i$ (belonging to the cell $\left\{m,n\right\}$) and generated by the elongation of all the  bars jointed at that node, provides the resultant  $\bR^{(m,n|i)}$, Fig. \ref{fig:CorrEForceNodal} (left). Considering the properties expressed by Eq.~(\ref{eq:39}),  the resultant forces at all of the lattice nodes are given through the three primary resultants $\bR^{(m,n|0)}$, 
 $\bR^{(m,n|1)}$,  $\bR^{(m,n|2)}$ as
\begin{equation}
\begin{array}{lll}
\bR^{(m,n|0)}=\sum\limits_{j=1}^6  \bF^{(m,n|0,j)},
\\[6mm]
\bR^{(m,n|1)}=\bF^{(m,n|1,0)}+\bF^{(m,n|1,2)}+\bF^{(m,n|1,6)}+\bF^{(m,n+1|3,0)}+\bF^{(m-1,n+1|5,6)}+\bF^{(m-1,n+1|5,0)},
\\[6mm]
\bR^{(m,n|2)}=\bF^{(m,n|2,0)}+\bF^{(m,n|2,1)}+\bF^{(m,n|2,3)}+\bF^{(m-1,n|6,0)}+\bF^{(m-1,n+1|4,3)}+\bF^{(m-1,n+1|4,0)}.
\end{array}
\label{eq:ResultantGeneral}
\end{equation}

Assuming quasi-static conditions, from property ~(\ref{eq:39}) the equilibrium of the whole lattice is attained  when the three primary resultants $\bR^{(m,n|0)}$, $\bR^{(m,n|1)}$, and $\bR^{(m,n|2)}$ vanish for every cell $\left\{m,n\right\}$ 
\begin{equation}
\label{equilibriobrio}
\bR^{(m,n|0)}=\bR^{(m,n|1)}=\bR^{(m,n|2)}=\b0,\qquad \forall \left\{m,n\right\}.
\end{equation}
The elastic energy $\mathsf{U}_{lat}^{(m,n)}$ stored within the cell $\left\{m,n\right\}$ (instrumental to later identify the energetically equivalent microstructured solid) is provided by
\begin{equation}
\mathsf{U}_{lat}^{(m,n)} =\dfrac{1}{2}\sum_{i=1}^{6}
k^{(i,0)} \left[E^{(m,n|i,0)}\right]^2
+ 
\dfrac{1}{4}\sum_{i=1}^{6}
k^{(i,i+1-6\delta_{i6})} \left[E^{(m,n|i,i+1-6\delta_{i6})}\right]^2 ,
\label{eq:EnergyGeneral}
\end{equation}
where only one half of the energy stored within the  bars along the hexagon perimeter has been considered, so that the total energy of the infinite lattice is obtained by summing the energy of each cell
\begin{equation}
\mathsf{U}_{lat} =\sum_{m,n\in \mathbb{Z}}\mathsf{U}_{lat}^{(m,n)}.
\end{equation}

\subsection{Definition of an average operator for the displacement gradient in the  lattice structure}
\label{secAvarage}

With reference to a generic field $\textbf{f}(x_1,x_2)$  over  a domain $\Omega$ of a continuous body, its gradient and the related average are respectively given by
\begin{equation}\label{strnz0}
f_{j,k}(x_1, x_2)=\frac{\partial f_j(x_1, x_2)}{\partial x_k}, \qquad
\langle f_{j,k}\rangle = \dfrac{1}{||\Omega||}\int_{\Omega}f_{j,k} \mbox{d}\Omega,
\end{equation}
where $||\Omega||$ is the measure of $\Omega$. By means of the divergence theorem, the gradient average can be rewritten as
\begin{equation}
\label{strnz}
	\langle f_{j,k}\rangle = \dfrac{1}{||\Omega||}\int_{\partial\Omega} f_j n_k \mbox{ds}, 
\end{equation}
where only the evaluation of the field $\textbf{f}(x_1,x_2)$ along the cell perimeter is needed. In order to compute the displacement gradient average, the displacement field along the cell perimeter can be linearly interpolated as
\begin{equation}
\bu(s;m,n,i) =\bu^{(m,n|i)} + \left( \bu^{(m,n|i+1-6\delta_{i6})} - \bu^{(m,n|i)}\right) \frac{s}{\ell}, \qquad
i=1,...,6,
\label{eq:MeanDisp}
\end{equation}
where $s$ is the curvilinear coordinate  along the bar of the cell $\{m,n\}$ connecting the node $i$ to node $i+1-6\delta_{i6}$ and measuring the distance from the former ($i$=1,...,6).
Considering this interpolating field $\bu(s;m,n,i)$ and identifying $\Omega$ with the hexagonal domain, the   average of the displacement gradient for the lattice structure (identified with the subscript \lq ${\textsf{lat}}$' to highlight its relation with the lattice, and not with the continuum) can be obtained by substituting Eq.~(\ref{eq:MeanDisp}) into Eq.~(\ref{strnz}) as
\begin{equation}
	\langle{u}_{j,k}\rangle^{(m,n)}_{\textsf{lat}} = 
	\dfrac{1}{3\sqrt{3} \ell}\sum_{i=1}^{6} \left(u^{(m,n|i)}_j + u^{(m,n|i+1-6\delta_{i6})}_j \right)n^{(i)}_{k},
\label{eq:GradMedia}
\end{equation}
which, when the normal vectors $n^{(i)}_{k}$ are expressed with respect to the unit vectors $g^{(i)}_{k}$, reduces to
\begin{equation}
\langle{u}_{j,k}\rangle^{(m,n)}_{\textsf{lat}} = 
\dfrac{1}{9\ell}\sum_{i=1}^{6} \left(u^{(m,n|i)}_j + u^{(m,n|i+1-6\delta_{i6})}_j \right)\left(g^{(i)}_{k}+g^{(i+1-6\delta_{i6})}_{k}\right).
\label{eq:GradMediaRVE}
\end{equation}
More specifically, the four components of $\langle{u}_{j,k}\rangle^{(m,n)}_{\textsf{lat}}$ can be expressed in the reference system $\be_1$--$\be_2$ as
\begin{equation}
\begin{split}
\langle{\nabla \bu(\bx)}\rangle^{(m,n)}_{\textsf{lat}}
=\dfrac{1}{\ell}
\resizebox{0.8\textwidth}{!}{$
\begin{bmatrix}
	\dfrac{-u_{1}^{(m,n|2)}-u_{1}^{(m,n|3)}+u_{1}^{(m,n|5)}+u_{1}^{(m,n|6)}}{2 \sqrt{3}} &
	\dfrac{2 u_{1}^{(m,n|1)}+u_{1}^{(m,n|2)}-u_{1}^{(m,n|3)}-2 u_{1}^{(m,n|4)}-u_{1}^{(m,n|5)}+u_{1}^{(m,n|6)}}{6}\\[3mm]
	\dfrac{-u_{2}^{(m,n|2)}-u_{2}^{(m,n|3)}+u_{2}^{(m,n|5)}+u_{2}^{(m,n|6)}}{2 \sqrt{3}} &
	\dfrac{2 u_{2}^{(m,n|1)}+u_{2}^{(m,n|2)}-u_{2}^{(m,n|3)}-2u_{2}^{(m,n|4)}-u_{2}^{(m,n|5)}+u_{2}^{(m,n|6)}}{6}\\
\end{bmatrix}
$}
.
\label{eq:mean3}
\end{split}
\end{equation}

An  alternative but equivalent way for deriving the average of the displacement gradient, Eq. (\ref{eq:mean3}), can be obtained with reference to the piecewise description of the displacement field along each one of the six equilateral triangles,  subdomains of the hexagonal cells and enclosed by the three different bars. Such a piecewise  description of the field $\bu^{(m,n,j)}(\bx)$ follows from the linear interpolation of the  displacements  of the central node and the two consecutive perimeter nodes $j$ and $j+1-6\delta_{j6}$ (with $j=1,...,6$), corresponding to the three vertices of the $j$-th triangle  composing the $\{m,n\}$ hexagonal cell, as
\begin{equation}
\bu^{(m,n,j)}(\bx)=\bA^{(m,n,j)}\bx+\bc^{(m,n,j)} \qquad\mbox{with}\qquad j=1,...,6 \qquad m,n\in\mathbb{Z}
\label{eq:sposttriangular}
\end{equation}
where matrix $\bA^{(m,n,j)}$ and the vector $\bc^{(m,n,j)}$ 
are
\begin{equation}
\resizebox{0.98\textwidth}{!}{$
\begin{array}{lll}
&A^{(m,n,j)}_{11}=\dfrac{2 \cos \left(\frac{\pi  j}{3}\right) (u_{1}^{(m,n|j)}-u_{1}^{(m,n|0)})+2 \cos \left(\frac{\pi  (j-1)}{3} \right) (u_{1}^{(m,n|0)}-u_{1}^{(m,n|j+1)})}{\sqrt{3} \ell},
&A^{(m,n,j)}_{12}=\dfrac{2 \cos \left(\frac{\pi  j}{3}\right) (u_{2}^{(m,n|j)}-u_{2}^{(m,n|0)})+2 \cos \left(\frac{\pi  (j-1)}{3} \right) (u_{2}^{(m,n|0)}-u_{2}^{(m,n|j+1)})}{\sqrt{3} \ell},\\[3mm]
&A^{(m,n,j)}_{21}=\dfrac{2 \sin \left(\frac{\pi  j}{3}\right) (u_{1}^{(m,n|j)}-u_{1}^{(m,n|0)})+2 \sin \left(\frac{\pi  (i-1)}{3} \right) (u_{1}^{(m,n|0)}-u_{1}^{(m,n|j+1)})}{\sqrt{3} \ell},&A^{(m,n,j)}_{22}=\dfrac{2 \sin \left(\frac{\pi  j}{3}\right) (u_{2}^{(m,n|j)}-u_{2}^{(m,n|0)})+2 \sin \left(\frac{\pi  (j-1)}{3} \right) (u_{2}^{(m,n|0)}-u_{2}^{(m,n|j+1)})}{\sqrt{3} \ell},\\[3mm]
&c^{(m,n,j)}_{1}=u_{1}^{(m,n|0)},&c^{(m,n,j)}_{2}=u_{2}^{(m,n|0)}.
\end{array}
$}
\label{eq:CoeffMedie2}
\end{equation}
The average of the displacement gradient within the unit cell $\{m,n\}$ follows from Eq. (\ref{strnz0}) as
\begin{equation}
\langle{\nabla \bu(\bx)}\rangle^{(m,n)}_{\textsf{lat}}=\frac{1}{||\Omega||}\sum_{j=1}^{6}\int_{\Omega^{(m,n,j)}} \nabla \bu^{(m,n,j)}(\bx) \mbox{d} \Omega .
\label{eq:mean2}
\end{equation}
which, considering the piecewise description of displacement (\ref{eq:sposttriangular}), Eq. (\ref{eq:mean2}) can be rewritten as 
\begin{equation}
\begin{split}
&\langle{\nabla \bu(\bx)}\rangle^{(m,n)}_{\textsf{lat}}
=\frac{1}{6}\sum\limits_{j=0}^{6}
\begin{bmatrix}
	A^{(m,n,j)}_{11} & A^{(m,n,j)}_{12} \\[5mm]
	A^{(m,n,j)}_{21} & A^{(m,n,j)}_{22} 
\end{bmatrix}
,
\end{split}
\end{equation}
and that, recalling Eq. (\ref{eq:CoeffMedie2}), reduces to the same expression given by Eq.~(\ref{eq:GradMediaRVE}).

\section{Second-order displacement boundary condition}\label{qqua}

The key for the identification procedure performed in the next Section is the imposition to the infinite lattice 
of a linear and a quadratic nodal displacement fields (as in \cite{bacca2013mindlin}, \cite{mattia2013mindlin}, \cite{BACIGALUPO2017}, \cite{bigoni2007analytical}), together with an \lq additional field' $\bDelta \bu^{(m,n|i)}$, namely, 
\begin{equation}
u_r^{(m,n|i)}= \alpha_{sr} x_s^{(m,n|i)} + \beta_{str} x_s^{(m,n|i)} x_t^{(m,n|i)} + \Delta u_r^{(m,n|i)},\qquad\mbox{with}\qquad r,s,t=1,2
\label{eq:campospost}
\end{equation}
where $\alpha_{sr}$ and $\beta_{str}$ are tensors defining the displacement amplitudes and satisfying the symmetry properties $\alpha_{sr}=\alpha_{rs}$ and $\beta_{str}= \beta_{tsr}$, so that they have in general three and six independent components, respectively.
The presence of the additional term $\Delta u_r^{(m,n|i)}$ is 
necessary, as shown further on, for attaining the quasi-static equilibrium for every $\alpha_{sr}$ and $\beta_{str}$ as defined by Eq.~(\ref{equilibriobrio}).
The displacement field expressed through Eq.~(\ref{eq:campospost}) can equivalently be written as 
\begin{equation}
\bu^{(m,n|i)} = \balpha \bx^{(m,n|i)}  + \left(\bx^{(m,n|i)}  \otimes \bx^{(m,n|i)}\right) \scalpp  \bbeta
+\bDelta \bu^{(m,n|i)},
\label{eq:campospost2}
\end{equation}
where the second-order tensor $\balpha$ and the third-order tensor $\bbeta$ have components $\alpha_{sr}=(\balpha)_{sr}$ and $\beta_{str}=(\bbeta)_{str}$.
In Eq.~(\ref{eq:campospost2}), the dyadic product $\otimes$ and double scalar product $\scalpp$ are introduced, respectively  defined as $\left(\ba  \otimes \bb\right)_{st}=a_s b_t$ and $\left(\bA  \scalpp \boldsymbol{\mathcal{B}}\right)_{r}=A_{st} \mathcal{B}_{str}$.
Considering the displacement field (\ref{eq:campospost2}), the elongation of the bars can be computed  from  Eq.~(\ref{eq:ElonGeneral}) as
\begin{equation}
\begin{split}
E^{(m,n|i,j)} =& \ell \left\{ \balpha \left(\bg^{(i)} - \bg^{(j)}\right) 
+ 2  \left[\bx^{(m,n|0)} \otimes \left(\bg^{(i)} - \bg^{(j)}\right)\right] \scalpp \bbeta
\right.\\
& \left.+  \ell \left[\left(\bg^{(i)} + \bg^{(j)}\right) \otimes \left(\bg^{(i)} - \bg^{(j)}\right)\right] \scalpp \bbeta \right\}\scalp
\left(\bg^{(i)} - \bg^{(j)}\right)+\Delta E^{(m,n|i,j)}, \qquad i\neq j,
\end{split}
\label{eq:44}
\end{equation}
so that the corresponding  force at the $i$-th node can be evaluated from Eq.~(\ref{eq:ForceGeneral}) as 
\begin{equation}
\begin{split}
\bF^{(m,n|i,j)}  =& -k^{(i,j)}\ell~\bG^{(i,j)}
\left\{ \balpha \left(\bg^{(i)} - \bg^{(j)}\right) 
+ 2  \left[\bx^{(m,n|0)} \otimes \left(\bg^{(i)} - \bg^{(j)}\right)\right] \scalpp \bbeta
\right.\\
& \left.+  \ell \left[\left(\bg^{(i)} + \bg^{(j)}\right) \otimes \left(\bg^{(i)} - \bg^{(j)}\right)\right]\scalpp \bbeta \right\} +\bDelta \bF^{(m,n|i,j)}, \qquad i\neq j,
\end{split}
\label{eq:Forces}
\end{equation}
where
\begin{equation}
\label{eq:ForcCorretti}
\begin{array}{lll}
\Delta E^{(m,n|i,j)}=
\left(\bDelta\bu^{(m,n|i)} - \bDelta\bu^{(m,n|j)}\right)\scalp
\left(\bg^{(i)} - \bg^{(j)}\right), \\[4mm]
\bDelta \bF^{(m,n|i,j)}=-
k^{(i,j)}\,\Delta E^{(m,n|i,j)}\left(\bg^{(i)} - \bg^{(j)}\right), 
\earr
\qquad i\neq j,
\end{equation}
and
\begin{equation}
\bG^{(i,j)}=\left(\bg^{(i)} - \bg^{(j)}\right)\otimes \left(\bg^{(i)} - \bg^{(j)}\right).
\end{equation}
In combination with Eqs.~(\ref{eq:Forces}) and (\ref{eq:ForcCorretti})$_2$, the three primary resultants $\bR^{(m,n|0)}$, 
 $\bR^{(m,n|1)}$,  $\bR^{(m,n|2)}$, Eqs.~(\ref{eq:ResultantGeneral}), reduce to 
\begin{equation}
\begin{split}
\bR^{(m,n|0)}=&\left ( \widehat{k} - \widetilde{k} \right )  \ell  \sum_{i=1,3,5}  \left(\bg^{(i)}  \scalp \balpha \bg^{(i)}\right) \bg^{(i)}  
+ \left ( \widehat{k} + \widetilde{k} \right ) \ell^2 \sum_{i=1,3,5}  \left [ \left( \bg^{(i)} \otimes \bg^{(i)}\right) \scalpp \bbeta \scalp \bg^{(i)} \right ] \bg^{(i)} \\
& + 2 \left ( \widehat{k} - \widetilde{k} \right ) \ell \sum_{i=1,3,5}  \left [ \left(\bx^{(m,n|0)} \otimes \bg^{(i)} \right) \scalpp \bbeta \scalp \bg^{(i)} \right ] \bg^{(i)}
+ \sum_{j=1}^{6} k^{(0,j)} \bG^{(0,j)}  \left(\bDelta \bu^{(m,n|0)} - \bDelta \bu^{(m,n|j)}\right),
\end{split}
\label{eqEq0Big}
\end{equation}
\begin{equation}
\begin{split}
\bR^{(m,n|1)}=&\left ( \overline{k} - \widehat{k} \right )  \ell  \sum_{i=1,3,5} \left(\bg^{(i)}  \scalp \balpha \bg^{(i)}\right) \bg^{(i)} 
 + \left ( \overline{k} + \widehat{k} \right ) \ell^2 \sum_{i=1,3,5} \left [ \left(  \bg^{(i)} \otimes \bg^{(i)}\right) \scalpp \bbeta \scalp \bg^{(i)} \right ] \bg^{(i)}  \\
&+ 2 \left ( \overline{k} - \widehat{k} \right ) \ell \sum_{i=1,3,5}  \left [ \left(\bx^{(m,n|0)} \otimes \bg^{(i)} \right) \scalpp \bbeta \scalp \bg^{(i)}\right ]\bg^{(i)}
+ 2  \left(\overline{k} - \widehat{k} \right ) \ell^2 \sum_{i=1,3,5}  \left [ \left(\bg^{(1)} \otimes \bg^{(i)} \right) \scalpp \bbeta \scalp \bg^{(i)} \right ] \bg^{(i)}\\
&+\overline{k} 
\left[\bG^{(1,0)} 
\left(\bDelta \bu^{(m,n|1)} - \bDelta \bu^{(m,n|0)}\right) 
+ \bG^{(5,0)} 
\left(\bDelta \bu^{(m,n|1)} - \bDelta \bu^{(m-1,n+1|0)}\right)\right.\\
&+\left. \bG^{(3,0)} \left(\bDelta \bu^{(m,n|1)} - \bDelta \bu^{(m,n+1|0)}\right)\right]-\widehat{k} \left[\bG^{(3,0)}  \left(\bDelta \bu^{(m,n|1)} - \bDelta \bu^{(m,n|2)}\right)  \right.\\
&+ \left. \bG^{(5,0)} \left(\bDelta \bu^{(m,n|1)} - \bDelta \bu^{(m,n|6)}\right) + \bG^{(1,0)}  \left(\bDelta \bu^{(m,n|1)} - \bDelta \bu^{(m,n+1|2)}\right)\right],
\end{split}
\label{eqEq1Big}
\end{equation}
\begin{equation}
\begin{split}
\bR^{(m,n|2)}=&\left ( \widetilde{k} - \overline{k} \right )  \ell  \sum_{i=1,3,5} \left(\bg^{(i)}  \scalp \balpha \bg^{(i)}\right) \bg^{(i)} 
+ \left ( \widetilde{k} + \overline{k} \right ) \ell^2 \sum_{i=1,3,5}  \left [ \left( \bg^{(i)} \otimes \bg^{(i)}\right) \scalpp \bbeta \scalp \bg^{(i)} \right ] \bg^{(i)} \\
& + 2 \left ( \widetilde{k} - \overline{k} \right ) \ell \sum_{i=1,3,5} \left [ \left(\bx^{(m,n|0)} \otimes \bg^{(i)} \right) \scalpp \bbeta \scalp \bg^{(i)}\right ] \bg^{(i)}
+ 2 \left ( \widetilde{k} - \overline{k} \right ) \ell^2 \sum_{i=1,3,5} \left [ \left(\bg^{(2)} \otimes \bg^{(i)} \right) \scalpp \bbeta \scalp \bg^{(i)}\right ] \bg^{(i)}\\
&+
\widetilde{k} \left[
\bG^{(2,0)}  \left(\bDelta \bu^{(m,n|2)} - \bDelta \bu^{(m,n|0)}\right)
+ \bG^{(4,0)}  \left(\bDelta \bu^{(m,n|2)} - \bDelta \bu^{(m-1,n+1|0)}\right)\right.\\
&+ \left.
\bG^{(6,0)}  \left(\bDelta \bu^{(m,n|2)} - \bDelta \bu^{(m-1,n|0)}\right)\right]-
\overline{k} 
\left[\bG^{(4,0)}  \left(\bDelta \bu^{(m,n|2)} - \bDelta \bu^{(m,n|3)}\right)\right.\\
&\left.
+ \bG^{(6,0)}  \left(\bDelta \bu^{(m,n|2)} - \bDelta \bu^{(m,n|1)}\right)
+ \bG^{(2,0)}  \left(\bDelta \bu^{(m,n|2)} - \bDelta \bu^{(m-1,n|1)}\right
)\right].
\end{split}
\label{eqEq2Big}
\end{equation}

It follows from the above that all of the  resultant forces $\bR^{(m,n|i)}$ may be annihilated only when  the additional field
$\bDelta \bu^{(m,n|i)}$ assumes a linear expression which, under the constraint given by equations (\ref{eq:39}), is provided in the following  general form (Fig. \ref{fig:CorrEForceNodal}, right)
\begin{equation}
\bDelta \bu^{(m,n|i)}=
\left\{\begin{array}{lllll}
\bDelta \ba^{(m,n|0)}=\bZ\bx^{(m,n|0)} + \bz, \\[3mm]
\bDelta \bb^{(m,n|i)}=\bV \bx^{(m,n|i)} + \bv,\qquad \qquad &i\,\, \mbox{odd},\\[3mm]
\bDelta \bc^{(m,n|i)}=\bW \bx^{(m,n|i)} + \bw,\qquad \qquad &i\neq 0\,\, \mbox{and}\,\, \mbox{even},
\end{array}
\right.
\label{eq:EspreCampCorr}
\end{equation}
which implies that the average of the displacement gradient (\ref{eq:GradMediaRVE})  in the lattice is
\begin{equation}
\langle{\bnabla\bu}\rangle^{(m,n)}_{\textsf{lat}}=\balpha+\ell  \bbeta \scalp 
\left[
\begin{array}{ccc}
\sqrt{3}(2m+n)\\
3n
\end{array}
\right] 
+\dfrac{\bV + \bW }{2}.
\label{eq:MeanGradCorr}
\end{equation}

Considering the additional field , Eq. (\ref{eq:EspreCampCorr}),  the three primary resultants $\bR^{(m,n|0)}$, 
 $\bR^{(m,n|1)}$,  $\bR^{(m,n|2)}$, Eqs.~(\ref{eqEq0Big})--(\ref{eqEq2Big}), reduce to
\begin{equation}
\begin{split}
	\bR^{(m,n|0)}=&\left ( \widehat{k} - \widetilde{k} \right )  \ell  \sum_{i=1,3,5}  \left(\bg^{(i)}  \scalp \balpha \bg^{(i)}\right) \bg^{(i)}  
	+ \left ( \widehat{k} + \widetilde{k} \right ) \ell^2 \sum_{i=1,3,5}  \left [ \left( \bg^{(i)} \otimes \bg^{(i)}\right) \scalpp \bbeta \scalp \bg^{(i)} \right ] \bg^{(i)} \\
	& + 2 \left ( \widehat{k} - \widetilde{k} \right ) \ell \sum_{i=1,3,5}  \left [ \left(\bx^{(m,n|0)} \otimes \bg^{(i)} \right) \scalpp \bbeta \scalp \bg^{(i)} \right ] \bg^{(i)}+\\
	&+\sum_{i=1,3,5}   \left(\bg^{(i)} \otimes \bg^{(i)} \right) \left[ \widehat{k}   \left( \ell \bV \bg^{(i)} + \left(\bV - \bZ \right) \bx^{(m,n|0)} + \bv - \bz \right) \right.\\
	& \left. + \widetilde{k}   \left( - \ell \bW \bg^{(i)} + \left(\bW - \bZ \right) \bx^{(m,n|0)} + \bw - \bz \right)  \right] ,
\end{split}
\label{eqEq0BigCorr}
\end{equation}
\begin{equation}
\begin{split}
	\bR^{(m,n|1)}=&\left ( \overline{k} - \widehat{k} \right )  \ell  \sum_{i=1,3,5} \left(\bg^{(i)}  \scalp \balpha \bg^{(i)}\right) \bg^{(i)} 
	+ \left ( \overline{k} + \widehat{k} \right ) \ell^2 \sum_{i=1,3,5} \left [ \left(  \bg^{(i)} \otimes \bg^{(i)}\right) \scalpp \bbeta \scalp \bg^{(i)} \right ] \bg^{(i)}  \\
	&+ 2 \left ( \overline{k} - \widehat{k} \right ) \ell \sum_{i=1,3,5}  \left [ \left(\bx^{(m,n|0)} \otimes \bg^{(i)} \right) \scalpp \bbeta \scalp \bg^{(i)}\right ]\bg^{(i)} \\
	&+ 2 \left ( \overline{k} - \widehat{k} \right ) \ell^2 \sum_{i=1,3,5}  \left [ \left(\bg^{(1)} \otimes \bg^{(i)} \right) \scalpp \bbeta \scalp \bg^{(i)} \right ] \bg^{(i)}+\\
	&+\sum_{i=1,3,5}  \left(\bg^{(i)} \otimes \bg^{(i)} \right) \left[ \widehat{k}   \left( - \ell\bV \bg^{(1)} + \left(\bZ - \bV\right) \bx^{(m,n|0)} + \ell\bZ \left(\bg^{(1)} - \bg^{(i)}\right) + \bz - \bv  \right) \right.\\
	& \left.  + \overline{k} \left(  - \ell\bV \bg^{(1)} + \left(\bW - \bV \right)\bx^{(m,n|0)} + \ell\bW  \left(\bg^{(i)} + \bg^{(1)}\right) + \bw - \bv  \right) \right] ,
\end{split}
\label{eqEq1BigCorr}
\end{equation}
\begin{equation}
\begin{split}
	\bR^{(m,n|2)}=&\left ( \widetilde{k} - \overline{k} \right )  \ell  \sum_{i=1,3,5} \left(\bg^{(i)}  \scalp \balpha \bg^{(i)}\right) \bg^{(i)} 
	+ \left ( \widetilde{k} + \overline{k} \right ) \ell^2 \sum_{i=1,3,5}  \left [ \left( \bg^{(i)} \otimes \bg^{(i)}\right) \scalpp \bbeta \scalp \bg^{(i)} \right ] \bg^{(i)} \\
	& + 2 \left ( \widetilde{k} - \overline{k} \right ) \ell \sum_{i=1,3,5} \left [ \left(\bx^{(m,n|0)} \otimes \bg^{(i)} \right) \scalpp \bbeta \scalp \bg^{(i)}\right ] \bg^{(i)} \\
	& + 2 \left ( \widetilde{k} - \overline{k} \right ) \ell^2 \sum_{i=1,3,5} \left [ \left(\bg^{(2)} \otimes \bg^{(i)} \right) \scalpp \bbeta \scalp \bg^{(i)}\right ] \bg^{(i)}+\\
	&+\sum_{i=1,3,5} \left(\bg^{(i)} \otimes \bg^{(i)} \right) \left[ \widetilde{k}   \left( - \ell\bW \bg^{(2)} + \left(\bZ - \bW\right) \bx^{(m,n|0)} + \ell\bZ \left(\bg^{(i)} + \bg^{(2)}\right) + \bz - \bw \right) \right.\\
	& \left.  + \overline{k} \left( -\ell\bW \bg^{(2)} + \left(\bV - \bW \right)\bx^{(m,n|0)} - \ell\bV  \left(\bg^{(i)} - \bg^{(2)}\right) + \bv - \bw  \right) \right] .
\end{split}
\label{eqEq2BigCorr}
\end{equation}

The annihilation of the three resultant forces $\textbf{R}^{(m,n|0)}$,
 $\textbf{R}^{(m,n|1)}$, and $\textbf{R}^{(m,n|2)}$ for every unit cell $\{m,n\}$ is equivalent to a  system of 30 linear equations in 
the 18 unknown components of the vectors $\textbf{v}$, $\textbf{w}$, and $\textbf{z}$, and of the matrices $\textbf{V}$, $\textbf{W}$, and $\textbf{Z}$ (Eqs. (\ref{eq:EspreCampCorr})).
 Solving this system leads to two results, namely, (i.) the determination of 12 out of the 18 additional field components, which depend on the components of $\textbf{z}$ and $\textbf{Z}$  assumed as free parameters as 
\begin{equation}
\begin{array}{ccc}
\bv
=
\mathcal{K}^{[1]}
\left\{
\begin{bmatrix}
	\alpha_{12}	\\ 
	\dfrac{\alpha_{11}-\alpha_{22}}{2}
\end{bmatrix}
+
\begin{bmatrix}
	\dfrac{Z_{12} + Z_{12}}{2}	\\ 
	\dfrac{Z_{11}-Z_{22}}{2}
\end{bmatrix}
\right\}\ell
+
\left\{
\mathcal{K}^{[3]}
\begin{bmatrix}
	\beta_{111}+\beta_{122}	\\ 
	\beta_{222}+\beta_{211}	
\end{bmatrix}
+
\mathcal{K}^{[5]}
\begin{bmatrix}
	\beta_{221}-\beta_{122}	\\ 
	\beta_{112}-\beta_{211}	
\end{bmatrix}
\right\}\ell^2
+\bz,
\vspace{4mm}
\\
\bw
=
\mathcal{K}^{[2]}
\left\{
\begin{bmatrix}
	\alpha_{12}	\\ 
	\dfrac{\alpha_{11}-\alpha_{22}}{2}
\end{bmatrix}
+
\begin{bmatrix}
	\dfrac{Z_{12} + Z_{12}}{2}	\\ 
	\dfrac{Z_{11}-Z_{22}}{2}
\end{bmatrix}
\right\}\ell
+
\left\{
\mathcal{K}^{[4]}
\begin{bmatrix}
	\beta_{111}+\beta_{122}	\\ 
	\beta_{222}+\beta_{211}	
\end{bmatrix}
+
\mathcal{K}^{[6]}
\begin{bmatrix}
	\beta_{221}-\beta_{122}	\\ 
	\beta_{112}-\beta_{211}	
\end{bmatrix}
\right\}\ell^2
+\bz ,
\vspace{4mm}
\\
\bV =
\mathcal{K}^{[1]}
\begin{bmatrix}
		\beta_{112} + \beta_{211} & \beta_{122} + \beta_{221}\\
		\\
		\beta_{111} - \beta_{122} & \beta_{211} - \beta_{222}
\end{bmatrix}\ell+\bZ
,
\qquad
\bW =
\mathcal{K}^{[2]}
\begin{bmatrix}
	\beta_{112} + \beta_{211} & \beta_{122} + \beta_{221}\\
	\\
	\beta_{111} - \beta_{122} & \beta_{211} - \beta_{222}
\end{bmatrix}\ell+\bZ,
\label{eq:CorrVWvw}
\end{array}
\end{equation}
and (ii.)  two  linear equations for the six components of $\bbeta$. 

It follows from these two equations that tensor $\bbeta$ is constrained to have  only four independent components and will be  henceforth referred as $\bbeta^{\textsf{lat}}$, a symbol defining the set of generic quadratic amplitude tensors $\bbeta$, for which the lattice structure is in equilibrium in the absence of external nodal forces. Considering  $\beta_{111}$, $\beta_{221}$, $\beta_{112}$, $\beta_{222}$ as  the four independent components, tensor $\bbeta^{\textsf{lat}}$ is defined by the six components $\beta_{111}, \beta_{221}, \beta_{112}, \beta_{222}, \beta_{211}^{\textsf{lat}},$ and $
\beta_{122}^{\textsf{lat}}$, where the last two are 
\begin{equation}
\left[
\begin{array}{cc}
\beta_{211}^{\textsf{lat}}\\
\beta_{122}^{\textsf{lat}}
\end{array}
\right]=
-\left(1+\dfrac{9I_{[3]}}{2I_{[1]}I_{[2]}}\right)
\left[
\begin{array}{cc}
\beta_{222}\\
\beta_{111}
\end{array}
\right]
-\dfrac{9I_{[3]}}{2I_{[1]}I_{[2]}}
\left[
\begin{array}{cc}
\beta_{112}\\
\beta_{221}
\end{array}
\right].
\label{eq:EquiConstrainBeta}
\end{equation}
In Eqs.~(\ref{eq:CorrVWvw}) and (\ref{eq:EquiConstrainBeta}), the coefficients  $I_{[j]}$ ($j=1,2,3$) are the three invariants of the diagonal matrix $\bK$
\begin{equation}
\bK = \left[
\begin{array}{cccc}
	\overline{k} & 0 & 0 \\ 
	0 & \widehat{k} & 0 \\
	0 & 0 & \widetilde{k}
\end{array}\right],
\label{eq:59}
\end{equation}
so that
\begin{equation}
I_{[1]} = \text{tr} \bK= \overline{k} +  \widehat{k} +\widetilde{k}, \qquad
I_{[2]} = \dfrac{1}{2}\left[\left(\text{tr}\bK\right)^2-\text{tr}\bK^2 \right]=
\overline{k} \, \widehat{k}\,+\overline{k} \,  \widetilde{k}+ \widehat{k}\, \widetilde{k},\qquad
 I_{[3]} = \det(\bK)=\overline{k} \, \widehat{k}\, \widetilde{k},
\end{equation}
while the coefficients  $\mathcal{K}^{[j]}$ ($j=1,...,6$) are given by
\begin{equation}
\begin{array}{cccc}
&\mathcal{K}^{[1]}=\dfrac{\overline{k} (\widehat{k}-\widetilde{k})+\widetilde{k} (\widehat{k}-\overline{k})}{I_{[2]}}, \qquad &\mathcal{K}^{[2]}=\dfrac{\widehat{k} (\overline{k}-\widetilde{k})+\overline{k} (\widehat{k}-\widetilde{k})}{I_{[2]}}, \qquad &\mathcal{K}^{[3]}=\dfrac{3 \widehat{k} (\overline{k}+\widetilde{k})+4 \overline{k} (\overline{k}+2 \widetilde{k})}{4 I_{[2]}},
\vspace{4mm}\\
&\mathcal{K}^{[4]}=\dfrac{3 \widetilde{k} (k+\widehat{k})+4 \overline{k} (k+2 \widehat{k})}{4 I_{[2]}}, \qquad &\mathcal{K}^{[5]}=\dfrac{I_{[2]}+3 \overline{k} \widetilde{k}}{4 I_{[2]}}, \qquad &\mathcal{K}^{[6]}=\dfrac{I_{[2]}+3 \overline{k} \widehat{k}}{4 I_{[2]}}.
\end{array}
\label{eq:CoeffCampCorr}
\end{equation}
Imposing that the additional field $\bDelta \bu^{(m,n|i)}$ does not affect the mean value of the displacement gradient $\langle\nabla\bu\rangle^{(m,n)}_{\textsf{lat}}$, Eq.~(\ref{eq:MeanGradCorr}), leads to the condition
\begin{equation}
\bV + \bW=\b0,
\label{eq:MeanGradCorr2}
\end{equation}
which, considering Eq.~(\ref{eq:CorrVWvw}), implies the following expression for $\bZ$
\begin{equation}
\bZ =
-\frac{\mathcal{K}^{[1]}+\mathcal{K}^{[2]}}{2}
\normalsize{\begin{bmatrix}
		\beta_{112} + \beta_{211}^{\mathsf{lat}} & \beta_{122}^{\mathsf{lat}} + \beta_{221}\\
		\\
		\beta_{111} - \beta_{122}^{\mathsf{lat}} & \beta_{211}^{\mathsf{lat}} - \beta_{222}
\end{bmatrix}}\ell,
\label{eq:CorrZ}
\end{equation}
while the vector $\bz$ appearing in Eqs.~(\ref{eq:CorrVWvw})  remains indeterminate because it only produces a rigid-body translation.

It is worth noting that:
\begin{itemize}
\item in the case of bars  with same stiffness ($\overline{k}=\widetilde{k}=\widehat{k}$), enforcing Eqs.(\ref{eq:EquiConstrainBeta}) automatically provides the equilibrium Eqs. (\ref{eqEq0BigCorr})--(\ref{eqEq2BigCorr}) for the generic purely quadratic displacement field augmented by a rigid translation $\bz$,
\begin{equation}
\bar{k}=\widetilde{k}=\widehat{k} 
\quad
\Longrightarrow
\quad
\begin{cases}
\bv=\bw=\bz,\\
\bV=\bW=\bZ=\b0,
\end{cases}
\end{equation}
so that the additional field reduces to a rigid-body translation, $\bDelta \bu=\bz$;
\item in the case when $\bbeta=\b0$, it follows that $\bV=\bW=\bZ=\b0$ but the additional field is in general non-null when two over the three stiffnesses are different from each other. Indeed, the additional field is annihilated only when $\bg^{(1)} \scalp \balpha \bg^{(1)}=\bg^{(3)} \scalp \balpha \bg^{(3)}=\bg^{(5)} \scalp \balpha \bg^{(5)}$ (or equivalently, $\alpha_{11}=\alpha_{22}$ and $\alpha_{12}=0$), except in the particular case of bars having same stiffness ($\overline{k}=\widetilde{k}=\widehat{k}$), in which case the additional field is always null;
\item  the second-order tensors $\bV$, $\bW$, and $\bZ$ of the additional field  display the following permutation properties 
\begin{equation}
	\bV\left(\kappa_1, \kappa_2, \kappa_3\right) = \bV\left(\kappa_1, \kappa_3, \kappa_2\right), \qquad
	\bW\left(\kappa_1, \kappa_2, \kappa_3\right) = \bW\left(\kappa_1, \kappa_3, \kappa_2\right), \qquad
	\bZ\left(\kappa_1, \kappa_2, \kappa_3\right) = - \bZ\left(\kappa_1, \kappa_3, \kappa_2\right). 
\end{equation}

In the case $\bbeta=\b0$, the above equations are also complemented by  following properties for the vectors $\bv$, $\bw$ of the additional field 
\begin{equation}
	\bv\left(\kappa_1, \kappa_2, \kappa_3\right) = \bv\left(\kappa_3,\kappa_2, \kappa_1\right),\qquad
	\bw\left(\kappa_1, \kappa_2, \kappa_3\right) = \bw\left(\kappa_2, \kappa_1, \kappa_3\right), \qquad \mbox{when}\,\,\bbeta=\b0.
\end{equation}
\end{itemize}

\begin{figure}[H]
	\centering
	\includegraphics[width=0.8\linewidth]{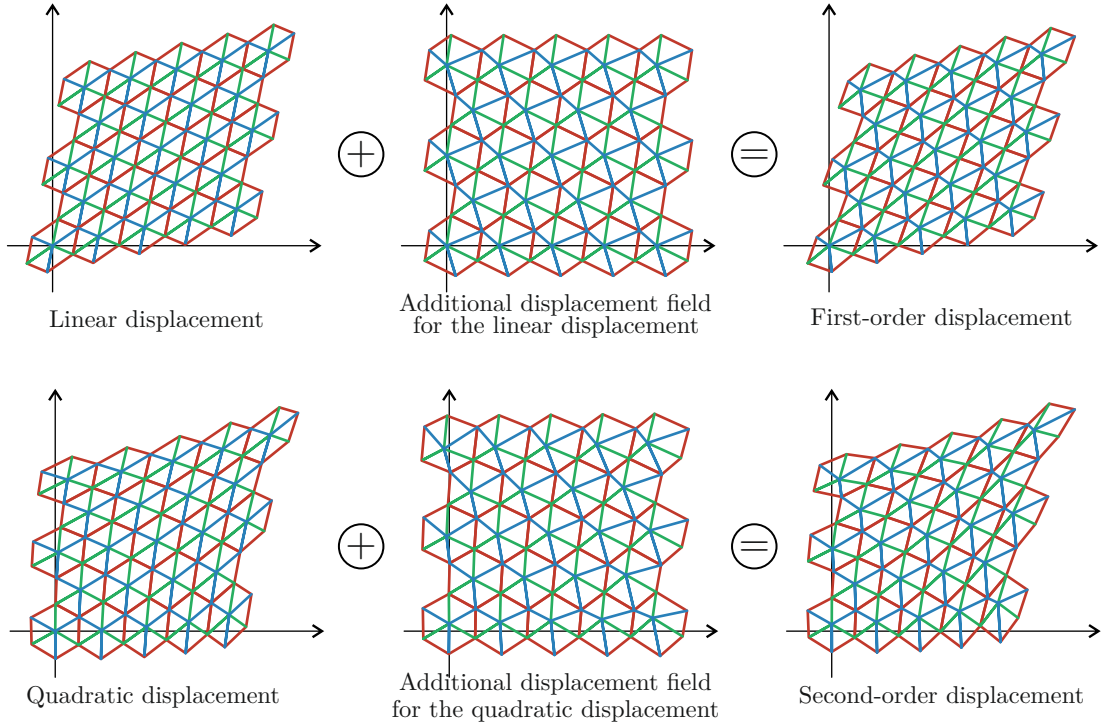}
	\caption{(Upper part) Deformed configurations for a lattice with bars of stiffness $\overline{k}=\widehat{k}=10\widetilde{k}$ subject to  (left) a purely linear displacement condition  with $\{\alpha_{11},\alpha_{22},\alpha_{12}\}=\{0,0,1/5\}$,  (center) its additional field, and (right) the  sum of these two. (Lower part) As in the upper part, but for a purely quadratic displacement condition with $\{\beta_{111},\beta_{221},\beta_{112},\beta_{222},\beta_{211}^{\mathsf{lat}},\beta_{122}^{\mathsf{lat}}\}=\{-1,1,1,-1,1,1\}1/(80\ell)$.}
	\label{fig:DefoLinQuadCorr}
\end{figure}

At this stage, the additional field $\bDelta \bu^{(m,n|i)}$, Eq. (\ref{eq:EspreCampCorr}), results completely defined through Eqs. (\ref{eq:CorrVWvw}), (\ref{eq:EquiConstrainBeta}), and (\ref{eq:CorrZ}). 
With the purpose of highlighting the contribution of the additional field $\bDelta\bu$ to the considered second-order displacement, Eq.~(\ref{eq:campospost}), 
three deformed configurations of the lattice are shown in Fig.~\ref{fig:DefoLinQuadCorr}.

Looking to the upper row of the figure, the first image on the left shows the displacement produced by a purely linear ($\bbeta=\b0$) 
didplacement, while the second image depicts the corresponding additional field only. Finally the image on the right is the composition of the two. 
The lower row shows respectively a purely quadratic ($\balpha=\b0$) displacement, its additional field $\bDelta \bu^{(m,n|i)}$, and the composition of the two. 
In the figure, the following stiffnesses of the lattice have been considered: $\overline{k}=\widehat{k}=10\widetilde{k}$. 

\section{Identification of the higher-order solid equivalent to the lattice structure}\label{identi}

Considering the second-order displacement field Eq.~(\ref{eq:campospost}) defined by the tensors $\balpha$ and $\bbeta^{\mathsf{lat}}$ 
Eqs.~(\ref{eq:EquiConstrainBeta}) and by the \lq additional field' $\bDelta \bu^{(m,n|i)}$,  Eqs.~(\ref{eq:CorrVWvw}) and Eq.~(\ref{eq:CorrZ}), the elastic energy stored within the lattice cell $\{m,n\}$ is computed. This elastic energy is shown to display the same mathematical structure of the elastic energy stored within a unit cell made up of a  homogeneous elastic second-gradient solid ($\mathsf{SGE}$) when subject to a quadratic displacement field, defined by the tensors $\balpha$ and $\bbeta^{\mathsf{SGE}}$ (note that $\bbeta^{\mathsf{SGE}}$ defines the coefficients of all quadratic fields which generate equilibrated stresses in a second-gradient elastic material without body forces).
Therefore, imposing the elastic energy matching between the lattice and the $\mathsf{SGE}$ solid allows for the identification of the constitutive parameters of the latter and shows that the self-equilibrium condition provides the same constrained boundary condition for the two materials, so that  $\bbeta^{\mathsf{lat}}=\bbeta^{\mathsf{SGE}}$.

It is instrumental to represent the components of the tensors $\balpha$ and $\bbeta^{(\boldsymbol{\cdot})}$ (where the superscript $(\boldsymbol{\cdot})$ denotes either $(\mathsf{lat})$ or $(\mathsf{SGE})$) using a vectorial notation through the vectors  $\boldsymbol{\mathsf{a}}$ and $\boldsymbol{\mathsf{b}}^{(\boldsymbol{\cdot})}$  as 
\begin{equation}
\boldsymbol{\mathsf{a}}
=\begin{bmatrix}
	\alpha_{11}\\
	\alpha_{22}\\
	2\alpha_{12}
\end{bmatrix}, \qquad
\boldsymbol{\mathsf{b}}^{(\boldsymbol{\cdot})}
=\begin{bmatrix}
\beta_{111}\\
\beta_{221}\\
\beta_{112}\\
\beta_{222}\\
2\beta_{211}\\
2\beta_{122}
\end{bmatrix},
\label{eq:contrazioBD}
\end{equation}
and to collect the four components of $\bbeta^{(\boldsymbol{\cdot})}$ not constrained by the equilibrium Eq.~(\ref{eq:EquiConstrainBeta}) in the vector $\boldsymbol{\mathsf{b}}^{*}$ 
\begin{equation}
\boldsymbol{\mathsf{b}}^{*}
=\begin{bmatrix}
	\beta_{111}\\
	\beta_{221}\\
	\beta_{112}\\
	\beta_{222}
\end{bmatrix},
\end{equation}
so that vector $\boldsymbol{\mathsf{b}}^{(\boldsymbol{\cdot})}$ can be obtained as
 \begin{equation}
\boldsymbol{\mathsf{b}}^{(\boldsymbol{\cdot})}=
\boldsymbol{\mathsf{T}}^{(\boldsymbol{\cdot})}
\boldsymbol{\mathsf{b}}^{*}
\end{equation}
where the matrix $\boldsymbol{\mathsf{T}}^{(\boldsymbol{\cdot})}$ is the transformation matrix enforcing the equilibrium conditions in the lattice (in which case it will be denoted as $\boldsymbol{\mathsf{T}}^{\mathsf{lat}}$) or in the second-gradient elastic solid (in which case it will be denoted as $\boldsymbol{\mathsf{T}}^{\mathsf{SGE}}$).

\subsection{Energy stored within the lattice structure}

Considering the second-order displacement field Eq.~(\ref{eq:campospost}) defined by the tensors $\balpha$ and $\bbeta^{\textsf{lat}}$ under the equilibrium constraint Eqs.~(\ref{eq:EquiConstrainBeta}) and with the additional displacement given by Eqs.~(\ref{eq:CorrVWvw}) and (\ref{eq:CorrZ}), the elastic strain energy $\mathsf{U}_{\mathsf{lat}}^{(m,n)}\left(\balpha,\bbeta^{\textsf{lat}}\right)$, stored within the lattice unit cell $\{m,n\}$ can be written in terms of vectors $\boldsymbol{\mathsf{a}}$ and $\boldsymbol{\mathsf{b}}^{\mathsf{lat}}$, as
\begin{equation}
\mathsf{U}_{\mathsf{lat}}^{(m,n)}\left(\boldsymbol{\mathsf{a}}, \boldsymbol{\mathsf{b}}^{\mathsf{lat}}\right)=
\mathsf{U}_{\mathsf{lat}}^{(m,n)}\left(\boldsymbol{\mathsf{a}}, \boldsymbol{\mathsf{T}}^{\mathsf{lat}}\boldsymbol{\mathsf{b}}^{*}\right),
\end{equation}
so that $\boldsymbol{\mathsf{b}}^{\mathsf{lat}}=
\boldsymbol{\mathsf{T}}^{\mathsf{lat}}
\boldsymbol{\mathsf{b}}^{*}$ with the definition 
\begin{equation}
\boldsymbol{\mathsf{T}}^{\mathsf{lat}}
=
\left(
\begin{array}{cccc}
	1 & 0 & 0 & 0 \\
	0 & 1 & 0 & 0 \\
	0 & 0 & 1 & 0 \\
	0 & 0 & 0 & 1 \\
	0 & 0 & -\frac{9 I_{[3]}}{I_{[1]} I_{[2]}} & -\frac{9 I_{[3]}}{I_{[1]} I_{[2]}}-2 \\
	-\frac{9 I_{[3]}}{I_{[1]} I_{[2]}}-2 & -\frac{9 I_{[3]}}{I_{[1]} I_{[2]}} & 0 & 0 \\
\end{array}
\right).
\label{eq:passaggioAStarMic}
\end{equation}

Therefore, from Eq.~(\ref{eq:EnergyGeneral}), the elastic energy of the lattice can be expressed as $\mathsf{U}_{\mathsf{lat}}^{(m,n)}\left(\boldsymbol{\mathsf{a}}, \boldsymbol{\mathsf{b}}^{*}\right)$ and therefore can be represented as the following quadratic form in $\boldsymbol{\mathsf{a}}$ and $\boldsymbol{\mathsf{b}}^*$
\begin{equation}
\begin{split}
	\mathsf{U}_{\mathsf{lat}}^{(m,n)}\left(\boldsymbol{\mathsf{a}}, \boldsymbol{\mathsf{b}}^{*}\right)=
	\ell^2&\left\{ \boldsymbol{\mathsf{a}} \scalp \boldsymbol{\mathsf{H}}^{[1]}\kern-0.3em\left(\overline{k},\widehat{k},\widetilde{k}\right)\,\boldsymbol{\mathsf{a}}
	+ 2\ell \boldsymbol{\mathsf{a}}\scalp\left[ m \boldsymbol{\mathsf{H}}^{[2]}\kern-0.3em\left(\overline{k},\widehat{k},\widetilde{k}\right)\, + n \boldsymbol{\mathsf{H}}^{[3]}\kern-0.3em\left(\overline{k},\widehat{k},\widetilde{k}\right)\, + \boldsymbol{\mathsf{H}}^{[4]}\kern-0.3em\left(\overline{k},\widehat{k},\widetilde{k}\right)\,\right]\boldsymbol{\mathsf{b}}^{*}\right.\\
	&\left. + \ell^2 \boldsymbol{\mathsf{b}}^{*} \scalp\left[m^2 \boldsymbol{\mathsf{H}}^{[5]}\kern-0.3em\left(\overline{k},\widehat{k},\widetilde{k}\right)\, + n^2 \boldsymbol{\mathsf{H}}^{[6]}\kern-0.3em\left(\overline{k},\widehat{k},\widetilde{k}\right)\, + m\,n \boldsymbol{\mathsf{H}}^{[7]}\kern-0.3em\left(\overline{k},\widehat{k},\widetilde{k}\right)\, + m \boldsymbol{\mathsf{H}}^{[8]}\kern-0.3em\left(\overline{k},\widehat{k},\widetilde{k}\right)\,+ \right.\right.\\
	&\left.\left. + n \boldsymbol{\mathsf{H}}^{[9]}\kern-0.3em\left(\overline{k},\widehat{k},\widetilde{k}\right)\, + 		\boldsymbol{\mathsf{H}}^{[10]}\kern-0.3em\left(\overline{k},\widehat{k},\widetilde{k}\right)\, \right] \boldsymbol{\mathsf{b}}^{*}\right\},
\end{split}
\label{eq:Ene2ordRVE}
\end{equation}
where the matrices $\boldsymbol{\mathsf{H}}^{[r]}$ ($r=1,...,10$) depend on 
the values of the three stiffnesses $\overline{k},\widehat{k}$, and $\widetilde{k}$. These matrices have different dimensions ($3\times 3$ for $r=1$, $3\times 4$ for $r=2,3,4$, and $4 \times 4$ in the other cases) and their components $\mathsf{H}_{ij}^{[r]}$ are reported in Appendix A.
From Eq.~(\ref{eq:Ene2ordRVE}) it is evident that  the strain energy depends on the cell position whenever $\boldsymbol{\mathsf{b}}^*\neq0$, so that it becomes independent of  indexes $m$ and $n$ only when $\boldsymbol{\mathsf{b}}^*=0$, a condition corresponding to $\boldsymbol{\mathsf{b}}^{\mathsf{lat}}=0$ and also implying $\bbeta^{\mathsf{lat}}= 0$.

\subsection{Energy stored within a second-gradient elastic solid}
\label{Sec:SecGradCont}

With reference to the \lq form I'  elastic material introduced by Mindlin \cite{mindlin1964micro}, a second-gradient elastic ($\textsf{SGE}$) solid  has a quadratic strain energy density $\mathcal{U}_{\mathbb{SGE}}$ function of the strain $\bepsilon$ and the curvature $\bchi$, which can be derived from the displacement field $\bu$  as 
\begin{equation}
\epsilon_{ij}=\frac{u_{i,j}+u_{j,i}}{2},
\qquad
\chi_{ijk}=u_{k,ij},
\end{equation}
displaying the symmetry properties $\epsilon_{ij}=\epsilon_{ji}$ and $\chi_{ijk}=\chi_{jik}$.
The quadratic  strain energy density $\mathcal{U}_{\mathbb{SGE}}$ can be decomposed as
\begin{equation}
\mathcal{U}_{\mathbb{SGE}}\left(\bepsilon,\bchi\right)= \mathcal{U}_{\mathbb{C}}\left(\bepsilon\right) + \mathcal{U}_{\mathbb{M}}\left(\bepsilon,\bchi\right) + \mathcal{U}_{\mathbb{A}}\left(\bchi\right) ,
\label{eq:EneCont1}
\end{equation}
where $\mathcal{U}_{\mathbb{C}}\left(\bepsilon\right)$ is a \lq purely local' (Cauchy) energy term and $\mathcal{U}_{\mathbb{A}}\left(\bchi\right)$ a 
\lq completely non-local' energy term, while the mutual energy  term  $\mathcal{U}_{\mathbb{M}}\left(\bepsilon,\bchi\right)$  expresses the coupling between strain and curvature, 
\begin{equation}
 \mathcal{U}_{\mathbb{C}}\left(\bepsilon\right)=\frac{1}{2}\mathbb{C}_{ijkl}\epsilon_{ij}\epsilon_{kl}, \qquad
 \mathcal{U}_{\mathbb{M}}\left(\bepsilon,\bchi\right)=\mathbb{M}_{ijklm}\chi_{ijk}\epsilon_{lm},
	\qquad
  \mathcal{U}_{\mathbb{A}}\left(\bchi\right)= \frac{1}{2}\mathbb{A}_{ijklmn}\chi_{ijk}\chi_{lmn},
\label{eq:EneCont2}
\end{equation}
being $\boldmath{\mathbb{C}}$,  $\boldmath{\mathbb{M}}$, and $\boldmath{\mathbb{A}}$ the  fourth-, fifth-, and sixth-order constitutive tensors,  respectively, 
possessing the following symmetries
\begin{equation}
\begin{array}{cc}
	\mathbb{C}_{ijkl}=\mathbb{C}_{jikl}=\mathbb{C}_{ijlk}=\mathbb{C}_{klij}, \qquad
		\mathbb{M}_{ijklm}=\mathbb{M}_{ijkml}=\mathbb{M}_{jiklm}, \qquad
	\mathbb{A}_{ijklmn}=\mathbb{A}_{jiklmn}=\mathbb{A}_{ijkmln}=\mathbb{A}_{lmnijk}.
\end{array}
\end{equation} 

The tensors work-conjugate to the fundamental kinematic fields $\bepsilon$ and $\bchi$ are respectively the stress $\bsigma$ and double 
stress $\btau$, defined as
\begin{equation}
\sigma_{ij} = \mathbb{C}_{ijlm}\epsilon_{lm} + \mathbb{M}_{ijlmn}\chi_{lmn},\qquad
\tau_{kji} = \mathbb{A}_{kjilmn}\chi_{lmn} + \mathbb{M}_{lmkji}\epsilon_{lm},
\label{eq:23}
\end{equation}
which are restricted to satisfy the equilibrium equations, that in the absence of body-forces are expressed by
\begin{equation}
\sigma_{ij,j} - \tau_{kji,jk} = 0.
\label{eq:selfequiEqui2}
\end{equation}
The vectorial representations for the strain $\bepsilon$ and the curvature $\bchi$ are introduced  through the  strain $\boldsymbol{\mathsf{p}}$ and curvature $\boldsymbol{\mathsf{q}}$ vectors  as
\begin{equation}
\boldsymbol{\mathsf{p}}
=
\begin{bmatrix}
	\epsilon_{11}\\
	\epsilon_{22}\\
	2\epsilon_{12}
\end{bmatrix},
\qquad
\boldsymbol{\mathsf{q}}
=\begin{bmatrix}
	\chi_{111}\\
	\chi_{221}\\
	\chi_{112}\\
	\chi_{222}\\
	2\chi_{211}\\
	2\chi_{122}
\end{bmatrix},
\label{eq:curv_def_vett1}
\end{equation}
so that the elastic energy densities (\ref{eq:EneCont2}) can be rewritten as
\begin{equation}
\mathcal{U}_{\mathbb{C}}\left(\bepsilon\right)= \mathcal{U}_{\boldsymbol{\mathsf{C}}}\left(\boldsymbol{\mathsf{p}}\right),
\quad
\mathcal{U}_{\mathbb{M}}\left(\bepsilon,\bchi\right)=\mathcal{U}_{\boldsymbol{\mathsf{M}}}\left(\boldsymbol{\mathsf{p}},\boldsymbol{\mathsf{q}}\right),
\quad
\mathcal{U}_{\mathbb{A}}\left(\bchi\right)=\mathcal{U}_{\boldsymbol{\mathsf{A}}}\left(\boldsymbol{\mathsf{q}}\right),
\label{eq:EneCont3}
\end{equation}
where
\begin{equation}
 \mathcal{U}_{\boldsymbol{\mathsf{C}}}\left(\boldsymbol{\mathsf{p}}\right)=\frac{1}{2}\mathsf{C}_{ij}\mathsf{p}_{i}\mathsf{p}_{j},
\quad
\mathcal{U}_{\boldsymbol{\mathsf{M}}}\left(\boldsymbol{\mathsf{p}},\boldsymbol{\mathsf{q}}\right)=\mathsf{M}_{jk} \mathsf{p}_{j}\mathsf{q}_k,
\quad
\mathcal{U}_{\boldsymbol{\mathsf{A}}}\left(\boldsymbol{\mathsf{q}}\right)=\frac{1}{2} \mathsf{A}_{kl}\mathsf{q}_k\mathsf{q}_l ,\quad i,j=1,2,3 \quad k,l=1,...,6,
\label{eq:EneCont4}
\end{equation}
with the matrices $\mathsf{C}_{ij}$, $\mathsf{M}_{jk}$, and $\mathsf{A}_{kl}$ respectively representing the constitutive tensors $\boldmath{\mathbb{C}}$, $\boldmath{\mathbb{M}}$, and $\boldmath{\mathbb{A}}$ in the Voigt notation. Note that matrices $\mathsf{C}_{ij}$ and $\mathsf{A}_{jk}$ are square and symmetric (the former of order 3 and the latter of order 6), while $\mathsf{M}_{jk}$ is a rectangular (3 $\times$ 6) matrix.  Considering this notation, the strain energy density $\mathcal{U}_{\mathsf{SGE}}\left(\boldsymbol{\mathsf{p}},\boldsymbol{\mathsf{q}}\right)$ can be introduced as
\begin{equation}
\mathcal{U}_{\mathsf{SGE}}\left(\boldsymbol{\mathsf{p}},\boldsymbol{\mathsf{q}}\right)=\mathcal{U}_{\boldsymbol{\mathsf{C}}}\left(\boldsymbol{\mathsf{p}}\right)+\mathcal{U}_{\boldsymbol{\mathsf{M}}}\left(\boldsymbol{\mathsf{p}},\boldsymbol{\mathsf{q}}\right)+\mathcal{U}_{\boldsymbol{\mathsf{A}}}\left(\boldsymbol{\mathsf{q}}\right),
\end{equation}
representing the strain energy density $\mathcal{U}_{\mathbb{SGE}}\left(\bepsilon,\bchi\right)$ in the Voigt notation, so that
\begin{equation}
\mathcal{U}_{\mathbb{SGE}}\left(\bepsilon,\bchi\right)=\mathcal{U}_{\mathsf{SGE}}\left(\boldsymbol{\mathsf{p}}\left(\bepsilon\right),\boldsymbol{\mathsf{q}}\left(\bchi\right)\right).
\label{eq:EquivaEner}
\end{equation}

It is assumed now that the second-gradient elastic material is subject to  remote quadratic displacement boundary conditions provided by the second-order  displacement field, Eq.~(\ref{eq:campospost}), in the absence of the additional field ($\bDelta \bu^{(m,n|i)}~=~\b0$, see also Sect. \ref{Homogenization}), 
\begin{equation}
\bu (\bx) = \balpha \bx + \left(\bx \otimes \bx\right) \scalpp \bbeta.
\label{eq:campospostCont}
\end{equation}
The quadratic displacement field (\ref{eq:campospostCont}) is restricted, at first order, by equilibrium, 
\begin{equation}
\mathbb{C}_{ljkh}~\beta_{jkh}=0,
\label{eq:selfequiEqui}
\end{equation}
an equation which introduces two relationships between the six coefficients $\beta_{ijk}$, so that two of them are dependent on the remaining four. Therefore, the coefficients $\beta_{ijk}$ are re-assembled in the vector $\bbeta^{\mathsf{SGE}}$, so that 
\begin{equation}
\bu (\bx) = \balpha \bx + \left(\bx \otimes \bx\right) \scalpp \bbeta^{\mathsf{SGE}} ,
\label{eq:spostselfequiEqui}
\end{equation}
where 
\begin{equation}
\begin{array}{c}
\beta^{\mathsf{SGE}}_{111} = \beta_{111},\quad
\beta^{\mathsf{SGE}}_{221} = \beta_{221},\quad
\beta^{\mathsf{SGE}}_{112} = \beta_{112},\quad
\beta^{\mathsf{SGE}}_{222} = \beta_{222}\\ [5 mm]
\beta^{\mathsf{SGE}}_{211}=\beta_{111} \mathcal{D}_{1}+\beta_{221} \mathcal{D}_{2}+\beta_{112} \mathcal{D}_{3}+\beta_{222} \mathcal{D}_{4}, ~~~~~
\beta^{\mathsf{SGE}}_{122}=\beta_{111} \mathcal{D}_{5}+\beta_{221} \mathcal{D}_{6}+\beta_{112} \mathcal{D}_{7}+\beta_{222} \mathcal{D}_{8},
\end{array}
\label{eq:selfequiResults}
\end{equation}
in which 
\begin{equation}
\begin{array}{ccc}
\mathcal{D}_{1} = \dfrac{2 \mathsf{C}_{13}^2-\mathsf{C}_{11} \left(\mathsf{C}_{12}+\mathsf{C}_{33}\right)}{\left(\mathsf{C}_{12}+\mathsf{C}_{33}\right)^2-4 \mathsf{C}_{13} \mathsf{C}_{23}}, ~~~~~
\mathcal{D}_{2} = \dfrac{2 \mathsf{C}_{13} \mathsf{C}_{23}-\mathsf{C}_{33} \left(\mathsf{C}_{12}+\mathsf{C}_{33}\right)}{\left(\mathsf{C}_{12}+\mathsf{C}_{33}\right)^2-4 \mathsf{C}_{13} \mathsf{C}_{23}}, ~~~~~
\mathcal{D}_{3} = \dfrac{\mathsf{C}_{13} \left(\mathsf{C}_{33}-\mathsf{C}_{12}\right)}{\left(\mathsf{C}_{12}+\mathsf{C}_{33}\right)^2-4 \mathsf{C}_{13} \mathsf{C}_{23}} ,
\vspace{4mm}\\
\mathcal{D}_{4} = \dfrac{2 \mathsf{C}_{13} \mathsf{C}_{22}-\mathsf{C}_{23} \left(\mathsf{C}_{12}+\mathsf{C}_{33}\right)}{\left(\mathsf{C}_{12}+\mathsf{C}_{33}\right)^2-4 \mathsf{C}_{13} \mathsf{C}_{23}}, ~~~~~
\mathcal{D}_{5} = \dfrac{2 \mathsf{C}_{11} \mathsf{C}_{23}-\mathsf{C}_{13} \left(\mathsf{C}_{12}+\mathsf{C}_{33}\right)}{\left(\mathsf{C}_{12}+\mathsf{C}_{33}\right)^2-4 \mathsf{C}_{13} \mathsf{C}_{23}}, ~~~~~
\mathcal{D}_{6} = \dfrac{\mathsf{C}_{23} \left(\mathsf{C}_{12}-\mathsf{C}_{33}\right)}{4 \mathsf{C}_{13} \mathsf{C}_{23}-\left(\mathsf{C}_{12}+\mathsf{C}_{33}\right)^2} ,
\vspace{4mm}\\
\mathcal{D}_{7} = \dfrac{\mathsf{C}_{33} \left(\mathsf{C}_{12}+\mathsf{C}_{33}\right)-2 \mathsf{C}_{13} \mathsf{C}_{23}}{4 \mathsf{C}_{13} \mathsf{C}_{23}-\left(\mathsf{C}_{12}+\mathsf{C}_{33}\right)^2}, ~~~~~
\mathcal{D}_{8} = \dfrac{2 \mathsf{C}_{23}^2-\mathsf{C}_{22} \left(\mathsf{C}_{12}+\mathsf{C}_{33}\right)}{\left(\mathsf{C}_{12}+\mathsf{C}_{33}\right)^2-4 \mathsf{C}_{13} \mathsf{C}_{23}}.
\end{array}
\label{eq:coeffD}
\end{equation}
From now on the constrained tensor $\bbeta$, due to Eqs.(\ref{eq:selfequiResults}), will be denoted by $\bbeta^{\mathsf{SGE}}$,
so that 
the strain $\boldsymbol{\mathsf{p}}$ and the curvature $\boldsymbol{\mathsf{q}}$ vectors can be rewritten as
\begin{equation}
\boldsymbol{\mathsf{p}}^{\mathsf{SGE}}=
\boldsymbol{\mathsf{p}}\left(\boldsymbol{\mathsf{a}},\boldsymbol{\mathsf{T}}^{\mathsf{SGE}}\boldsymbol{\mathsf{b}}^{*}\right),\qquad
\boldsymbol{\mathsf{q}}^{\mathsf{SGE}}=
2\boldsymbol{\mathsf{T}}^{\mathsf{SGE}}\boldsymbol{\mathsf{b}}^{*},
\end{equation}
where 
\begin{equation}
\boldsymbol{\mathsf{T}}^{\mathsf{SGE}}=
	\left(
	\begin{array}{cccc}
		1 & 0 & 0 & 0 \\
		0 & 1 & 0 & 0 \\
		0 & 0 & 1 & 0 \\
		0 & 0 & 0 & 1 \\
		2 \mathcal{D}_{5} & 2 \mathcal{D}_{6} & 2 \mathcal{D}_{7} & 2 \mathcal{D}_{8} \\
		2 \mathcal{D}_{1} & 2 \mathcal{D}_{2} & 2 \mathcal{D}_{3} & 2 \mathcal{D}_{4} \\
	\end{array}
	\right),
\end{equation}
and $\boldsymbol{\mathsf{p}}^{\mathsf{SGE}}$ can also be expressed as
\begin{equation}
\boldsymbol{\mathsf{p}}^{\mathsf{SGE}}\left(\boldsymbol{\mathsf{a}},\boldsymbol{\mathsf{b}}^{*}\right)=
\boldsymbol{\mathsf{a}}+2\left(\boldsymbol{\mathsf{P}}^{[1]}x_1^{(m,n)}+\boldsymbol{\mathsf{P}}^{[2]}x_2^{(m,n)}\right)\boldsymbol{\mathsf{b}}^{*}
\label{eq:curv_def_vett3}
\end{equation}
with
\begin{equation}
	\boldsymbol{\mathsf{P}}^{[1]}=
	\left(
	\begin{array}{cccc}
		1 & 0 & 0 & 0 \\
		\mathcal{D}_{1} & \mathcal{D}_{2} & \mathcal{D}_{3} & \mathcal{D}_{4} \\
		\mathcal{D}_{5} & \mathcal{D}_{6} & \mathcal{D}_{7}+1 & \mathcal{D}_{8} \\
	\end{array}
	\right),
	\qquad
	\boldsymbol{\mathsf{P}}^{[2]}=
	\left(
	\begin{array}{cccc}
		\mathcal{D}_{5} & \mathcal{D}_{6} & \mathcal{D}_{7} & \mathcal{D}_{8} \\
		0 & 0 & 0 & 1 \\
		\mathcal{D}_{1} & \mathcal{D}_{2}+1 & \mathcal{D}_{3} & \mathcal{D}_{4} \\
	\end{array}
	\right).
\label{eq:passaggioAStarSGM}
\end{equation}

From Eqs. (\ref{eq:selfequiResults}), the energy densities, Eqs.(\ref{eq:EneCont4}), become
\begin{equation}
\mathcal{U}_{\boldsymbol{\mathsf{C}}}\left(\boldsymbol{\mathsf{p}}^{\mathsf{SGE}}\right)=\frac{1}{2}\mathsf{C}_{ij}\mathsf{p}^{\mathsf{SGE}}_{i}\mathsf{p}^{\mathsf{SGE}}_{j},
\quad
\mathcal{U}_{\boldsymbol{\mathsf{M}}}\left(\boldsymbol{\mathsf{p}}^{\mathsf{SGE}},2\boldsymbol{\mathsf{T}}^{\mathsf{SGE}}\boldsymbol{\mathsf{b}}^{*}\right)=\mathsf{M}^{*}_{jk} \mathsf{p}^{\mathsf{SGE}}_{j}\mathsf{q}^{*}_k,
\quad
\mathcal{U}_{\boldsymbol{\mathsf{A}}}\left(2\boldsymbol{\mathsf{T}}^{\mathsf{SGE}}\boldsymbol{\mathsf{b}}^{*}\right)=\frac{1}{2} \mathsf{A}^{*}_{kl}\mathsf{q}^{*}_k\mathsf{q}^{*}_l ,\quad
\begin{array}{ll}
i,j=1,2,3, \\
k,l=1,2,3,4,
\end{array}
\label{eq:EneCont5}
\end{equation}
where $\boldsymbol{\mathsf{q}}^{*}=2\boldsymbol{\mathsf{b}}^{*}$ and
\begin{equation}
\boldsymbol{\mathsf{M}}^{*}=\boldsymbol{\mathsf{M}}\boldsymbol{\mathsf{T}}^{\mathsf{SGE}},
\qquad
\boldsymbol{\mathsf{A}}^{*}=\left(\boldsymbol{\mathsf{T}}^{\mathsf{SGE}}\right)^T\boldsymbol{\mathsf{A}}\boldsymbol{\mathsf{T}}^{\mathsf{SGE}}.
\end{equation}

Matrices $\boldsymbol{\mathsf{M}}^{*}$ and $\boldsymbol{\mathsf{A}}^{*}$ have reduced dimensions, so that the former is a rectangular 3 $\times$ 4 matrix and the latter a symmetric square matrix of order 4. They define the \emph{condensed representation} for the constitutive 
matrices $\boldsymbol{\mathsf{M}}$ and $\boldsymbol{\mathsf{A}}$, so that the strain energy density of the second-gradient elastic material can be seen as 
a function of $\boldsymbol{\mathsf{a}}$ and $\boldsymbol{\mathsf{b}}^{*}$, namely, $\mathcal{U}_{\mathsf{SGE}}\left(\boldsymbol{\mathsf{a}},\boldsymbol{\mathsf{b}}^{*}\right)$.

The  elastic energy stored in a hexagonal domain $\Omega^{(m,n)}$  made up of a second-gradient elastic continuum (with the same shape and location of the lattice's unit cell $\{m,n\}$) is obtained through volume integration of the strain energy density
\begin{equation}
\begin{split}
	\mathsf{U}_{\mathsf{SGE}}^{(m,n)}\left(\boldsymbol{\mathsf{a}}, \boldsymbol{\mathsf{b}}^{*}\right)&= \int_{\Omega^{(m,n)}}\mathcal{U}_{\mathsf{SGE}}\left(\boldsymbol{\mathsf{a}}, \boldsymbol{\mathsf{b}}^{*}\right) d \Omega ,
\end{split}
\end{equation}
which is evaluated as
\begin{equation}
\begin{split}
	\mathsf{U}_{\mathsf{SGE}}^{(m,n)}\left(\boldsymbol{\mathsf{a}}, \boldsymbol{\mathsf{b}}^{*}\right)&= \ell^2\left\{ \boldsymbol{\mathsf{a}} \scalp \boldsymbol{\mathsf{G}}^{[1]}\kern-0.3em\left(\mathsf{C}_{ij}\right)\boldsymbol{\mathsf{a}}
	+ 2\ell \boldsymbol{\mathsf{a}}\scalp\left[ m \boldsymbol{\mathsf{G}}^{[2]}\kern-0.3em\left(\mathsf{C}_{ij}\right) + n \boldsymbol{\mathsf{G}}^{[3]}\kern-0.3em\left(\mathsf{C}_{ij}\right) + \boldsymbol{\mathsf{G}}^{[4]}\kern-0.3em\left(\mathsf{M}^{*}_{ij}\right)\right]\boldsymbol{\mathsf{b}}^{*}\right.\\
	&\left. + \ell^2 \boldsymbol{\mathsf{b}}^{*} \scalp\left[m^2 \boldsymbol{\mathsf{G}}^{[5]}\kern-0.3em\left(\mathsf{C}_{ij}\right) + n^2 \boldsymbol{\mathsf{G}}^{[6]}\kern-0.3em\left(\mathsf{C}_{ij}\right) + m\,n \boldsymbol{\mathsf{G}}^{[7]}\kern-0.3em\left(\mathsf{C}_{ij}\right) + m\boldsymbol{\mathsf{G}}^{[8]}\kern-0.3em\left(\mathsf{M}^{*}_{ij}\right) \right.\right. \\
	&\left.\left. + n \boldsymbol{\mathsf{G}}^{[9]}\kern-0.3em\left(\mathsf{M}^{*}_{ij}\right) + \boldsymbol{\mathsf{G}}^{[10]}\kern-0.3em\left(\mathsf{C}_{ij},\mathsf{A}^{*}_{ij}\right) \right] \boldsymbol{\mathsf{b}}^{*}\right\},
\end{split}
\label{eq:Ene2ordCont2}
\end{equation}
where the coefficients of the matrices $\boldsymbol{\mathsf{G}}^{[r]}$ ($r=1,...,10$) are reported in Appendix A.

\subsection{Identification of the \lq condensed' second-gradient material equivalent to the lattice structure}
\label{Homogenization}

By imposing the elastic energy matching between the lattice, Eq.~(\ref{eq:Ene2ordRVE}), and for the moment unknown effective second-gradient material in the \lq condensed form', Eq.~(\ref{eq:Ene2ordCont2}), to hold for every unit cell $\{m,n\}$ and every pair of vectors $\boldsymbol{\mathsf{a}}$ and $\boldsymbol{\mathsf{b}}^{*}$
\begin{equation}
\mathsf{U}_{\mathsf{lat}}^{(m,n)}\left(\boldsymbol{\mathsf{a}}, \boldsymbol{\mathsf{b}}^{*}\right) = \mathsf{U}_{\mathsf{SGE}}^{(m,n)}\left(\boldsymbol{\mathsf{a}}, \boldsymbol{\mathsf{b}}^{*}\right), \qquad \forall \,\, m,n, \boldsymbol{\mathsf{a}}, \boldsymbol{\mathsf{b}}^{*},
\label{eq:EnergyEqui}
\end{equation}
the following identities are obtained
\begin{equation}
\boldsymbol{\mathsf{G}}^{[r]}=\boldsymbol{\mathsf{H}}^{[r]}
\qquad\forall\,r\in[1,10].
\label{eq:DiffeRela}
\end{equation}

It is highlighted that imposing the energy equivalence, Eq.(\ref{eq:EnergyEqui}), at first-order ($\bbeta=\b0$ and therefore $\boldsymbol{\mathsf{b}}^{*}=\b0$) implies
\begin{equation}
\boldsymbol{\mathsf{G}}^{[1]}=\boldsymbol{\mathsf{H}}^{[1]},
\label{eq:DiffeRela2}
\end{equation}
providing all the coefficients of the matrix $\boldsymbol{\mathsf{C}}$ as
	\begin{equation}
	\begin{array}{ccc}
		\mathsf{C}_{11} = \mathsf{C}_{22} = \dfrac{2I_{[1]}I_{[2]} + 9I_{[3]}}{4 \sqrt{3} I_{[2]}},\quad
		\mathsf{C}_{12} = \dfrac{2I_{[1]}I_{[2]} - 9I_{[3]}}{4 \sqrt{3} I_{[2]}},\quad
		\mathsf{C}_{13} = \mathsf{C}_{23} = 0,\quad
		\mathsf{C}_{33} = \dfrac{\mathsf{C}_{11}-\mathsf{C}_{12}}{2} = \dfrac{9I_{[3]}}{4 \sqrt{3} I_{[2]}},
	\end{array}
	\label{eq:1stcostant}
	\end{equation}
which coincide with the corresponding constants obtained in \cite{day1992elastic} through a different identification technique. From the first-order result,
Eq. (\ref{eq:1stcostant}), it follows that
 the two transformation matrices are the same for both the lattice and the equivalent material, namely,
\begin{equation}
\boldsymbol{\mathsf{T}}^{\mathsf{lat}}=\boldsymbol{\mathsf{T}}^{\mathsf{SGE}},
\end{equation}
so that $\boldsymbol{\mathsf{b}}^{\mathsf{lat}} = \boldsymbol{\mathsf{b}}^{\mathsf{SGE}}$ and therefore $\bbeta^{\mathsf{lat}}\left(\boldsymbol{\mathsf{b}}^{*}\right)=\bbeta^{\mathsf{SGE}}\left(\boldsymbol{\mathsf{b}}^{*}\right)$, meaning that the linear and quadratic components of the displacement field imposed to both the solid and the lattice coincide.

The non-local properties can now be identified from  Eq. (\ref{eq:DiffeRela}) for $r=2,...,10$. In particular,  the ten components of the matrix $\boldsymbol{\mathsf{A}}^{*}$ are identified as
\begin{equation}
\begin{split}
&\hspace{2mm} \mathsf{A}^{*}_{13} = 0, \quad
\mathsf{A}^{*}_{14} = 0, \quad
\mathsf{A}^{*}_{23} = 0, \quad
\mathsf{A}^{*}_{24} = 0
,\\
&\begin{array}{lcl}
\mathsf{A}^{*}_{11} &=& \dfrac{\sqrt{3} I_{[3]} \ell^2}{64  I_{[1]}^2 I_{[2]}^4}\left[-50 \overline{k}^5 \left(\widehat{k}+\widetilde{k}\right)^3-\overline{k}^4 \left(\widehat{k}+\widetilde{k}\right)^2 \left(100 \widehat{k}^2+359 \widehat{k} \widetilde{k}+100 \widetilde{k}^2\right) + \right.\\
&&\left. - \overline{k}^3 \left(\widehat{k}+\widetilde{k}\right) \left(50 \widehat{k}^4+419 \widehat{k}^3 \widetilde{k}+339 \widehat{k}^2 \widetilde{k}^2+419 \widehat{k} \widetilde{k}^3+50 \widetilde{k}^4\right) + \right.\\
&&\left. + 2 \overline{k}^2 \widehat{k} \widetilde{k} \left(24 \widehat{k}^4+459 \widehat{k}^3 \widetilde{k}+1853 \widehat{k}^2 \widetilde{k}^2+459 \widehat{k} \widetilde{k}^3+24 \widetilde{k}^4\right) + \right.\\
&&\left. + \overline{k} \widehat{k}^2 \widetilde{k}^2 \left(\widehat{k}+\widetilde{k}\right) \left(219 \widehat{k}^2+1283 \widehat{k} \widetilde{k}+219 \widetilde{k}^2\right)+121 \widehat{k}^3 \widetilde{k}^3 \left(\widehat{k}+\widetilde{k}\right)^2\right]
,\\
\mathsf{A}^{*}_{12} &=& \dfrac{\sqrt{3} I_{[3]} \ell^2}{64  I_{[1]}^2 I_{[2]}^4}\left[10 \overline{k}^5 \left(\widehat{k}+\widetilde{k}\right)^3+5 \overline{k}^4 \left(\widehat{k}+\widetilde{k}\right)^2 \left(4 \widehat{k}^2+5 \widehat{k} \widetilde{k}+4 \widetilde{k}^2\right) + \right.\\
&&\left. + \overline{k}^3 \left(\widehat{k}+\widetilde{k}\right) \left(10 \widehat{k}^4-71 \widehat{k}^3 \widetilde{k}-303 \widehat{k}^2 \widetilde{k}^2-71 \widehat{k} \widetilde{k}^3+10 \widetilde{k}^4\right) + \right.\\
&&\left. + 2 \overline{k}^2 \widehat{k} \widetilde{k} \left(6 \widehat{k}^4-9 \widehat{k}^3 \widetilde{k}+641 \widehat{k}^2 \widetilde{k}^2-9 \widehat{k} \widetilde{k}^3+6 \widetilde{k}^4\right) + \right.\\
&&\left. - \overline{k} \widehat{k}^2 \widetilde{k}^2 \left(\widehat{k}+\widetilde{k}\right) \left(33 \widehat{k}^2+\widehat{k} \widetilde{k}+33 \widetilde{k}^2\right)-35 \widehat{k}^3 \widetilde{k}^3 \left(\widehat{k}+\widetilde{k}\right)^2\right]
,\\
\mathsf{A}^{*}_{22} &=& \dfrac{\sqrt{3} I_{[3]} \ell^2}{64  I_{[1]}^2 I_{[2]}^4}\left[-10 \overline{k}^5 \left(\widehat{k}+\widetilde{k}\right)^3-\overline{k}^4 \left(\widehat{k}+\widetilde{k}\right)^2 \left(20 \widehat{k}^2-137 \widehat{k} \widetilde{k}+20 \widetilde{k}^2\right) + \right.\\
&&\left. - \overline{k}^3 \left(\widehat{k}+\widetilde{k}\right) \left(10 \widehat{k}^4-53 \widehat{k}^3 \widetilde{k}+219 \widehat{k}^2 \widetilde{k}^2-53 \widehat{k} \widetilde{k}^3+10 \widetilde{k}^4\right) + \right.\\
&&\left. + 2 \overline{k}^2 \widehat{k} \widetilde{k} \left(12 \widehat{k}^4-45 \widehat{k}^3 \widetilde{k}+349 \widehat{k}^2 \widetilde{k}^2-45 \widehat{k} \widetilde{k}^3+12 \widetilde{k}^4\right) + \right.\\
&&\left. + \overline{k} \widehat{k}^2 \widetilde{k}^2 \left(\widehat{k}+\widetilde{k}\right) \left(51 \widehat{k}^2-197 \widehat{k} \widetilde{k}+51 \widetilde{k}^2\right)+17 \widehat{k}^3 \widetilde{k}^3 \left(\widehat{k}+\widetilde{k}\right)^2\right],\\
\mathsf{A}^{*}_{33} &=& \dfrac{\sqrt{3} \ell^2}{192  I_{[1]}^2 I_{[2]}^4}\left[2 \overline{k}^6 \left(\widehat{k}+\widetilde{k}\right)^3 \left(4 \widehat{k}^2-7 \widehat{k} \widetilde{k}+4 \widetilde{k}^2\right) + \right.\\
&&\left. + \overline{k}^5 \left(\widehat{k}+\widetilde{k}\right)^2 \left(16 \widehat{k}^4-132 \widehat{k}^3 \widetilde{k}+181 \widehat{k}^2 \widetilde{k}^2-132 \widehat{k} \widetilde{k}^3+16 \widetilde{k}^4\right) + \right.\\
&&\left. + \overline{k}^4 \left(\widehat{k}+\widetilde{k}\right) \left(8 \widehat{k}^6-110 \widehat{k}^5 \widetilde{k}+301 \widehat{k}^4 \widetilde{k}^2+667 \widehat{k}^3 \widetilde{k}^3+301 \widehat{k}^2 \widetilde{k}^4-110 \widehat{k} \widetilde{k}^5+8 \widetilde{k}^6\right) + \right.\\
&&\left. + 2 \overline{k}^3 \widehat{k} \widetilde{k} \left(4 \widehat{k}^6+27 \widehat{k}^5 \widetilde{k}-101 \widehat{k}^4 \widetilde{k}^2-587 \widehat{k}^3 \widetilde{k}^3-101 \widehat{k}^2 \widetilde{k}^4+27 \widehat{k} \widetilde{k}^5+4 \widetilde{k}^6\right) + \right.\\
&&\left. - \overline{k}^2 \widehat{k}^2 \widetilde{k}^2 \left(\widehat{k}+\widetilde{k}\right) \left(6 \widehat{k}^4-121 \widehat{k}^3 \widetilde{k}-349 \widehat{k}^2 \widetilde{k}^2-121 \widehat{k} \widetilde{k}^3+6 \widetilde{k}^4\right) + \right.\\
&&\left. - \overline{k} \widehat{k}^3 \widetilde{k}^3 \left(\widehat{k}+\widetilde{k}\right)^2 \left(4 \widehat{k}^2+43 \widehat{k} \widetilde{k}+4 \widetilde{k}^2\right)+2 \widehat{k}^4 \widetilde{k}^4 \left(\widehat{k}+\widetilde{k}\right)^3\right]
\end{array}
\end{split}
\end{equation}
\begin{equation*}
\begin{array}{lcl}
\mathsf{A}^{*}_{34} &=& \dfrac{\sqrt{3} \ell^2}{64  I_{[1]}^2 I_{[2]}^4}\left[-2 \overline{k}^6 \left(\widehat{k}+\widetilde{k}\right)^3 \left(4 \widehat{k}^2+3 \widehat{k} \widetilde{k}+4 \widetilde{k}^2\right) + \right.\\
&&\left. - \overline{k}^5 \left(\widehat{k}+\widetilde{k}\right)^2 \left(16 \widehat{k}^4+4 \widehat{k}^3 \widetilde{k}-63 \widehat{k}^2 \widetilde{k}^2+4 \widehat{k} \widetilde{k}^3+16 \widetilde{k}^4\right) + \right.\\
&&\left. - \overline{k}^4 \left(\widehat{k}+\widetilde{k}\right) \left(8 \widehat{k}^6+6 \widehat{k}^5 \widetilde{k}-267 \widehat{k}^4 \widetilde{k}^2-173 \widehat{k}^3 \widetilde{k}^3-267 \widehat{k}^2 \widetilde{k}^4+6 \widehat{k} \widetilde{k}^5+8 \widetilde{k}^6\right) + \right.\\
&&\left. - 2 \overline{k}^3 \widehat{k} \widetilde{k} \left(4 \widehat{k}^6-15 \widehat{k}^5 \widetilde{k}+115 \widehat{k}^4 \widetilde{k}^2+461 \widehat{k}^3 \widetilde{k}^3+115 \widehat{k}^2 \widetilde{k}^4-15 \widehat{k} \widetilde{k}^5+4 \widetilde{k}^6\right) + \right.\\
&&\left. + \overline{k}^2 \widehat{k}^2 \widetilde{k}^2 \left(\widehat{k}+\widetilde{k}\right) \left(6 \widehat{k}^4+71 \widehat{k}^3 \widetilde{k}+43 \widehat{k}^2 \widetilde{k}^2+71 \widehat{k} \widetilde{k}^3+6 \widetilde{k}^4\right) + \right.\\
&&\left. + \overline{k} \widehat{k}^3 \widetilde{k}^3 \left(\widehat{k}+\widetilde{k}\right)^2 \left(4 \widehat{k}^2+35 \widehat{k} \widetilde{k}+4 \widetilde{k}^2\right)-2 \widehat{k}^4 \widetilde{k}^4 \left(\widehat{k}+\widetilde{k}\right)^3\right]
,\\
\mathsf{A}^{*}_{44} &=& \dfrac{\sqrt{3} \ell^2}{64  I_{[1]}^2 I_{[2]}^4}\left[2 \overline{k}^6 \left(\widehat{k}+\widetilde{k}\right)^3 \left(12 \widehat{k}^2-\widehat{k} \widetilde{k}+12 \widetilde{k}^2\right) + \right.\\
&&\left. + \overline{k}^5 \left(\widehat{k}+\widetilde{k}\right)^2 \left(48 \widehat{k}^4+260 \widehat{k}^3 \widetilde{k}+103 \widehat{k}^2 \widetilde{k}^2+260 \widehat{k} \widetilde{k}^3+48 \widetilde{k}^4\right) + \right.\\
&&\left. + \overline{k}^4 \left(\widehat{k}+\widetilde{k}\right) \left(24 \widehat{k}^6+286 \widehat{k}^5 \widetilde{k}+583 \widehat{k}^4 \widetilde{k}^2-255 \widehat{k}^3 \widetilde{k}^3+583 \widehat{k}^2 \widetilde{k}^4+286 \widehat{k} \widetilde{k}^5+24 \widetilde{k}^6\right) + \right.\\
&&\left. + 2 \overline{k}^3 \widehat{k} \widetilde{k} \left(12 \widehat{k}^6-3 \widehat{k}^5 \widetilde{k}-735 \widehat{k}^4 \widetilde{k}^2-1753 \widehat{k}^3 \widetilde{k}^3-735 \widehat{k}^2 \widetilde{k}^4-3 \widehat{k} \widetilde{k}^5+12 \widetilde{k}^6\right) + \right.\\
&&\left. - \overline{k}^2 \widehat{k}^2 \widetilde{k}^2 \left(\widehat{k}+\widetilde{k}\right) \left(18 \widehat{k}^4+309 \widehat{k}^3 \widetilde{k}+937 \widehat{k}^2 \widetilde{k}^2+309 \widehat{k} \widetilde{k}^3+18 \widetilde{k}^4\right) + \right.\\
&&\left. - \overline{k} \widehat{k}^3 \widetilde{k}^3 \left(\widehat{k}+\widetilde{k}\right)^2 \left(12 \widehat{k}^2+17 \widehat{k} \widetilde{k}+12 \widetilde{k}^2\right)+6 \widehat{k}^4 \widetilde{k}^4 \left(\widehat{k}+\widetilde{k}\right)^3\right],
\end{array}
\label{eq:2stcostant}
\end{equation*}
while the twelve components of the matrix $\boldsymbol{\mathsf{M}}^{*}$  as
\begin{equation}
\begin{array}{ccc}
\mathsf{M}^{*}_{11} = \mathsf{M}^{*}_{12} = \mathsf{M}^{*}_{21} = \mathsf{M}^{*}_{22} = \mathsf{M}^{*}_{31} = \mathsf{M}^{*}_{32} = \mathsf{M}^{*}_{33} = \mathsf{M}^{*}_{34} =0,
\vspace{4mm}\\
\mathsf{M}^{*}_{13} = \mathsf{M}^{*}_{23} = \dfrac{\left(\widehat{k}-\widetilde{k}\right)  \left(I_{[1]} I_{[2]}-9 I_{[3]}\right)  \left(\widehat{k} \widetilde{k}-2 \overline{k} \left(\widehat{k}+\widetilde{k}\right) \right) }{8 \sqrt{3}  I_{[1]} I_{[2]}^2}\ell,
\vspace{4mm}\\
\mathsf{M}^{*}_{14} = \mathsf{M}^{*}_{24} = -\dfrac{3 \left(\widehat{k}-\widetilde{k}\right) \left(I_{[1]} I_{[2]}+3 I_{[3]}\right) \left(\widehat{k} \widetilde{k}-2 \overline{k} \left(\widehat{k}+\widetilde{k}\right)\right)}{8 \sqrt{3}  I_{[1]} I_{[2]}^2}\ell.
\end{array}
\label{eq:3stcostant}
\end{equation}
It is worth to note that the result provided by Eqs.~(\ref{eq:1stcostant})--(\ref{eq:3stcostant}) shows that
the constitutive matrices are invariant with respect the following permutations of $\{\overline{k},\widehat{k},\widetilde{k}\}$:
\begin{equation}
\begin{array}{ccc}
\boldsymbol{\mathsf{C}}\left(\kappa_1, \kappa_2, \kappa_3\right) = \boldsymbol{\mathsf{C}}\left(\kappa_1, \kappa_3, \kappa_2\right) = 
	\boldsymbol{\mathsf{C}}\left(\kappa_2, \kappa_1, \kappa_3\right) = \boldsymbol{\mathsf{C}}\left(\kappa_2, \kappa_3, \kappa_1\right) = 
	\boldsymbol{\mathsf{C}}\left(\kappa_3, \kappa_1, \kappa_2\right) = \boldsymbol{\mathsf{C}}\left(\kappa_3, \kappa_2, \kappa_1\right)\\
	\boldsymbol{\mathsf{A}}^{*}\left(\kappa_1, \kappa_2, \kappa_3\right) = \boldsymbol{\mathsf{A}}^{*}\left(\kappa_1, \kappa_3, \kappa_2\right),\qquad
	\boldsymbol{\mathsf{M}}^{*}\left(\kappa_1, \kappa_2, \kappa_3\right) = -\boldsymbol{\mathsf{M}}^{*}\left(\kappa_1, \kappa_3, \kappa_2\right).
	\end{array}
	\end{equation}

It can be therefore concluded that 
\begin{center}
\begin{quote}
\it{
the effective response approaches that of a Cauchy elastic material only in the limit of vanishing length of the lattice's bars, $\ell \longrightarrow 0$, a condition for which $\mathsf{M}^{*}_{ij}=\mathsf{A}^{*}_{ij}=0$.
}
\end{quote}
\end{center}
Finally, 
from Eqs.~(\ref{eq:1stcostant})--(\ref{eq:3stcostant}) it is evident that the stiffness ratio between the bars may have a dramatic effect on the equivalent solid response, as further discussed in second part of this article \cite{rizzipt2}.

\subsection{Influence of the additional field $\bDelta \bu^{\{m,n|i\}}$}

It is remarked that, although $\bbeta^{\mathsf{SGE}}=\bbeta^{\textsf{lat}}
$, the displacement fields imposed  to the lattice differs from that imposed to the equivalent solid due to the presence of the additional field $\bDelta\bu^{(m,n|i)}$ in the former. From the practical point of view, however, the amplitude of such an additional field 
does not play an important role when compared to the amplitude of the quadratic part, so that the deformed configuration of the solid very well represents that of the lattice, even if in the latter the additional field is present. 

\begin{figure}[H]
	\centering
	\includegraphics[width=0.8\textwidth]{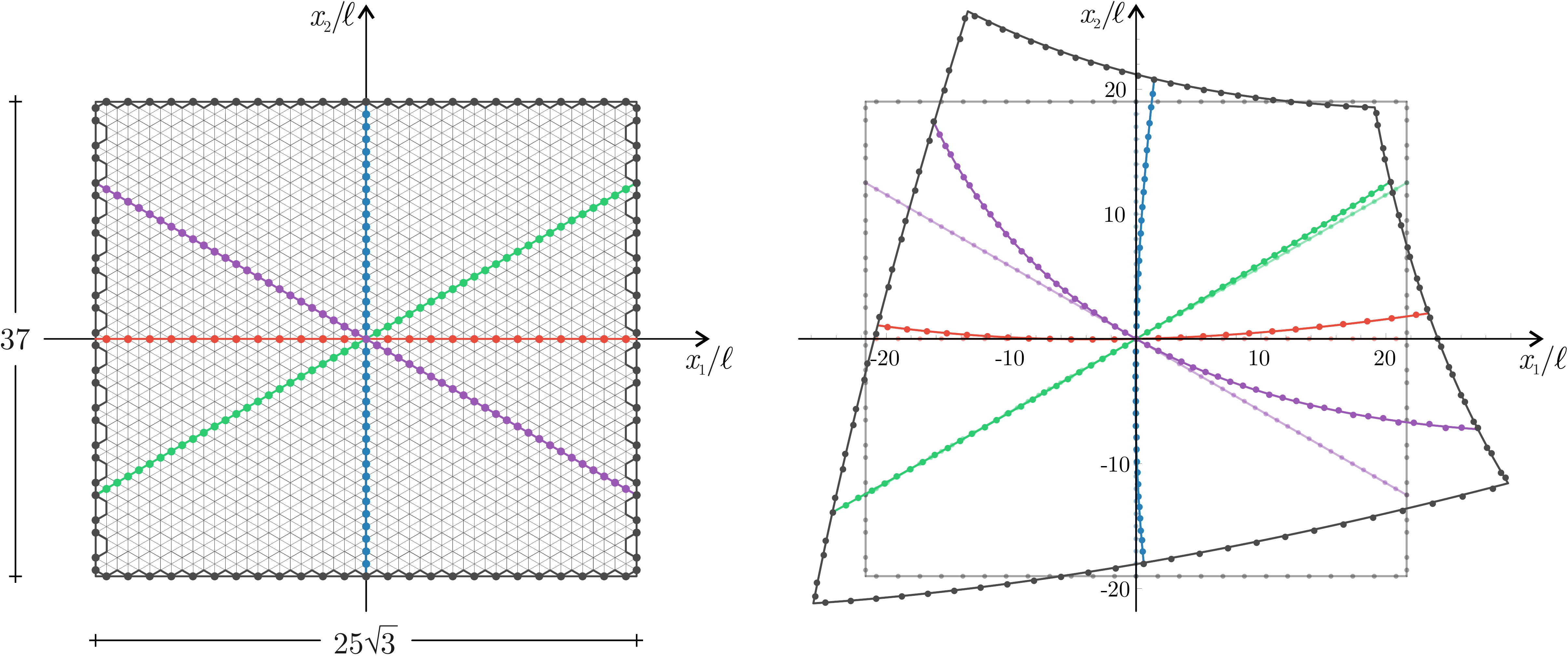}
	\caption{(Left) Rectangular domain (having sides $25\sqrt{3}\ell$ $\times$ $37\ell$)  occupied in one case by the lattice (625 hexagonal unit cells) and in the other case by the equivalent continuum (only its boundary is reported).
		(Right) Undeformed and deformed configurations for initially straight lines of the equivalent continuum, when subject to a linear and quadratic displacement field. The same displacement, plus the additional field $\bDelta \bu^{\{m,n|i\}}$, are applied to the lattice, of which the nodal positions are reported 
		as spots.
		The additional field $\bDelta \bu^{\{m,n|i\}}$ is observed to play only a marginal role in the overall deformation of the lattice.}
	\label{fig:CompaContDisc}
\end{figure}

To analyze the influence of the additional field on the kinematics of the lattice and of the equivalent solid, a rectangular domain (having sides $25\sqrt{3}\ell$ $\times$ $37\ell$) is considered, occupied in one case by the lattice, which is shown on the left in Fig.~\ref{fig:CompaContDisc}, (625 hexagonal unit cells, namely, 25 along each axis of the rectangle) and in the other case by the equivalent continuum with its boundary reported on the left in Fig.~\ref{fig:CompaContDisc}. The solid is subject to a displacement field characterized by tensors $\balpha$ and $\bbeta^{\mathsf{SGE}}$, while the lattice is subject to the same $\balpha$ and to $\bbeta^{\textsf{lat}}=\bbeta^{\mathsf{SGE}}$ plus the additional field $\bDelta \bu^{\{m,n|i\}}$. 
In particular, the following values have been selected to produce the figure $\alpha_{11}=0.0\overline{18}$, $\alpha_{22}=0.02$, $\alpha_{12}=0.0\overline{2}$ and $\beta_{111}^{\mathsf{SGE}}=\beta_{111}^{\textsf{lat}}=0.0029$, $\beta_{221}^{\mathsf{SGE}}=\beta_{221}^{\textsf{lat}}=0.0028\overline{6}$, $\beta_{112}^{\mathsf{SGE}}=\beta_{112}^{\textsf{lat}}=0.00\overline{3}$, $\beta_{222}^{\mathsf{SGE}}=\beta_{222}^{\textsf{lat}}=0.004$. Moreover, having selected $\widehat{k}/\overline{k}=2$ and $\widetilde{k}/\overline{k}=3$ as bars' stiffness ratios, the remaining two components of $\bbeta^{\mathsf{SGE}}$ result from Eq. (\ref{eq:EquiConstrainBeta}) and (\ref{eq:selfequiResults}) as $\beta_{211}^{\mathsf{SGE}}=\beta_{211}^{\textsf{lat}}=-0.007$ and $\beta_{122}^{\mathsf{SGE}}=\beta_{122}^{\textsf{lat}}=-0.0052$. 
The additional field $\bDelta \bu^{\{m,n|i\}}$ applied to the lattice has been calculated with the given values of $\balpha$ and $\bbeta^{\textsf{lat}}$ through Eq.~(\ref{eq:CorrVWvw}) and (\ref{eq:CorrZ}).

The undeformed and deformed configurations (visible as lines for the equivalent solid and as spots for the lattice) are reported in Fig.~\ref{fig:CompaContDisc}.
The positions of the undeformed lattice's nodes were chosen to lie on the undeformed lines of the continuum. The fact that, after deformation, the dots overlap the deformed lines demonstrates that the additional field (needed to enforce equilibrium in the lattice) affects only marginally the overall displacement of the lattice, in which the linear and quadratic displacement fields prevail.

\section{Discussion}

\begin{figure}[H]
	\centering
	\includegraphics[width=\textwidth]{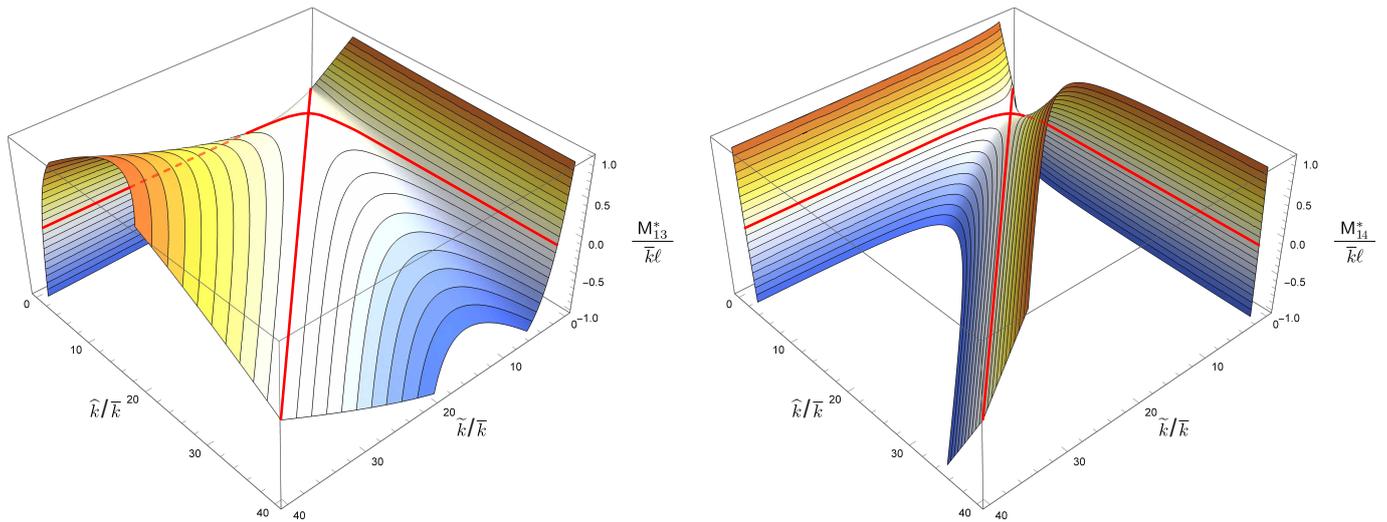}
	\caption{Nonlocal constitutive parameters $\mathsf{M}^{*}_{13}$ (left) and $\mathsf{M}^{*}_{14}$  (right) as functions of the bar stiffness ratios  $\widehat{k}/\overline{k}$ and $\widetilde{k}/\overline{k}$. The red lines represent the  stiffness ratios pairs for which a centrosymmetric response is attained, while in all the other cases the 
	solid equivalent to the hexagonal bars' lattice displays a non-centrosymmetric mechanical behaviour.}
	\label{fig:m13}
\end{figure}

An infinite hexagonal lattice of bars (only subject to axial forces and characterized by three different elastic stiffnesses) has been considered and solved, when loaded at infinity with a quadratic displacement field, enhanced with an additional displacement to 
comply with the periodicity constraint of the lattice.
 Its elastic energy has been shown to match with that of a second-gradient 
(\lq form I' Mindlin) elastic material, subject to the same quadratic field. In this way, a homogeneous continuum, enriched with an internal length, has been derived, 
which is equivalent to the discrete lattice. However, this continuum was only identified in a \lq condensed' form, so that not all constitutive parameters have been identified. 
For those appearing in the condensed version, 
closed form expressions have been given, showing the influence of the lattice properties (the hexagon side length $\ell$ and the bars stiffness $\overline{k}$, $\widehat{k}$, $\widetilde{k}$). As an example, the higher-order constitutive parameters $\mathsf{M}^{*}_{13}$ and $\mathsf{M}^{*}_{14}$ ruling the non-centrosymmetric behaviour (and made dimensionless through division by $\overline{k} \ell$) are portrayed in Fig. \ref{fig:m13} where  two stiffness ratios $\widehat{k}/\overline{k}$ and $\widetilde{k}/\overline{k}$ are varied. The red lines highlight the condition for which both parameters vanish, so that, correspondingly, centrosymmetric response is retrieved, while in all the other cases non-centrosymmetry characterizes the mechanical behaviour of the equivalent material. 

The fact that the equivalent material is only defined in a \lq condensed' form is a consequence of the fact that the elastic energy equivalence between the solid and the lattice has been so far restricted to self-equilibrated displacement fields. This means, in other words, that the mechanical tests applied both to the lattice and to the continuum are not enough in number to completely characterize the latter. 
Nevertheless, the presented results allow already to conclude that even a simple hexagonal lattice, which corresponds to an equivalent isotropic, local, and centrosymmetric material at a first-order of approximation, at a higher approximation displays anisotropic, non-local, and non-centrosymmetric effects. 
Therefore, the presented results provide a tool for advanced mechanical design of microstructured solids.
The complete derivation of all material constants of the second-gradient equivalent elastic solids is deferred to Part II of this study, together with 
the analysis of positive definitess and symmetry of the equivalent material and with an assessment of the validity of the second-gradient model.

\paragraph{Acknowledgements.} G.R., D.V., F.D.C. gratefully
acknowledge financial support from the grant ERC Advanced Grant \lq
Instabilities and nonlocal multiscale modelling of materials' ERC-2013-ADG-340561-INSTABILITIES. 
D.B. gratefully acknowledges financial support from PRIN 2015 \lq Multi-scale mechanical models for the design and optimization of micro-structured smart materials and metamaterials' 2015LYYXA8-006.
\bibliographystyle{plain}
\bibliography{Biblio1}

\begin{thebibliography}{10}

\bibitem{Seppecher2018}
H.~Abdoul-Anziz and P.~Seppecher.
\newblock Strain gradient and generalized continua obtained by homogenizing
  frame lattices.
\newblock {\em Mathematics and Mechanics of Complex Systems}, 6(3):213--250,
  2018.

\bibitem{askar1968structural}
A.~Askar and A.S. Cakmak.
\newblock A structural model of a micropolar continuum.
\newblock {\em International Journal of Engineering Science}, 6(10):583--589,
  1968.

\bibitem{bacca2013mindlin}
M.~Bacca, D.~Bigoni, F.~Dal~Corso, and D.~Veber.
\newblock Mindlin second-gradient elastic properties from dilute two-phase
  cauchy-elastic composites. part \mbox{I}: Closed form expression for the
  effective higher-order constitutive tensor.
\newblock {\em International Journal of Solids and Structures},
  50(24):4010--4019, 2013.

\bibitem{mattia2013mindlin}
M.~Bacca, D.~Bigoni, F.~Dal~Corso, and D.~Veber.
\newblock Mindlin second-gradient elastic properties from dilute two-phase
  cauchy-elastic composites part \mbox{II}: Higher-order constitutive
  properties and application cases.
\newblock {\em International Journal of Solids and Structures}, 2013.

\bibitem{bacigalupo2012computational}
A.~Bacigalupo and L.~Gambarotta.
\newblock Computational two-scale homogenization of periodic masonry:
  characteristic lengths and dispersive waves.
\newblock {\em Computer Methods in Applied Mechanics and Engineering},
  213:16--28, 2012.

\bibitem{bacigalupo2014second2}
A.~Bacigalupo and L.~Gambarotta.
\newblock Second-gradient homogenized model for wave propagation in
  heterogeneous periodic media.
\newblock {\em International Journal of Solids and Structures},
  51(5):1052--1065, 2014.

\bibitem{BACIGALUPO2017}
A.~Bacigalupo, F.~Paggi, M.and Dal~Corso, and D.~Bigoni.
\newblock Identification of higher-order continua equivalent to a cauchy
  elastic composite.
\newblock {\em Mechanics Research Communications}, 2017.

\bibitem{begley1998mechanics}
M.R. Begley and J.W. Hutchinson.
\newblock The mechanics of size-dependent indentation.
\newblock {\em Journal of the Mechanics and Physics of Solids},
  46(10):2049--2068, 1998.

\bibitem{BEVERIDGE2013246}
A.J. Beveridge, M.A. Wheel, and D.H. Nash.
\newblock The micropolar elastic behaviour of model macroscopically
  heterogeneous materials.
\newblock {\em International Journal of Solids and Structures}, 50(1):246 --
  255, 2013.

\bibitem{bigoni2007analytical}
D.~Bigoni and W.J. Drugan.
\newblock Analytical derivation of cosserat moduli via homogenization of
  heterogeneous elastic materials.
\newblock {\em Journal of Applied Mechanics}, 74(4):741--753, 2007.

\bibitem{born1954dynamical}
M.~Born and K.~Huang.
\newblock {\em Dynamical theory of crystal lattices}.
\newblock Clarendon press, 1954.

\bibitem{buechner2003size}
P.M. Buechner and R.S. Lakes.
\newblock Size effects in the elasticity and viscoelasticity of bone.
\newblock {\em Biomechanics and modeling in mechanobiology}, 1(4):295--301,
  2003.

\bibitem{cauchy1828}
A.L. Cauchy.
\newblock Sur l'equilibre et le mouvement d'un syst{\'e}me de points materiels
  sollicit{\'e}s par forces d'attraction ou de r{\'e}pulsion mutuelle.
\newblock {\em Ex. de Math.}, pages 187--213, 1828.

\bibitem{dal2011stability}
F.~Dal~Corso and J.R. Willis.
\newblock Stability of strain-gradient plastic materials.
\newblock {\em Journal of the Mechanics and Physics of Solids},
  59(6):1251--1267, 2011.

\bibitem{danas2012size}
K.~Danas, V.S. Deshpande, and N.A. Fleck.
\newblock Size effects in the conical indentation of an elasto-plastic solid.
\newblock {\em Journal of the Mechanics and Physics of Solids},
  60(9):1605--1625, 2012.

\bibitem{day1992elastic}
A.R. Day, K.A. Snyder, E.J. Garboczi, and M.F. Thorpe.
\newblock The elastic moduli of a sheet containing circular holes.
\newblock {\em Journal of the Mechanics and Physics of Solids},
  40(5):1031--1051, 1992.

\bibitem{Genoese2018}
A.~Genoese, A.~Genoese, N.~L. Rizzi, and Ginevra Salerno.
\newblock Force constants of bn, sic, aln and gan sheets through discrete
  homogenization.
\newblock {\em Meccanica}, 53(3):593--611, 2018.

\bibitem{gourgiotis2014steady}
P.A. Gourgiotis and A.~Piccolroaz.
\newblock Steady-state propagation of a mode \mbox{II} crack in couple stress
  elasticity.
\newblock {\em International Journal of Fracture}, 188(2):119--145, 2014.

\bibitem{gourgiotis2016analysis}
P.A. Gourgiotis, Th. Zisis, and K.P. Baxevanakis.
\newblock Analysis of the tilted flat punch in couple-stress elasticity.
\newblock {\em International Journal of Solids and Structures}, 85:34--43,
  2016.

\bibitem{keating1966effect}
P.N. Keating.
\newblock Effect of invariance requirements on the elastic strain energy of
  crystals with application to the diamond structure.
\newblock {\em Physical Review}, 145(2):637, 1966.

\bibitem{kirkwood1939skeletal}
J.G. Kirkwood.
\newblock The skeletal modes of vibration of long chain molecules.
\newblock {\em The Journal of Chemical Physics}, 7(7):506--509, 1939.

\bibitem{lakes1986experimental}
R.S. Lakes.
\newblock Experimental microelasticity of two porous solids.
\newblock {\em International Journal of Solids and Structures}, 22(1):55--63,
  1986.

\bibitem{latture}
R.M. Latture, M.R. Begley, and F.W. Zok.
\newblock Design and mechanical properties of elastically isotropic trusses.
\newblock {\em Journal of Materials Research}, 33(3):249--263, 2018.

\bibitem{le2013homogenization}
H.~Le~Dret and A.~Raoult.
\newblock Homogenization of hexagonal lattices.
\newblock {\em Networks and Heterogeneous Media}, 8(2):541--572, 2013.

\bibitem{mindlin1964micro}
R.D. Mindlin.
\newblock Micro-structure in linear elasticity.
\newblock {\em Archive for Rational Mechanics and Analysis}, 16(1):51--78,
  1964.

\bibitem{neumann1975equations}
H.P. Neumann.
\newblock Equations of state and phase transitions for some plane-lattice
  models.
\newblock {\em Physical Review A}, 11(3):1043, 1975.

\bibitem{ostoja2002lattice}
M.~Ostoja-Starzewski.
\newblock Lattice models in micromechanics.
\newblock {\em Applied Mechanics Reviews}, 55(1):35--60, 2002.

\bibitem{piccolroaz2012mode}
A.~Piccolroaz, G.~Mishuris, and E.~Radi.
\newblock Mode \mbox{III} interfacial crack in the presence of couple-stress
  elastic materials.
\newblock {\em Engineering Fracture Mechanics}, 80:60--71, 2012.

\bibitem{rizzipt2}
G.~Rizzi, D.~Veber, F.~Dal~Corso, and D.~Bigoni.
\newblock Identification of second-gradient elastic materials from planar
  hexagonal lattices. part \mbox{II}: Mechanical characteristics and model
  validation.
\newblock {\em International Journal of Solids and Structures},
  176--177:19--35, 2019.

\bibitem{Seppecher2011}
P.~Seppecher, J.-J. Alibert, and F.~Dell'Isola.
\newblock Linear elastic trusses leading to continua with exotic mechanical
  interactions.
\newblock {\em Journal of Physics: Conference Series}, 319:109--124, 09 2011.

\bibitem{Shi1995}
P.~Shi, G.and~Tong.
\newblock The derivation of equivalent constitutive equations of honeycomb
  structures by a two scale method.
\newblock {\em Computational Mechanics}, 15(5):395--407, 1995.

\bibitem{sluys1993wave}
L.J. Sluys, R.~De~Borst, and H.-B. M{\"u}hlhaus.
\newblock Wave propagation, localization and dispersion in a gradient-dependent
  medium.
\newblock {\em International Journal of Solids and Structures},
  30(9):1153--1171, 1993.

\bibitem{snyder1992elastic}
K.A. Snyder, E.J. Garboczi, and A.~R. Day.
\newblock The elastic moduli of simple two-dimensional isotropic composites:
  Computer simulation and effective medium theory.
\newblock {\em Journal of applied physics}, 72(12):5948--5955, 1992.

\bibitem{SPADONI2012156}
A.~Spadoni and M.~Ruzzene.
\newblock Elasto-static micropolar behavior of a chiral auxetic lattice.
\newblock {\em Journal of the Mechanics and Physics of Solids}, 60(1):156 --
  171, 2012.

\bibitem{suiker2001comparison}
A.S.J. Suiker, A.V. Metrikine, and R.~De~Borst.
\newblock Comparison of wave propagation characteristics of the cosserat
  continuum model and corresponding discrete lattice models.
\newblock {\em International Journal of Solids and Structures},
  38(9):1563--1583, 2001.

\bibitem{warren2002three}
W.~E. Warren and E.~Byskov.
\newblock Three-fold symmetry restrictions on two-dimensional micropolar
  materials.
\newblock {\em European Journal of Mechanics-A/Solids}, 21(5):779--792, 2002.

\bibitem{WASEEM2013148}
A.~Waseem, A.J. Beveridge, M.A. Wheel, and D.H. Nash.
\newblock The influence of void size on the micropolar constitutive properties
  of model heterogeneous materials.
\newblock {\em European Journal of Mechanics - A/Solids}, 40:148 -- 157, 2013.

\bibitem{zisis2015contact}
Th. Zisis, P.A. Gourgiotis, and F.~Dal~Corso.
\newblock A contact problem in couple stress thermoelasticity: The indentation
  by a hot flat punch.
\newblock {\em International Journal of Solids and Structures}, 63:226--239,
  2015.

\end{thebibliography}

\setcounter{equation}{0}
\renewcommand{\theequation}{{A.}\arabic{equation}}
\section*{Appendix A - Components of the matrices $\boldsymbol{\mathsf{H}}^{[r]}$ and $\boldsymbol{\mathsf{G}}^{[r]}$}

The coefficients of the matrices $\boldsymbol{\mathsf{H}}^{[r]}$ ($r=1,...,10$)  are
\begin{equation}
\begin{array}{cccc}
&\mathsf{H}_{11}^{[1]}=\dfrac{3}{16} \left(2 I_{[1]}+\dfrac{9 I_{[3]}}{I_{[2]}}\right),\quad
	\mathsf{H}_{12}^{[1]}=\dfrac{3}{16} \left(2 I_{[1]}-\dfrac{9 I_{[3]}}{I_{[2]}}\right),\quad
	\mathsf{H}_{13}^{[1]}=0,\\[5mm]
&\mathsf{H}_{22}^{[1]}=\dfrac{3}{16} \left(2 I_{[1]}+\dfrac{9 I_{[3]}}{I_{[2]}}\right),\quad
	\mathsf{H}_{23}^{[1]}=0,\quad
	\mathsf{H}_{33}^{[1]}=\dfrac{27 I_{[3]}}{16 I_{[2]}}
\end{array}
\end{equation}
\begin{equation}
\begin{array}{ccc}
&\mathsf{H}_{11}^{[2]}=\dfrac{27 \sqrt{3} I_{[3]} \left(2 I_{[1]} I_{[2]}+9 I_{[3]}\right)}{16 I_{[1]} I_{[2]}^2},\quad
	\mathsf{H}_{12}^{[2]}=\dfrac{27 \sqrt{3} I_{[3]} (9 I_{[3]}-2 I_{[1]} I_{[2]})}{16 I_{[1]} I_{[2]}^2},\quad
	\mathsf{H}_{13}^{[2]}=0,\\[5mm]
&	\mathsf{H}_{14}^{[2]}=0,\quad
	\mathsf{H}_{21}^{[2]}=-\dfrac{81 \sqrt{3} I_{[3]} \left(2 I_{[1]} I_{[2]}+3 I_{[3]}\right)}{16 I_{[1]} I_{[2]}^2},\quad
	\mathsf{H}_{22}^{[2]}=-\dfrac{27 \sqrt{3} I_{[3]} \left(2 I_{[1]} I_{[2]}+9 I_{[3]}\right)}{16 I_{[1]} I_{[2]}^2},\\[5mm]
&	\mathsf{H}_{23}^{[2]}=0,\quad
	\mathsf{H}_{24}^{[2]}=0,\quad
	\mathsf{H}_{31}^{[2]}=0,\\[5mm]
&	\mathsf{H}_{32}^{[2]}=0,\quad
	\mathsf{H}_{33}^{[2]}=\dfrac{27 \sqrt{3} I_{[3]} \left(2 I_{[1]} I_{[2]}-9 I_{[3]}\right)}{16 I_{[1]} I_{[2]}^2},\quad
	\mathsf{H}_{34}^{[2]}=-\dfrac{27 \sqrt{3} I_{[3]} \left(2 I_{[1]} I_{[2]}+9 I_{[3]}\right)}{16 I_{[1]} I_{[2]}^2}
\end{array}
\end{equation}
\begin{equation}
\begin{array}{ccc}
&	\mathsf{H}_{11}^{[3]}=\dfrac{27 \sqrt{3} I_{[3]} \left(2 I_{[1]} I_{[2]}+9 I_{[3]}\right)}{32 I_{[1]} I_{[2]}^2},\quad
	\mathsf{H}_{12}^{[3]}=\dfrac{27 \sqrt{3} I_{[3]} \left(9 I_{[3]}-2 I_{[1]} I_{[2]}\right)}{32 I_{[1]} I_{[2]}^2},\quad
	\mathsf{H}_{13}^{[3]}=-\dfrac{81 I_{[3]} \left(2 I_{[1]} I_{[2]}+9 I_{[3]}\right)}{32 I_{[1]} I_{[2]}^2},\\[5mm]
&	\mathsf{H}_{14}^{[3]}=-\dfrac{243 I_{[3]} \left(2 I_{[1]} I_{[2]}+3 I_{[3]}\right)}{32 I_{[1]} I_{[2]}^2},\quad
	\mathsf{H}_{21}^{[3]}=-\dfrac{81 \sqrt{3} I_{[3]} \left(2 I_{[1]} I_{[2]}+3 I_{[3]}\right)}{32 I_{[1]} I_{[2]}^2},\quad
	\mathsf{H}_{22}^{[3]}=-\dfrac{27 \sqrt{3} I_{[3]} \left(2 I_{[1]} I_{[2]}+9 I_{[3]}\right)}{32 I_{[1]} I_{[2]}^2},\\[5mm]
&	\mathsf{H}_{23}^{[3]}=\dfrac{81 I_{[3]} \left(9 I_{[3]}-2 I_{[1]} I_{[2]}\right)}{32 I_{[1]} I_{[2]}^2},\quad
	\mathsf{H}_{24}^{[3]}=\dfrac{81 I_{[3]} \left(2 I_{[1]} I_{[2]}+9 I_{[3]}\right)}{32 I_{[1]} I_{[2]}^2},\quad
	\mathsf{H}_{31}^{[3]}=-\dfrac{81 I_{[3]} \left(2 I_{[1]} I_{[2]}+9 I_{[3]}\right)}{32 I_{[1]} I_{[2]}^2},\\[5mm]
&	\mathsf{H}_{32}^{[3]}=\dfrac{81 I_{[3]} \left(2 I_{[1]} I_{[2]}-9 I_{[3]}\right)}{32 I_{[1]} I_{[2]}^2},\quad
	\mathsf{H}_{33}^{[3]}=\dfrac{27 \sqrt{3} I_{[3]} \left(2 I_{[1]} I_{[2]}-9 I_{[3]}\right)}{32 I_{[1]} I_{[2]}^2},\quad
	\mathsf{H}_{34}^{[3]}=-\dfrac{27 \sqrt{3} I_{[3]} \left(2 I_{[1]} I_{[2]}+9 I_{[3]}\right)}{32 I_{[1]} I_{[2]}^2}
\end{array}
\end{equation}
\begin{equation}
\begin{array}{ccc}
&	\mathsf{H}_{11}^{[4]}=0,\quad
	\mathsf{H}_{12}^{[4]}=0,\quad
	\mathsf{H}_{13}^{[4]}=\dfrac{3  \left(\widehat{k}-\widetilde{k}\right) \left(I_{[1]} I_{[2]}-9 I_{[3]}\right) \left(\widehat{k} \widetilde{k}-2 k \left(\widehat{k}+\widetilde{k}\right)\right)}{16 I_{[1]} I_{[2]}^2},\\[5mm]
&	\mathsf{H}_{14}^{[4]}=-\dfrac{9  \left(\widehat{k}-\widetilde{k}\right) \left(I_{[1]} I_{[2]}+3 I_{[3]}\right) \left(\widehat{k} \widetilde{k}-2 k \left(\widehat{k}+\widetilde{k}\right)\right)}{16 I_{[1]} I_{[2]}^2},\quad
	\mathsf{H}_{21}^{[4]}=0,\quad
	\mathsf{H}_{22}^{[4]}=0,\\[5mm]
&	\mathsf{H}_{23}^{[4]}=\dfrac{3  \left(\widehat{k}-\widetilde{k}\right) \left(I_{[1]} I_{[2]}-9 I_{[3]}\right) \left(\widehat{k} \widetilde{k}-2 k \left(\widehat{k}+\widetilde{k}\right)\right)}{16 I_{[1]} I_{[2]}^2},\quad
	\mathsf{H}_{24}^{[4]}=-\dfrac{9  \left(\widehat{k}-\widetilde{k}\right) \left(I_{[1]} I_{[2]}+3 I_{[3]}\right) \left(\widehat{k} \widetilde{k}-2 k \left(\widehat{k}+\widetilde{k}\right)\right)}{16 I_{[1]} I_{[2]}^2},\\[5mm]
&	
	\mathsf{H}_{31}^{[4]}=0,\quad
	\mathsf{H}_{32}^{[4]}=0,\quad
	\mathsf{H}_{33}^{[4]}=0,\quad
	\mathsf{H}_{34}^{[4]}=0
\end{array}
\end{equation}
\begin{equation}
\begin{array}{ccc}
&	\mathsf{H}_{11}^{[5]}=\dfrac{81 I_{[3]} \left(2 I_{[1]} I_{[2]}+9 I_{[3]}\right) \left(8 I_{[1]} I_{[2]}+9 I_{[3]}\right)}{16 I_{[1]}^2 I_{[2]}^3},\quad
	\mathsf{H}_{12}^{[5]}=\dfrac{2187 I_{[3]}^2 \left(2 I_{[1]} I_{[2]}+3 I_{[3]}\right)}{16 I_{[1]}^2 I_{[2]}^3},\quad
	\mathsf{H}_{13}^{[5]}=0,\\[5mm]
&	\mathsf{H}_{14}^{[5]}=0,\quad
	\mathsf{H}_{22}^{[5]}=\dfrac{729 I_{[3]}^2 \left(2 I_{[1]} I_{[2]}+9 I_{[3]}\right)}{16 I_{[1]}^2 I_{[2]}^3},\quad
	\mathsf{H}_{23}^{[5]}=0,\\[5mm]
&	\mathsf{H}_{24}^{[5]}=0,\quad
	\mathsf{H}_{33}^{[5]}=\dfrac{81 I_{[3]} \left(2 I_{[1]} I_{[2]}-9 I_{[3]}\right)^2}{16 I_{[1]}^2 I_{[2]}^3},\quad
	\mathsf{H}_{34}^{[5]}=\dfrac{81 I_{[3]} \left(\dfrac{81 I_{[3]}^2}{I_{[1]}^2}-4 I_{[2]}^2\right)}{16 I_{[2]}^3},\\[5mm]
&	\mathsf{H}_{44}^{[5]}=\dfrac{81 I_{[3]} \left(2 I_{[1]} I_{[2]}+9 I_{[3]}\right)^2}{16 I_{[1]}^2 I_{[2]}^3}
\end{array}
\end{equation}
\begin{equation}
\begin{array}{ccc}
&	\mathsf{H}_{11}^{[6]}=\dfrac{81 I_{[3]} \left(2 I_{[1]} I_{[2]}+9 I_{[3]}\right) \left(7 I_{[1]} I_{[2]}+18 I_{[3]}\right)}{32 I_{[1]}^2 I_{[2]}^3},\quad
	\mathsf{H}_{12}^{[6]}=\dfrac{243 I_{[3]} \left(-2 I_{[1]}^2 I_{[2]}^2+9 I_{[1]} I_{[2]} I_{[3]}+54 I_{[3]}^2\right)}{32 I_{[1]}^2 I_{[2]}^3},\\[5mm]
&	\mathsf{H}_{13}^{[6]}=-\dfrac{81 \sqrt{3} I_{[3]} \left(2 I_{[1]} I_{[2]}+9 I_{[3]}\right)}{32 I_{[1]} I_{[2]}^2},\quad
	\mathsf{H}_{14}^{[6]}=-\dfrac{243 \sqrt{3} I_{[3]} \left(2 I_{[1]} I_{[2]}+3 I_{[3]}\right)}{32 I_{[1]} I_{[2]}^2},\\[5mm]
&	\mathsf{H}_{22}^{[6]}=\dfrac{243 I_{[3]} \left(2 I_{[1]}^2 I_{[2]}^2-15 I_{[1]} I_{[2]} I_{[3]}+54 I_{[3]}^2\right)}{32 I_{[1]}^2 I_{[2]}^3},\quad
	\mathsf{H}_{23}^{[6]}=\dfrac{81 \sqrt{3} I_{[3]} \left(2 I_{[1]} I_{[2]}-9 I_{[3]}\right)}{32 I_{[1]} I_{[2]}^2},\\[5mm]
&	\mathsf{H}_{24}^{[6]}=-\dfrac{81 \sqrt{3} I_{[3]} \left(2 I_{[1]} I_{[2]}+9 I_{[3]}\right)}{32 I_{[1]} I_{[2]}^2},\quad
	\mathsf{H}_{33}^{[6]}=\dfrac{81 I_{[3]} \left(2 I_{[1]}^2 I_{[2]}^2+9 I_{[1]} I_{[2]} I_{[3]}+162 I_{[3]}^2\right)}{32 I_{[1]}^2 I_{[2]}^3},\\[5mm]
&	\mathsf{H}_{34}^{[6]}=\dfrac{81 I_{[3]} \left(-2 I_{[1]}^2 I_{[2]}^2+81 I_{[1]} I_{[2]} I_{[3]}+162 I_{[3]}^2\right)}{32 I_{[1]}^2 I_{[2]}^3},\quad
	\mathsf{H}_{44}^{[6]}=\dfrac{81 I_{[3]} \left(2 I_{[1]} I_{[2]}+9 I_{[3]}\right) \left(13 I_{[1]} I_{[2]}+18 I_{[3]}\right)}{32 I_{[1]}^2 I_{[2]}^3}
\end{array}
\end{equation}
\begin{equation}
\begin{array}{ccc}
&	\mathsf{H}_{11}^{[7]}=\dfrac{81I_{[3]} \left(2I_{[1]}I_{[2]}+9I_{[3]}\right) \left(8I_{[1]}I_{[2]}+9I_{[3]}\right)}{16I_{[1]}^2I_{[2]}^3},\quad
	&\mathsf{H}_{12}^{[7]}=\dfrac{2187I_{[3]}^2 \left(2I_{[1]}I_{[2]}+3I_{[3]}\right)}{16I_{[1]}^2I_{[2]}^3},\\[5mm]
		&\mathsf{H}_{13}^{[7]}=-\dfrac{81 \sqrt{3}I_{[3]} \left(2I_{[1]}I_{[2]}+9I_{[3]}\right)}{16I_{[1]}I_{[2]}^2},\quad
&	\mathsf{H}_{14}^{[7]}=-\dfrac{243 \sqrt{3}I_{[3]} \left(2I_{[1]}I_{[2]}+3I_{[3]}\right)}{16I_{[1]}I_{[2]}^2},
\\[5mm]
	&\mathsf{H}_{22}^{[7]}=\dfrac{729I_{[3]}^2 \left(2I_{[1]}I_{[2]}+9I_{[3]}\right)}{16I_{[1]}^2I_{[2]}^3},\quad
	&\mathsf{H}_{23}^{[7]}=\dfrac{81 \sqrt{3}I_{[3]} \left(2I_{[1]}I_{[2]}-9I_{[3]}\right)}{16I_{[1]}I_{[2]}^2},\\[5mm]
&	\mathsf{H}_{24}^{[7]}=-\dfrac{81 \sqrt{3}I_{[3]} \left(2I_{[1]}I_{[2]}+9I_{[3]}\right)}{16I_{[1]}I_{[2]}^2},\quad
	&\mathsf{H}_{33}^{[7]}=\dfrac{81I_{[3]} \left(2I_{[1]}I_{[2]}-9I_{[3]}\right)^2}{16I_{[1]}^2I_{[2]}^3},\\[5mm]
	&	\mathsf{H}_{34}^{[7]}=\dfrac{81I_{[3]} \left(\dfrac{81I_{[3]}^2}{\text{I1}^2}-4I_{[2]}^2\right)}{16I_{[2]}^3},\quad
&	\mathsf{H}_{44}^{[7]}=\dfrac{81I_{[3]} \left(2I_{[1]}I_{[2]}+9I_{[3]}\right)^2}{16I_{[1]}^2I_{[2]}^3}
\end{array}
\end{equation}
\begin{equation}
\begin{array}{lcl}
&	\mathsf{H}_{11}^{[8]}=0,\quad
	\mathsf{H}_{12}^{[8]}=0,\quad
	\mathsf{H}_{13}^{[8]}=-\dfrac{27 \sqrt{3} I_{[3]}  \left(\widehat{k}-\widetilde{k}\right) \left(I_{[1]} I_{[2]}-9 I_{[3]}\right)\left(\widehat{k} \widetilde{k}-2 k \left(\widehat{k}+\widetilde{k}\right)\right)}{16 I_{[1]}^2 I_{[2]}^3},\\[5mm]
&	\mathsf{H}_{14}^{[8]}=\dfrac{81 \sqrt{3} I_{[3]}  \left(\widehat{k}-\widetilde{k}\right) \left(I_{[1]} I_{[2]}+3 I_{[3]}\right) \left(\widehat{k} \widetilde{k}-2 k \left(\widehat{k}+\widetilde{k}\right)\right)}{16 I_{[1]}^2 I_{[2]}^3},\quad
	\mathsf{H}_{22}^{[8]}=0,\\[5mm]
&	\mathsf{H}_{23}^{[8]}=-\dfrac{27 \sqrt{3} I_{[3]}  \left(\widehat{k}-\widetilde{k}\right) \left(I_{[1]} I_{[2]}-9 I_{[3]}\right) \left(\widehat{k} \widetilde{k}-2 k \left(\widehat{k}+\widetilde{k}\right)\right)}{16 I_{[1]}^2 I_{[2]}^3},\\
	&\mathsf{H}_{24}^{[8]}=\dfrac{81 \sqrt{3} I_{[3]}  \left(\widehat{k}-\widetilde{k}\right) \left(I_{[1]} I_{[2]}+3 I_{[3]}\right) \left(\widehat{k} \widetilde{k}-2 k \left(\widehat{k}+\widetilde{k}\right)\right)}{16 I_{[1]}^2 I_{[2]}^3},\\[5mm]
&	\mathsf{H}_{33}^{[8]}=0,\quad
	\mathsf{H}_{34}^{[8]}=0,\quad
	\mathsf{H}_{44}^{[8]}=0
\end{array}
\end{equation}
\begin{equation}
\begin{array}{ccc}
&	\mathsf{H}_{11}^{[9]}=0,\quad
	\mathsf{H}_{12}^{[9]}=0,\quad
	\mathsf{H}_{13}^{[9]}=-\dfrac{27 \sqrt{3} I_{[3]}  \left(\widehat{k}-\widetilde{k}\right) \left(I_{[1]} I_{[2]}-9 I_{[3]}\right) \left(\widehat{k} \widetilde{k}-2 k \left(\widehat{k}+\widetilde{k}\right)\right)}{32 I_{[1]}^2 I_{[2]}^3},\\[5mm]
&	\mathsf{H}_{14}^{[9]}=\dfrac{81 \sqrt{3} I_{[3]}  \left(\widehat{k}-\widetilde{k}\right) \left(I_{[1]} I_{[2]}+3 I_{[3]}\right) \left(\widehat{k} \widetilde{k}-2 k \left(\widehat{k}+\widetilde{k}\right)\right)}{32 I_{[1]}^2 I_{[2]}^3},\quad
	\mathsf{H}_{22}^{[9]}=0,\\[5mm]
&	\mathsf{H}_{23}^{[9]}=-\dfrac{27 \sqrt{3} I_{[3]}  \left(\widehat{k}-\widetilde{k}\right) \left(I_{[1]} I_{[2]}-9 I_{[3]}\right) \left(\widehat{k} \widetilde{k}-2 k \left(\widehat{k}+\widetilde{k}\right)\right)}{32 I_{[1]}^2 I_{[2]}^3},\\
&	\mathsf{H}_{24}^{[9]}=\dfrac{81 \sqrt{3} I_{[3]}  \left(\widehat{k}-\widetilde{k}\right) \left(I_{[1]} I_{[2]}+3 I_{[3]}\right) \left(\widehat{k} \widetilde{k}-2 k \left(\widehat{k}+\widetilde{k}\right)\right)}{32 I_{[1]}^2 I_{[2]}^3},\\[5mm]
&	\mathsf{H}_{33}^{[9]}=-\dfrac{81 I_{[3]}  \left(\widehat{k}-\widetilde{k}\right) \left(I_{[1]} I_{[2]}-9 I_{[3]}\right) \left(\widehat{k} \widetilde{k}-2 k \left(\widehat{k}+\widetilde{k}\right)\right)}{16 I_{[1]}^2 I_{[2]}^3},\\
&
	\mathsf{H}_{34}^{[9]}=\dfrac{81 I_{[3]}  \left(\widehat{k}-\widetilde{k}\right) \left(I_{[1]} I_{[2]}+9 I_{[3]}\right) \left(\widehat{k} \widetilde{k}-2 k \left(\widehat{k}+\widetilde{k}\right)\right)}{16 I_{[1]}^2 I_{[2]}^3},\\[5mm]
&	\mathsf{H}_{44}^{[9]}=\dfrac{243 I_{[3]}  \left(\widehat{k}-\widetilde{k}\right) \left(I_{[1]} I_{[2]}+3 I_{[3]}\right) \left(\widehat{k} \widetilde{k}-2 k \left(\widehat{k}+\widetilde{k}\right)\right)}{16 I_{[1]}^2 I_{[2]}^3}
\end{array}
\end{equation}

\begin{equation*}
	\mathsf{H}_{13}^{[10]}= \mathsf{H}_{14}^{[10]} = \mathsf{H}_{23}^{[10]}= \mathsf{H}_{24}^{[10]}=0
\end{equation*}
\begin{equation}
\begin{array}{rcl}
	\mathsf{H}_{11}^{[10]}&=&\dfrac{9 I_{[3]}}{64 I_{[1]}^2 I_{[2]}^4} \left[\left(-50 \left(\widehat{k}+\widetilde{k}\right)^3 \overline{k}^5-\left(\widehat{k}+\widetilde{k}\right)^2 \left(100 \widehat{k}^2+359 \widetilde{k} \widehat{k}+100 \widetilde{k}^2\right) \overline{k}^4 + \right.\right. \\[5mm]
&&\left.\left. - \left(\widehat{k}+\widetilde{k}\right) \left(50 \widehat{k}^4+419 \widetilde{k} \widehat{k}^3+339 \widetilde{k}^2 \widehat{k}^2+419 \widetilde{k}^3 \widehat{k}+50 \widetilde{k}^4\right) \overline{k}^3 + \right.\right.\\[5mm]
&&\left.\left. + 2 \widehat{k} \widetilde{k} \left(24 \widehat{k}^4+459 \widetilde{k} \widehat{k}^3+1853 \widetilde{k}^2 \widehat{k}^2+459 \widetilde{k}^3 \widehat{k}+24 \widetilde{k}^4\right) \overline{k}^2 + \right.\right.\\[5mm]
&&\left.\left. + \widehat{k}^2 \widetilde{k}^2 \left(\widehat{k}+\widetilde{k}\right) \left(219 \widehat{k}^2+1283 \widetilde{k} \widehat{k}+219 \widetilde{k}^2\right) \overline{k}+121 \widehat{k}^3 \widetilde{k}^3 \left(\widehat{k}+\widetilde{k}\right)^2\right) \ell^2\right] + \\[5mm]
&& + \left[45 I_{[3]} \left(2 I_{[1]} I_{[2]}+9 I_{[3]}\right) \left(5 I_{[1]} I_{[2]}+9 I_{[3]}\right)\right]/\left(64 I_{[1]}^2 I_{[2]}^3\right),\\[5mm]
	\mathsf{H}_{12}^{[10]}&=&\dfrac{9 I_{[3]}}{64 I_{[1]}^2 I_{[2]}^4} \left[\left(10 \left(\widehat{k}+\widetilde{k}\right)^3 \overline{k}^5+5 \left(\widehat{k}+\widetilde{k}\right)^2 \left(4 \widehat{k}^2+5 \widetilde{k} \widehat{k}+4 \widetilde{k}^2\right) \overline{k}^4 + \right.\right. \\[5mm]
&&\left.\left. + \left(\widehat{k}+\widetilde{k}\right) \left(10 \widehat{k}^4-71 \widetilde{k} \widehat{k}^3-303 \widetilde{k}^2 \widehat{k}^2-71 \widetilde{k}^3 \widehat{k}+10 \widetilde{k}^4\right) \overline{k}^3 + \right.\right. \\[5mm]
&&\left.\left. + 2 \widehat{k} \widetilde{k} \left(6 \widehat{k}^4-9 \widetilde{k} \widehat{k}^3+641 \widetilde{k}^2 \widehat{k}^2-9 \widetilde{k}^3 \widehat{k}+6 \widetilde{k}^4\right) \overline{k}^2 + \right.\right. \\
\end{array}
\end{equation}
\begin{equation*}
\begin{array}{rcl}
	\mathsf{H}_{34}^{[10]}&=&-\dfrac{9}{64 I_{[1]}^2 I_{[2]}^4} \left[\left(2 \left(\widehat{k}+\widetilde{k}\right)^3 \left(4 \widehat{k}^2+3 \widetilde{k} \widehat{k}+4 \widetilde{k}^2\right) \overline{k}^6 + \right.\right.\\[5mm]
&&\left.\left. + \left(\widehat{k}+\widetilde{k}\right)^2 \left(16 \widehat{k}^4+4 \widetilde{k} \widehat{k}^3-63 \widetilde{k}^2 \widehat{k}^2+4 \widetilde{k}^3 \widehat{k}+16 \widetilde{k}^4\right) \overline{k}^5 + \right.\right.\\[5mm]
&&\left.\left. + \left(\widehat{k}+\widetilde{k}\right) \left(8 \widehat{k}^6+6 \widetilde{k} \widehat{k}^5-267 \widetilde{k}^2 \widehat{k}^4-173 \widetilde{k}^3 \widehat{k}^3-267 \widetilde{k}^4 \widehat{k}^2+6 \widetilde{k}^5 \widehat{k}+8 \widetilde{k}^6\right) \overline{k}^4 + \right.\right.\\[5mm]
&&\left.\left. + 2 \widehat{k} \widetilde{k} \left(4 \widehat{k}^6-15 \widetilde{k} \widehat{k}^5+115 \widetilde{k}^2 \widehat{k}^4+461 \widetilde{k}^3 \widehat{k}^3+115 \widetilde{k}^4 \widehat{k}^2-15 \widetilde{k}^5 \widehat{k}+4 \widetilde{k}^6\right) \overline{k}^3 + \right.\right.\\[5mm]
&&\left.\left. - \widehat{k}^2 \widetilde{k}^2 \left(\widehat{k}+\widetilde{k}\right) \left(6 \widehat{k}^4+71 \widetilde{k} \widehat{k}^3+43 \widetilde{k}^2 \widehat{k}^2+71 \widetilde{k}^3 \widehat{k}+6 \widetilde{k}^4\right) \overline{k}^2 + \right.\right.\\[5mm]
&&\left.\left. - \widehat{k}^3 \widetilde{k}^3 \left(\widehat{k}+\widetilde{k}\right)^2 \left(4 \widehat{k}^2+35 \widetilde{k} \widehat{k}+4 \widetilde{k}^2\right) \overline{k} + 2 \widehat{k}^4 \widetilde{k}^4 \left(\widehat{k}+\widetilde{k}\right)^3\right) \ell^2 \right]+ \\[5mm]
&& + \left[45 I_{[3]} \left(-2 I_{[1]}^2 I_{[2]}^2+27 I_{[1]} I_{[2]} I_{[3]}+81 I_{[3]}^2\right)\right]/\left(64 I_{[1]}^2 I_{[2]}^3\right),\\[5mm]
	\mathsf{H}_{44}^{[10]}&=&\dfrac{9}{64 I_{[1]}^2 I_{[2]}^4} \left[\left(2 \left(\widehat{k}+\widetilde{k}\right)^3 \left(12 \widehat{k}^2-\widetilde{k} \widehat{k}+12 \widetilde{k}^2\right) \overline{k}^6 + \right.\right.\\
\end{array}
\end{equation*}

The coefficients of the matrices $\boldsymbol{\mathsf{G}}^{[r]}$ ($r=1,...,10$) are
\begin{equation}
	\mathsf{G}_{ij}^{[1]}=\dfrac{3}{4} \sqrt{3}\mathsf{C}_{ij}
\end{equation}
\begin{equation}
\begin{array}{ccc}
	\mathsf{G}_{11}^{[2]}=\dfrac{9}{2}\left(\mathsf{C}_{11}+\mathsf{C}_{12} \mathcal{D}_{1}+\mathsf{C}_{13} \mathcal{D}_{5}\right),\quad
	\mathsf{G}_{12}^{[2]}=\dfrac{9}{2}\left(\mathsf{C}_{12} \mathcal{D}_{2}+\mathsf{C}_{13} \mathcal{D}_{6}\right),\quad
	\mathsf{G}_{13}^{[2]}=\dfrac{9}{2}\left(\mathsf{C}_{12} \mathcal{D}_{3}+\mathsf{C}_{13} \mathcal{D}_{7}+\mathsf{C}_{13}\right),\\[5mm]
	\mathsf{G}_{14}^{[2]}=\dfrac{9}{2}\left(\mathsf{C}_{12} \mathcal{D}_{4}+\mathsf{C}_{13} \mathcal{D}_{8}\right),\quad
	\mathsf{G}_{21}^{[2]}=\dfrac{9}{2}\left(\mathsf{C}_{12}+\mathsf{C}_{22} \mathcal{D}_{1}+\mathsf{C}_{23}\mathcal{D}_{5}\right),\quad
	\mathsf{G}_{22}^{[2]}=\dfrac{9}{2}\left(\mathsf{C}_{22} \mathcal{D}_{2}+\mathsf{C}_{23} \mathcal{D}_{6}\right),\\[5mm]
	\mathsf{G}_{23}^{[2]}=\dfrac{9}{2}\left(\mathsf{C}_{22} \mathcal{D}_{3}+\mathsf{C}_{23} \mathcal{D}_{7}+\mathsf{C}_{23}\right),\quad
	\mathsf{G}_{24}^{[2]}=\dfrac{9}{2}\left(\mathsf{C}_{22} \mathcal{D}_{4}+\mathsf{C}_{23} \mathcal{D}_{8}\right),\quad
	\mathsf{G}_{31}^{[2]}=\dfrac{9}{2}\left(\mathsf{C}_{13}+\mathsf{C}_{23} \mathcal{D}_{1}+\mathsf{C}_{33} \mathcal{D}_{5}\right),\\[5mm]
	\mathsf{G}_{32}^{[2]}=\dfrac{9}{2}\left(\mathsf{C}_{23} \mathcal{D}_{2}+\mathsf{C}_{33}
	\mathcal{D}_{6}\right),\quad
	\mathsf{G}_{33}^{[2]}=\dfrac{9}{2}\left(\mathsf{C}_{23} \mathcal{D}_{3}+\mathsf{C}_{33} \mathcal{D}_{7}+\mathsf{C}_{33}\right),\quad
	\mathsf{G}_{34}^{[2]}=\dfrac{9}{2}\left(\mathsf{C}_{23} \mathcal{D}_{4}+\mathsf{C}_{33} \mathcal{D}_{8}\right).
\end{array}
\end{equation}
\begin{equation}
\begin{array}{ccc}
	\mathsf{G}_{11}^{[3]}=\dfrac{9}{4}\left(\sqrt{3} \mathsf{C}_{11} \mathcal{D}_{5}+\mathsf{C}_{11}+\mathsf{C}_{12} \mathcal{D}_{1}+\mathsf{C}_{13} \left(\sqrt{3} \mathcal{D}_{1}+\mathcal{D}_{5}\right)\right),\quad
	\mathsf{G}_{12}^{[3]}=\dfrac{9}{4}\left(\sqrt{3} \mathsf{C}_{11} \mathcal{D}_{6}+\mathsf{C}_{12} \mathcal{D}_{2}+\mathsf{C}_{13} \left(\sqrt{3} \mathcal{D}_{2}+\mathcal{D}_{6}+\sqrt{3}\right)\right),\\[5mm]
	\mathsf{G}_{13}^{[3]}=\dfrac{9}{4}\left(\sqrt{3}\mathsf{C}_{11} \mathcal{D}_{7}+\mathsf{C}_{12} \mathcal{D}_{3}+\mathsf{C}_{13} \left(\sqrt{3} \mathcal{D}_{3}+\mathcal{D}_{7}+1\right)\right),\quad
	\mathsf{G}_{14}^{[3]}=\dfrac{9}{4}\left(\sqrt{3} \mathsf{C}_{11} \mathcal{D}_{8}+\mathsf{C}_{12} \left(\mathcal{D}_{4}+\sqrt{3}\right)+\mathsf{C}_{13} \left(\sqrt{3} \mathcal{D}_{4}+\mathcal{D}_{8}\right)\right),\\[5mm]
	\mathsf{G}_{21}^{[3]}=\dfrac{9}{4}\left(\sqrt{3} \mathsf{C}_{12}\mathcal{D}_{5}+\mathsf{C}_{12}+\mathsf{C}_{22} \mathcal{D}_{1}+\mathsf{C}_{23} \left(\sqrt{3} \mathcal{D}_{1}+\mathcal{D}_{5}\right)\right),\quad
	\mathsf{G}_{22}^{[3]}=\dfrac{9}{4}\left(\sqrt{3} \mathsf{C}_{12} \mathcal{D}_{6}+\mathsf{C}_{22} \mathcal{D}_{2}+\mathsf{C}_{23} \left(\sqrt{3} \mathcal{D}_{2}+\mathcal{D}_{6}+\sqrt{3}\right)\right),\\[5mm]
	\mathsf{G}_{23}^{[3]}=\dfrac{9}{4}\left(\sqrt{3} \mathsf{C}_{12}\mathcal{D}_{7}+\mathsf{C}_{22} \mathcal{D}_{3}+\mathsf{C}_{23} \left(\sqrt{3} \mathcal{D}_{3}+\mathcal{D}_{7}+1\right)\right),\quad
	\mathsf{G}_{24}^{[3]}=\dfrac{9}{4}\left(\sqrt{3} \mathsf{C}_{12} \mathcal{D}_{8}+\mathsf{C}_{22} \left(\mathcal{D}_{4}+\sqrt{3}\right)+\mathsf{C}_{23} \left(\sqrt{3} \mathcal{D}_{4}+\mathcal{D}_{8}\right)\right),\\
\end{array}
\end{equation}
\begin{equation}
\mathsf{G}_{ij}^{[4]}=\dfrac{3}{2\ell} \sqrt{3}\mathsf{M}^{*}_{ij}
\end{equation}
\begin{equation}
\begin{array}{ccc}
	\mathsf{G}_{11}^{[5]}=9 \sqrt{3} \left(\mathsf{C}_{11}+2 \mathsf{C}_{12} \mathcal{D}_{1}+2 \mathcal{D}_{5} \left(\mathsf{C}_{13}+\mathsf{C}_{23} \mathcal{D}_{1}\right)+\mathsf{C}_{22} \mathcal{D}_{1}^2+\mathsf{C}_{33} \mathcal{D}_{5}^2\right),\\[5mm]
	\mathsf{G}_{12}^{[5]}=9 \sqrt{3} \left(\mathcal{D}_{2} \left(\mathsf{C}_{12}+\mathsf{C}_{22} \mathcal{D}_{1}+\mathsf{C}_{23} \mathcal{D}_{5}\right)+\mathcal{D}_{6}\left(\mathsf{C}_{13}+\mathsf{C}_{23} \mathcal{D}_{1}+\mathsf{C}_{33} \mathcal{D}_{5}\right)\right),\\[5mm]
	\mathsf{G}_{13}^{[5]}=9 \sqrt{3} \left(\mathcal{D}_{3} \left(\mathsf{C}_{12}+\mathsf{C}_{22} \mathcal{D}_{1}\right)+\mathsf{C}_{13} \left(\mathcal{D}_{7}+1\right)+\mathsf{C}_{23} \left(\mathcal{D}_{1} \mathcal{D}_{7}+\mathcal{D}_{1}+\mathcal{D}_{3} \mathcal{D}_{5}\right)+\mathsf{C}_{33}\mathcal{D}_{5} \left(\mathcal{D}_{7}+1\right)\right),\\[5mm]
	\mathsf{G}_{14}^{[5]}=9\sqrt{3} \left(\mathcal{D}_{4} \left(\mathsf{C}_{12}+\mathsf{C}_{22} \mathcal{D}_{1}+\mathsf{C}_{23} \mathcal{D}_{5}\right)+\mathcal{D}_{8} \left(\mathsf{C}_{13}+\mathsf{C}_{23} \mathcal{D}_{1}+\mathsf{C}_{33} \mathcal{D}_{5}\right)\right),\quad
	\mathsf{G}_{22}^{[5]}=9 \sqrt{3} \left(\mathsf{C}_{22} \mathcal{D}_{2}^2+\mathcal{D}_{6} \left(2 \mathsf{C}_{23} \mathcal{D}_{2}+\mathsf{C}_{33}\mathcal{D}_{6}\right)\right),\\[5mm]
	\mathsf{G}_{23}^{[5]}=9 \sqrt{3} \left(\mathsf{C}_{22} \mathcal{D}_{2} \mathcal{D}_{3}+\mathsf{C}_{23} \left(\mathcal{D}_{2} \mathcal{D}_{7}+\mathcal{D}_{2}+\mathcal{D}_{3} \mathcal{D}_{6}\right)+\mathsf{C}_{33} \mathcal{D}_{6} \left(\mathcal{D}_{7}+1\right)\right),\\
\end{array}
\end{equation}
\begin{equation*}
\begin{array}{rcl}
	\mathsf{G}_{11}^{[6]}&=&\dfrac{9}{4} \left(\left(3 \sqrt{3} \mathsf{C}_{11}+6 \mathsf{C}_{13}+\sqrt{3} \mathsf{C}_{33}\right) \mathcal{D}_{5}^2+2 \left(3 \mathsf{C}_{11}+\left(3 \mathsf{C}_{12}+\sqrt{3} \mathsf{C}_{23}+3 \mathsf{C}_{33}\right) \mathcal{D}_{1}+\sqrt{3} \mathsf{C}_{13} \left(3 \mathcal{D}_{1}+1\right)\right)
	\mathcal{D}_{5} + \right.\\[5mm]
	&&\left. + \sqrt{3} \mathsf{C}_{11}+\mathcal{D}_{1} \left(2 \sqrt{3} \mathsf{C}_{12}+6 \mathsf{C}_{13}+\left(\sqrt{3} \mathsf{C}_{22}+6 \mathsf{C}_{23}+3 \sqrt{3} \mathsf{C}_{33}\right) \mathcal{D}_{1}\right)\right),\\[5mm]
	\mathsf{G}_{12}^{[6]}&=&\dfrac{9}{4} \left(\sqrt{3} \mathsf{C}_{12} \mathcal{D}_{2}+3 \mathsf{C}_{11}\mathcal{D}_{6}+\mathsf{C}_{13} \left(3 \mathcal{D}_{2} \left(\sqrt{3} \mathcal{D}_{5}+1\right)+\sqrt{3} (3 \mathcal{D}_{1}+1) \mathcal{D}_{6}+3 \mathcal{D}_{5} \left(2 \mathcal{D}_{6}+\sqrt{3}\right)+3\right) + \right.\\[5mm]
	&&\left. + \mathcal{D}_{1} \left(\sqrt{3} \mathsf{C}_{22} \mathcal{D}_{2}+3 \mathsf{C}_{12} \mathcal{D}_{6}+3 \mathsf{C}_{33}\left(\sqrt{3} \mathcal{D}_{2}+\mathcal{D}_{6}+\sqrt{3}\right)+\mathsf{C}_{23} \left(6 \mathcal{D}_{2}+\sqrt{3} \mathcal{D}_{6}+3\right)\right) + \right.\\[5mm]
	&&\left. + \mathcal{D}_{5} \left(3 \mathsf{C}_{12} \mathcal{D}_{2}+\mathsf{C}_{33} \left(3 \mathcal{D}_{2}+\sqrt{3} \mathcal{D}_{6}+3\right)+\sqrt{3} (\mathsf{C}_{23} \mathcal{D}_{2}+3\mathsf{C}_{11} \mathcal{D}_{6})\right)\right),
	\\[5mm]
	\mathsf{G}_{13}^{[6]}&=&\dfrac{9}{4} \left(\sqrt{3} \mathsf{C}_{12} \mathcal{D}_{3}+3 \mathsf{C}_{11} \mathcal{D}_{7}+\mathcal{D}_{5} \left(\left(3 \mathsf{C}_{12}+\sqrt{3} \mathsf{C}_{23}+3 \mathsf{C}_{33}\right) \mathcal{D}_{3}+\sqrt{3} (\mathcal{D}_{7} \mathsf{C}_{33}+\mathsf{C}_{33}+3\mathsf{C}_{11} \mathcal{D}_{7})\right) + \right.\\
\end{array}
\end{equation*}
\begin{equation}
\begin{array}{rcl}
	\mathsf{G}_{33}^{[6]}&=&\dfrac{9}{4} \left(\sqrt{3} \mathsf{C}_{22} \mathcal{D}_{3}^2+2 \mathsf{C}_{23} \left(3 \mathcal{D}_{3}+\sqrt{3} (\mathcal{D}_{7}+1)\right) \mathcal{D}_{3}+\mathsf{C}_{33} \left(3 \sqrt{3} \mathcal{D}_{3}^2+6 (\mathcal{D}_{7}+1) \mathcal{D}_{3}+\sqrt{3}\left(\mathcal{D}_{7}+1\right)^2\right) + \right.\\[5mm]
	&&\left. + 3\mathcal{D}_{7}\left(2\mathsf{C}_{12}\mathcal{D}_{3}+\sqrt{3}\mathsf{C}_{11}\mathcal{D}_{7}+2\mathsf{C}_{13}\left(\sqrt{3}\mathcal{D}_{3}+\mathcal{D}_{7}+1\right)\right)\right),\\[5mm]
	\mathsf{G}_{34}^{[6]}&=&\dfrac{9}{4} \left(\mathsf{C}_{22} \mathcal{D}_{3} \left(\sqrt{3} \mathcal{D}_{4}+3\right)+3\sqrt{3} \mathsf{C}_{12} \mathcal{D}_{7}+3 \mathcal{D}_{4} \left(\left(\mathsf{C}_{12}+\sqrt{3} \mathsf{C}_{13}\right) \mathcal{D}_{7}+\mathsf{C}_{33} \left(\sqrt{3} \mathcal{D}_{3}+\mathcal{D}_{7}+1\right)\right) + \right.\\[5mm]
	&&\left. +\left(3 (\mathsf{C}_{12}+\mathsf{C}_{33}) \mathcal{D}_{3}+3 \mathsf{C}_{13} \left(\sqrt{3}\mathcal{D}_{3}+2 \mathcal{D}_{7}+1\right)+\sqrt{3} (\mathcal{D}_{7} \mathsf{C}_{33}+\mathsf{C}_{33}+3 \mathsf{C}_{11} \mathcal{D}_{7})\right) \mathcal{D}_{8} + \right.\\[5mm]
	&&\left. + \mathsf{C}_{23} \left(\left(\sqrt{3} \mathcal{D}_{4}+3\right)\left(\mathcal{D}_{7}+1\right)+\mathcal{D}_{3} \left(6\mathcal{D}_{4}+\sqrt{3}(\mathcal{D}_{8}+3)\right)\right)\right),\\[5mm]
		\mathsf{G}_{44}^{[6]}&=&\dfrac{9}{4} \left(3 \sqrt{3} \mathsf{C}_{33} \mathcal{D}_{4}^2+6 \left(\mathsf{C}_{12}+\sqrt{3} \mathsf{C}_{13}+\mathsf{C}_{33}\right) \mathcal{D}_{8} \mathcal{D}_{4}+\left(3 \sqrt{3} \mathsf{C}_{11}+6 \mathsf{C}_{13}+\sqrt{3} \mathsf{C}_{33}\right)\mathcal{D}_{8}^2 + \right.\\[5mm]
	&&\left. + \mathsf{C}_{22} \left(\mathcal{D}_{4} \left(\sqrt{3} \mathcal{D}_{4}+6\right)+3 \sqrt{3}\right)+6 \sqrt{3} \mathsf{C}_{12} \mathcal{D}_{8}+2 \mathsf{C}_{23} \left(3 \mathcal{D}_{4}^2+\sqrt{3} \left(\mathcal{D}_{8}+3\right) \mathcal{D}_{4}+3 \mathcal{D}_{8}\right)\right).
\end{array}
\end{equation}
\begin{equation*}
\begin{array}{rcl}
	\mathsf{G}_{11}^{[7]}&=&9 \left(\mathcal{D}_{1} \left(2 \sqrt{3} \mathsf{C}_{12}+3 \mathsf{C}_{13}+\sqrt{3} \mathsf{C}_{22} \mathcal{D}_{1}+3 \mathsf{C}_{23} \mathcal{D}_{1}\right)+\mathsf{C}_{11} \left(3 \mathcal{D}_{5}+\sqrt{3}\right) + \right.\\[5mm]
	&&\left. + \mathcal{D}_{5} \left(3 \mathcal{D}_{5} \mathsf{C}_{13}+2 \sqrt{3} \mathsf{C}_{13}+3 \mathsf{C}_{12}	\mathcal{D}_{1}+2 \sqrt{3} \mathsf{C}_{23} \mathcal{D}_{1}+3 \mathsf{C}_{33} \mathcal{D}_{1}+\sqrt{3} \mathsf{C}_{33} \mathcal{D}_{5}\right)\right),\\[5mm]
	\mathsf{G}_{12}^{[7]}&=&\dfrac{9}{2} \left(2 \sqrt{3} \mathsf{C}_{12} \mathcal{D}_{2}+3 \mathsf{C}_{33} \mathcal{D}_{5}+3 \mathsf{C}_{11} \mathcal{D}_{6}+\mathcal{D}_{5} \left(3 \mathsf{C}_{12}
	\mathcal{D}_{2}+2 \sqrt{3} \mathsf{C}_{23} \mathcal{D}_{2}+3 \mathsf{C}_{33} \mathcal{D}_{2}+2 \sqrt{3} \mathsf{C}_{33} \mathcal{D}_{6}\right) + \right.\\[5mm]
	&&\left. + \mathsf{C}_{13} \left(3 \mathcal{D}_{2}+6 \mathcal{D}_{5} \mathcal{D}_{6}+2 \sqrt{3} \mathcal{D}_{6}+3\right)+\mathcal{D}_{1} \left(2 \sqrt{3} \mathsf{C}_{22} \mathcal{D}_{2}+3
	\left(\mathsf{C}_{12}+\mathsf{C}_{33}\right) \mathcal{D}_{6}+\mathsf{C}_{23} \left(6 \mathcal{D}_{2}+2 \sqrt{3} \mathcal{D}_{6}+3\right)\right)\right),\\[5mm]
	\mathsf{G}_{13}^{[7]}&=&\dfrac{9}{2} \left(2 \sqrt{3} \mathsf{C}_{12} \mathcal{D}_{3}+3 \mathsf{C}_{11} \mathcal{D}_{7}+\mathcal{D}_{1} \left(2 \sqrt{3} \mathsf{C}_{22} \mathcal{D}_{3}+3 \mathsf{C}_{12}
	\mathcal{D}_{7}+3 \mathsf{C}_{33} (\mathcal{D}_{7}+1)\right) + \right.\\[5mm]
	&&\left. + \mathcal{D}_{5} \left(3 \left(\mathsf{C}_{12}+\mathsf{C}_{33}\right) \mathcal{D}_{3}+2 \sqrt{3} \mathsf{C}_{33} \left(\mathcal{D}_{7}+1\right)\right)+2 \mathsf{C}_{23} \left(\mathcal{D}_{3} \left(3 \mathcal{D}_{1}+\sqrt{3} \mathcal{D}_{5}\right)+\sqrt{3} \mathcal{D}_{1}(\mathcal{D}_{7}+1)\right) + \right.\\
\end{array}
\end{equation*}
\begin{equation}
\begin{array}{rcl}
	\mathsf{G}_{34}^{[7]}&=&\dfrac{9}{2} \left(\mathsf{C}_{22} \mathcal{D}_{3} \left(2 \sqrt{3} \mathcal{D}_{4}+3\right)+3 \mathcal{D}_{4} \left(\mathsf{C}_{33}+\left(\mathsf{C}_{12}+\mathsf{C}_{33}\right) \mathcal{D}_{7}\right)+3 \mathsf{C}_{13} \mathcal{D}_{8} + \right.\\[5mm]
	&&\left. + \left(3 \mathsf{C}_{12} \mathcal{D}_{3}+6 \mathsf{C}_{13}
	\mathcal{D}_{7}+\mathsf{C}_{33} \left(3 \mathcal{D}_{3}+2 \sqrt{3} (\mathcal{D}_{7}+1)\right)\right) \mathcal{D}_{8} + \right.\\[5mm]
	&&\left. + \mathsf{C}_{23} \left(3 \mathcal{D}_{7}+2 \mathcal{D}_{4} \left(3 \mathcal{D}_{3}+\sqrt{3} \left(\mathcal{D}_{7}+1\right)\right)+2 \sqrt{3} \mathcal{D}_{3} \mathcal{D}_{8}+3\right)\right),\\[5mm]
	\mathsf{G}_{44}^{[7]}&=&9 \left(\mathsf{C}_{22}
	\mathcal{D}_{4} \left(\sqrt{3} \mathcal{D}_{4}+3\right)+\mathcal{D}_{8} \left(3 \left(\mathsf{C}_{12}+\mathsf{C}_{33}\right) \mathcal{D}_{4}+3 \mathsf{C}_{13} \mathcal{D}_{8}+\sqrt{3} \mathsf{C}_{33} \mathcal{D}_{8}\right) + \right.\\[5mm]
	&&\left. + \mathsf{C}_{23} \left(3 \mathcal{D}_{4}^2+2 \sqrt{3} \mathcal{D}_{8} \mathcal{D}_{4}+3\mathcal{D}_{8}\right)\right).
\end{array}
\end{equation}

\begin{equation}
\begin{array}{ccc}
	\mathsf{G}_{11}^{[8]}=\dfrac{18}{\ell} \left(\mathcal{D}_{1} \mathsf{M}^{*}_{21}+\mathcal{D}_{5} \mathsf{M}^{*}_{31}+\mathsf{M}^{*}_{11}\right),\quad
	\mathsf{G}_{12}^{[8]}=\dfrac{9}{\ell} \left(\mathcal{D}_{1} \mathsf{M}^{*}_{22}+\mathcal{D}_{2} \mathsf{M}^{*}_{21}+\mathcal{D}_{5} \mathsf{M}^{*}_{32}+\mathcal{D}_{6} \mathsf{M}^{*}_{31}+\mathsf{M}^{*}_{12}\right),\\[5mm]
	\mathsf{G}_{13}^{[8]}=\dfrac{9}{\ell} \left(\mathcal{D}_{1} \mathsf{M}^{*}_{23}+\mathcal{D}_{3} \mathsf{M}^{*}_{21}+\mathcal{D}_{5} \mathsf{M}^{*}_{33}+\mathcal{D}_{7} \mathsf{M}^{*}_{31}+\mathsf{M}^{*}_{13}+\mathsf{M}^{*}_{31}\right),\quad
	\mathsf{G}_{14}^{[8]}=\dfrac{9}{\ell} \left(\mathcal{D}_{1} \mathsf{M}^{*}_{24}+\mathcal{D}_{4} \mathsf{M}^{*}_{21}+\mathcal{D}_{5} \mathsf{M}^{*}_{34}+\mathcal{D}_{8} \mathsf{M}^{*}_{31}+\mathsf{M}^{*}_{14}\right),\\[5mm]
	\mathsf{G}_{22}^{[8]}=\dfrac{18}{\ell} \left(\mathcal{D}_{2} \mathsf{M}^{*}_{22}+\mathcal{D}_{6} \mathsf{M}^{*}_{32}\right),\quad
	\mathsf{G}_{23}^{[8]}=\dfrac{9}{\ell} \left(\mathcal{D}_{2} \mathsf{M}^{*}_{23}+\mathcal{D}_{3} \mathsf{M}^{*}_{22}+\mathcal{D}_{6} \mathsf{M}^{*}_{33}+\mathcal{D}_{7} \mathsf{M}^{*}_{32}+\mathsf{M}^{*}_{32}\right),\\[5mm]
	\mathsf{G}_{24}^{[8]}=\dfrac{9}{\ell} \left(\mathcal{D}_{2} \mathsf{M}^{*}_{24}+\mathcal{D}_{4} \mathsf{M}^{*}_{22}+\mathcal{D}_{6} \mathsf{M}^{*}_{34}+\mathcal{D}_{8} \mathsf{M}^{*}_{32}\right),\quad
	\mathsf{G}_{33}^{[8]}=\dfrac{18}{\ell} \left(\mathcal{D}_{3} \mathsf{M}^{*}_{23}+\mathcal{D}_{7} \mathsf{M}^{*}_{33}+\mathsf{M}^{*}_{33}\right),\\[5mm]
	\mathsf{G}_{34}^{[8]}=\dfrac{9}{\ell} \left(\mathcal{D}_{3} \mathsf{M}^{*}_{24}+\mathcal{D}_{4} \mathsf{M}^{*}_{23}+\mathcal{D}_{7} \mathsf{M}^{*}_{34}+\mathcal{D}_{8} \mathsf{M}^{*}_{33}+\mathsf{M}^{*}_{34}\right),\quad
	\mathsf{G}_{44}^{[8]}=\dfrac{18}{\ell} \left(\mathcal{D}_{4} \mathsf{M}^{*}_{24}+\mathcal{D}_{8} \mathsf{M}^{*}_{34}\right)
\end{array}
\end{equation}

\begin{equation}
\begin{array}{ccc}
	\mathsf{G}_{11}^{[9]}=\dfrac{9}{\ell} \left(\mathcal{D}_{1} \left(\mathsf{M}^{*}_{21}+\sqrt{3} \mathsf{M}^{*}_{31}\right)+\sqrt{3} \mathcal{D}_{5} \mathsf{M}^{*}_{11}+\mathcal{D}_{5} \mathsf{M}^{*}_{31}+\mathsf{M}^{*}_{11}\right),\\[5mm]
	\mathsf{G}_{12}^{[9]}=\dfrac{9}{2} \left(\mathcal{D}_{1} \mathsf{M}^{*}_{22}+\sqrt{3} \left(\mathcal{D}_{1} \mathsf{M}^{*}_{32}+\mathsf{M}^{*}_{31}\right)+\mathcal{D}_{2}
	\left(\mathsf{M}^{*}_{21}+\sqrt{3} \mathsf{M}^{*}_{31}\right)+\sqrt{3} \mathcal{D}_{5} \mathsf{M}^{*}_{12}+\mathcal{D}_{5} \mathsf{M}^{*}_{32}+\mathcal{D}_{6} \left(\sqrt{3} \mathsf{M}^{*}_{11}+\mathsf{M}^{*}_{31}\right)+\mathsf{M}^{*}_{12}\right),\\[5mm]
	\mathsf{G}_{13}^{[9]}=\dfrac{9}{2\ell} \left(\mathcal{D}_{1} \mathsf{M}^{*}_{23}+\sqrt{3} \mathcal{D}_{1}
	\mathsf{M}^{*}_{33}+\mathcal{D}_{3} \mathsf{M}^{*}_{21}+\sqrt{3} \mathcal{D}_{3} \mathsf{M}^{*}_{31}+\sqrt{3} \mathcal{D}_{5} \mathsf{M}^{*}_{13}+\mathcal{D}_{5} \mathsf{M}^{*}_{33}+\mathcal{D}_{7} \left(\sqrt{3} \mathsf{M}^{*}_{11}+\mathsf{M}^{*}_{31}\right)+\mathsf{M}^{*}_{13}+\mathsf{M}^{*}_{31}\right),\\[5mm]
	\mathsf{G}_{14}^{[9]}=\dfrac{9}{2\ell} \left(\sqrt{3} \left(\mathcal{D}_{1}
	\mathsf{M}^{*}_{34}+\mathcal{D}_{4} \mathsf{M}^{*}_{31}\right)+\mathcal{D}_{1} \mathsf{M}^{*}_{24}+\left(\mathcal{D}_{4}+\sqrt{3}\right) \mathsf{M}^{*}_{21}+\sqrt{3} \mathcal{D}_{5} \mathsf{M}^{*}_{14}+\mathcal{D}_{5} \mathsf{M}^{*}_{34}+\mathcal{D}_{8} \left(\sqrt{3} \mathsf{M}^{*}_{11}+\mathsf{M}^{*}_{31}\right)+\mathsf{M}^{*}_{14}\right),\\[5mm]
	\mathsf{G}_{22}^{[9]}=\dfrac{9}{\ell} \left(\mathcal{D}_{2}
	\mathsf{M}^{*}_{22}+\sqrt{3} \left(\mathcal{D}_{2}+1\right) \mathsf{M}^{*}_{32}+\mathcal{D}_{6} \left(\sqrt{3} \mathsf{M}^{*}_{12}+\mathsf{M}^{*}_{32}\right)\right),\\
\end{array}
\end{equation}

\begin{equation}
\begin{array}{rcl}
	\mathsf{G}_{11}^{[10]}&=&\dfrac{3 \sqrt{3}}{\ell^{2}}\mathsf{A}^{*}_{11} + \dfrac{5}{8} \sqrt{3} \left(\mathcal{D}_{5}^2 \left(\mathsf{C}_{11}+\mathsf{C}_{33}\right)+\mathsf{C}_{11}+2 \mathsf{C}_{12} \mathcal{D}_{1}+2 \mathcal{D}_{5} \left(\mathcal{D}_{1} \left(\mathsf{C}_{13}+\mathsf{C}_{23}\right)+\mathsf{C}_{13}\right)+\mathcal{D}_{1}^2 \left(\mathsf{C}_{22}+\mathsf{C}_{33}\right)\right),\\[5mm]
	\mathsf{G}_{12}^{[10]}&=&\dfrac{3 \sqrt{3}}{\ell^{2}}\mathsf{A}^{*}_{12} + \dfrac{5}{8} \sqrt{3} \left(\mathcal{D}_{6}
	\left(\mathsf{C}_{11} \mathcal{D}_{5}+\mathcal{D}_{1} \left(\mathsf{C}_{13}+\mathsf{C}_{23}\right)+\mathsf{C}_{13}\right)+\mathcal{D}_{2} \left(\mathsf{C}_{12}+\mathcal{D}_{5} \left(\mathsf{C}_{13}+\mathsf{C}_{23}\right)+\mathsf{C}_{22} \mathcal{D}_{1}\right)+\mathsf{C}_{13} \mathcal{D}_{5} + \right.\\[5mm]
	&&\left. + \mathsf{C}_{33} \left(\mathcal{D}_{1} \mathcal{D}_{2}+\mathcal{D}_{1}+\mathcal{D}_{5} \mathcal{D}_{6}\right)\right),\\[5mm]
	\mathsf{G}_{13}^{[10]}&=&\dfrac{3 \sqrt{3}}{\ell^{2}}\mathsf{A}^{*}_{13} + \dfrac{5}{8}\sqrt{3} \left(\mathcal{D}_{7} \left(\mathcal{D}_{5} \left(\mathsf{C}_{11}+\mathsf{C}_{33}\right)+\mathcal{D}_{1} \left(\mathsf{C}_{13}+\mathsf{C}_{23}\right)+\mathsf{C}_{13}\right)+\mathsf{C}_{12} \mathcal{D}_{3}+\mathsf{C}_{13} \mathcal{D}_{3} \mathcal{D}_{5}+\mathsf{C}_{13} + \right.\\[5mm]
	&&\left. +\mathcal{D}_{1} \mathcal{D}_{3} \left(\mathsf{C}_{22}+\mathsf{C}_{33}\right)+ \mathsf{C}_{23} \mathcal{D}_{1}+\mathsf{C}_{23}
	\mathcal{D}_{3} \mathcal{D}_{5}+\mathsf{C}_{33} \mathcal{D}_{5}\right),\\
\end{array}
\end{equation}

\end{document}